\documentclass[twocolumn,superscriptaddress,amssymb]{revtex4-2}

\usepackage[colorlinks=true,linkcolor=magenta,citecolor=magenta,urlcolor=magenta,linktocpage=true]{hyperref}
\usepackage[usenames,dvipsnames]{color}
\usepackage{psfrag,graphicx}
\usepackage{dcolumn}
\usepackage{bm}
\usepackage{amsfonts,amssymb,amsmath}
\usepackage{slashed}
\usepackage{mathtools}
\usepackage{amsmath}
\usepackage{todonotes}

\newcommand{\ket}[1]{\left |#1 \right\rangle}
\newcommand{\bra}[1]{\left \langle #1 \right |}
\newcommand{\ketbra}[2]{\left|#1\right\rangle \left\langle #2\right|}
\newcommand{\braket}[2]{\left \langle #1|#2\right\rangle }
\newcommand{\tr}[1]{\text{tr}(#1)}

\begin{document}
\title{Modeling the unphysical pseudomode model with physical ensembles: \\simulation, mitigation, and restructuring of non-Markovian quantum noise}
\author{Mauro Cirio}
\email{cirio.mauro@gmail.com}
\affiliation{Graduate School of China Academy of Engineering Physics, Haidian District, Beijing, 100193, China}
\author{Si Luo}
\affiliation{Graduate School of China Academy of Engineering Physics, Haidian District, Beijing, 100193, China}
\author{Pengfei Liang}
\affiliation{Graduate School of China Academy of Engineering Physics, Haidian District, Beijing, 100193, China}
\author{Franco Nori}
\affiliation{Theoretical Physics Laboratory, Cluster for Pioneering Research, RIKEN, Wakoshi, Saitama 351-0198, Japan}
\affiliation{Quantum Computing Center, RIKEN, Wakoshi, Saitama, 351-0198, Japan}
\affiliation{Physics Department, University of Michigan, Ann Arbor, MI 48109-1040, USA}
\author{Neill Lambert}
\email{nwlambert@gmail.com}
\affiliation{Theoretical Physics Laboratory, Cluster for Pioneering Research, RIKEN, Wakoshi, Saitama 351-0198, Japan}

\date{\today}
\begin{abstract}
{The influence of a Gaussian environment on a quantum system can be  described by effectively replacing the continuum with a discrete set of ancillary quantum and classical degrees of freedom. This defines a pseudomode model which can be used to classically simulate the reduced system dynamics. Here, we consider an alternative point of view and analyze the potential benefits of an analog or digital quantum simulation of the pseudomode model itself. Superficially, such a direct experimental implementation is, in general, impossible due to the unphysical properties of the effective degrees of freedom involved.  However, we show that the effects of the unphysical pseudomode model can still be reproduced using measurement results over an ensemble of physical systems involving ancillary harmonic modes and an optional stochastic driving field. This is done by introducing an extrapolation technique whose efficiency  is  limited by stability against imprecision in the measurement data. We examine how such a simulation would allow us to (i) perform accurate quantum simulation of the effects of complex non-perturbative and non-Markovian environments in regimes that are challenging for classical simulation, (ii)  conversely, mitigate potential unwanted non-Markovian noise present in quantum devices, and (iii) restructure some of some of the properties of a given physical bath, such as its temperature.}
\end{abstract}
\maketitle
\section{Introduction}
The definition of a physical system requires a distinction between \emph{internal} and \emph{external} degrees of freedom. The interaction between these constituents causes the internal, or \emph{closed}, system dynamics to become \emph{open}, i.e., affected by the external environment. For example, information stored in the system can propagate towards a measurement device or it can simply be lost, or \emph{dissipated}, in the continuum of a thermal bath. Interestingly, this distinction between internal and external degrees of freedom becomes more ambiguous as the coupling between the system and the environment increases. In this case, information can be coherently exchanged between the system and the bath before the dynamics eventually stabilizes into a state characterizing \emph{hybridization} properties of the whole system-environment. 

As a consequence, a mathematical model for these, so-called, non-Markovian regimes \cite{Petruccione,Gardiner}, must also include a characterization of the external continuum alongside the system. This can be achieved by selecting the most relevant physical environmental degrees of freedom engaging in the interaction using, for example,  the polaron transformation \cite{Holstein1,Holstein2,Jackson,Silbey,Silbey2,Weiss,Jang,Jang2,PhysRevLett.103.146404,McCutcheon,Jang3,PhysRevB.83.165101,Kolli,PollockNazir,PollockThesis,Xu}, chain mappings \cite{Chin_2010,PhysRevLett.105.050404,Woods,PhysRevB.101.155134}, or the reaction coordinate model \cite{Garg,Martinazzo,iles2014environmental,Woods,PhysRevB.97.205405,Melina}. In parallel, it is also possible to follow an \emph{effective} philosophy, i.e. to focus on the \emph{influence} of the environment on the system rather than on the full system-bath dynamics. This leads to effective master equations for the system dynamics which can include deterministic \cite{Nakajima,Zwanzig} and stochastic  \cite{Diosi_2,Diosi_1,PhysRevA.58.1699,PhysRevLett.82.1801} \emph{memory kernels}, or use the path-integral \cite{Vernon,Caldeira_Leggett_1,Caldeira_Leggett_2,Caldeira_0,Bonig,PhysRevB.78.235311,Jin_Matisse,PhysRevLett.109.170402,Xiong2015} or the canonical formalism \cite{Petruccione, Ma_2012, Aurell,JianMa,Cirio2021}  to derive time-local differential equations involving additional ancillary degrees of freedom in Liouville space \cite{Stockburger_JCP1999,Stockburger_CP2001,PhysRevLett.88.170407,Shao,STOCKBURGER2004159,PhysRevLett.100.230402,Stockburger_2016,PhysRevLett.123.050601,Chernyak1,Chernyak2,PhysRevE.102.062134,Yun-AnShao} or within the hierarchical equation of motion formalism (HEOM) \cite{Tanimura_3,Tanimura_1,YAN2004216,Ishizaki_1,Tanimura_2,Ishizaki_2,PhysRevLett.104.250401,doi:10.1143/JPSJ.81.063301,Moix,Tanimura_2014,Yan_Shao,Hsieh_1,Hsieh_2,Lambert_Bofin,Tanimura_2020,Tanimura_2021}. Among this last category of methods is the pseudomode model \cite{PhysRevA.50.3650,PhysRevA.55.2290,PhysRevLett.110.086403,PhysRevB.89.165105,PhysRevB.92.245125,Schwarz,Dorda,Mascherpa,Lemmer_2018,Tamascelli,Lambert,PhysRevLett.123.090402,PhysRevResearch.2.043058,PhysRevA.101.052108,PhysRevResearch.5.033011} which encodes all non-perturbative effects of a Gaussian environment (such as Bosonic and Fermionic baths initially at thermal equilibrium) using ancillary harmonic modes and, possibly, classical stochastic processes \cite{LuoSi}. 

An interesting characteristic  of the pseudomode model ($\text{PM}$) is that,  while its ancillary degrees of freedom  superficially resemble simple physical harmonic modes and driving fields, in reality they can take on unphysical features and exhibit unphysical dynamics. This combination of superficial simplicity and unintuitive unphysicality is a consequence of the \emph{effective} nature of the model which focuses on \emph{reproducing the reduced system dynamics after averaging over the ancillary degrees of freedom}. 
This point of view can be more formally described as a map $E_\text{phys}\mapsto \text{PM}_\text{can}$ between a \emph{physical environment} $E_\text{phys}$ and its corresponding \emph{canonical pseudomode model} $\text{PM}_\text{can}$, whose unphysical parameters allow us to reduce the number of pseudomodes, and hence computational resources, needed to represent a given environment. In other words, the collection of these models directly correspond to real physical environments.
In this context, pseudomodes have primarily been seen as a numerical tool that allows us to simulate non-Markovian and non-perturbative environments \cite{Chen2015,Fruchtman} in a transparent and simple way.  \emph{Here, we enlarge the domain of applicability of the method by considering how pseudomodes can be used as ancilla degrees of freedom in experiments as a means to simulate, mitigate, or restructure physical environments.}

To develop a framework encompassing all these applications, it is useful to consider pseudomode models \text{PM} with fully \emph{unconstrained} parameters, i.e., corresponding to a map $E_\text{PM}\mapsto \text{PM}$ which generalizes the canonical case, $\text{PM}_{\text{can}}$, to simulate a potential unphysical \emph{pseudo-environment} $E_\text{PM}$.

To give meaning to this analysis, it is necessary to  analyze what applications would be made available given  the possibility to engineer these generalized environments in experiments (either via analog or digital techniques).

\begin{figure}[t!]
\centering
\includegraphics[width=\linewidth]{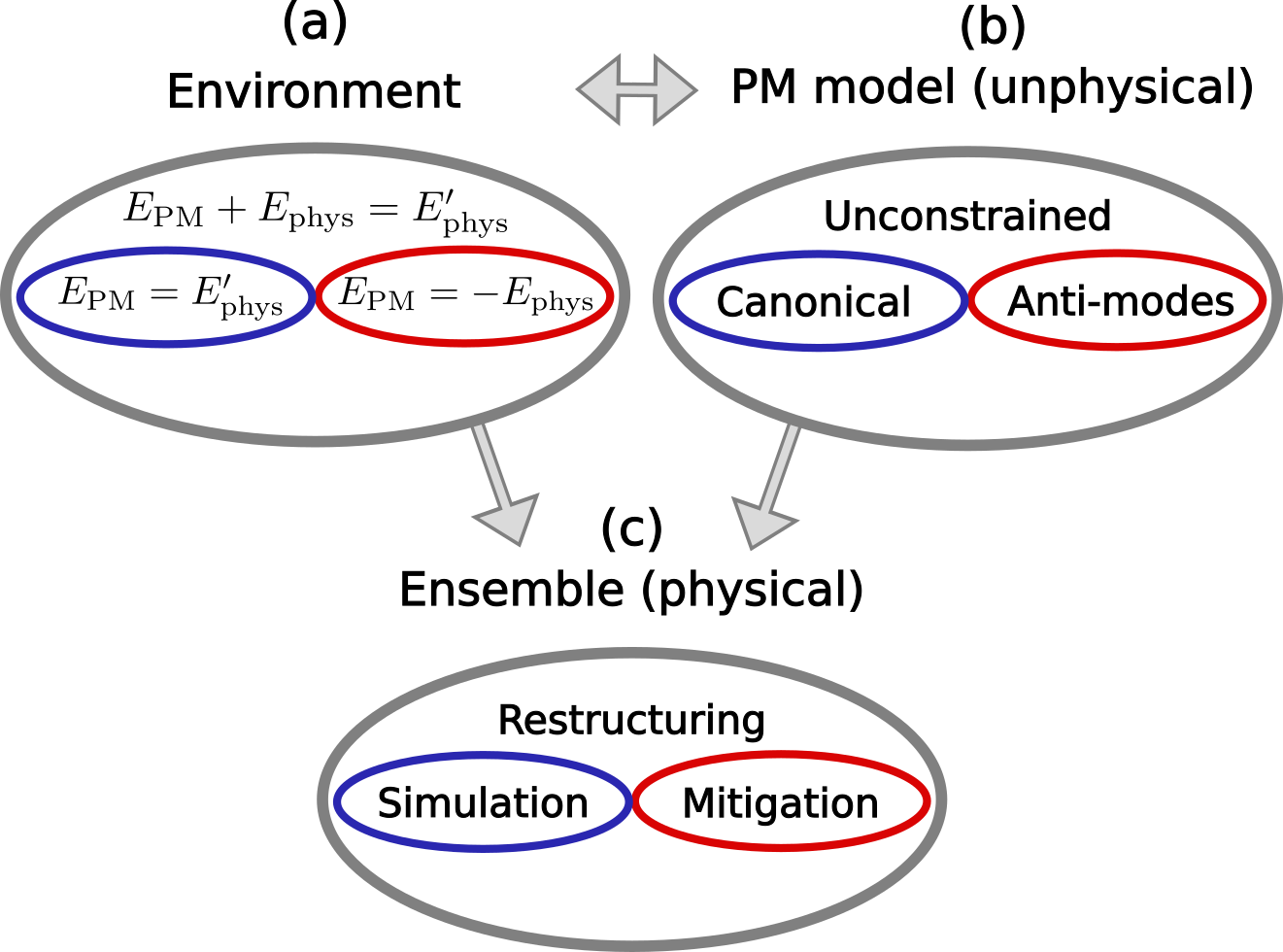}
 \caption{\label{fig:1}Logical structure of the presented framework. 
The parameters of a pseudomode model, $\text{PM}$, can be extended to describe environments, $E_\text{PM}$, which are physical, $E'_\text{phys}$, when coupled to a system alongside another physical environment $E_\text{phys}$ as in (a). In the most general case, this pseudo-environment $E_\text{PM}$ corresponds to an abstract unconstrained pseudomode model (b) which can be realized with a physical ensemble leading to a ``restructuring'' of the original bath $E_\text{phys}$ (c).  More specific applications are found (blue ellipses) 
 by imposing  $E_\text{phys}=0$ (a) which lead to a simulation of the environment $E'_\text{phys}$ (b) through the canonical pseudomode model (c). Similarly (red ellipses), imposing $E'_\text{phys}=0$ (a), corresponds to a mitigation of the effects of the bath $E_\text{phys}$ (b) on the system through what we call ``anti-modes'' (c).}
\end{figure}

As an intuitive example, it is useful to consider what happens when such environments are coupled to the system alongside some originally present environmental bath $E_\text{phys}$. We can then restrict the analysis to the situation in which the resulting `total' environment $E'_\text{phys}$ is physical, i.e., 
\begin{equation}
\label{eq:1}
\begin{array}{ccccc}
    E_\text{PM}&+&E_\text{phys}&=&E'_\text{phys}\\
    \downarrow&&\downarrow&&\downarrow\\
      \text{pseudomode bath}&&\text{original bath}&&\text{``restructured''}\\  
\text{(unphysical)}&&\text{(physical)}&&\text{(physical)}
    \end{array}
\end{equation}
as exemplified in Fig.~\ref{fig:1}.  This simple setting can be  used to show different benefits of a possible simulation of the unphysical pseudomode environment $E_\text{PM}$ (henceforth dubbed \emph{pseudo-environment}).
Specifically, we will analyze three different applications.

\begin{itemize}
\item \emph{Quantum simulation.} First, we note that, by imposing $E_\text{phys}=0$ in the equation above, we recover the canonical pseudomode model $\text{PM}_\text{can}$ in which the pseudo-environment directly correspond to a physical one. As a consequence,  a protocol able to engineer the canonical pseudo-environment would correspond to performing a \emph{quantum simulation of non-Markovian quantum noise} by using physical analogs or digital simulations \cite{PhysRevA.75.032329,BalutaNori,PhysRevA.86.023837,PhysRevX.4.031043,RevModPhys.86.153,PhysRevLett.115.240501,PhysRevB.92.174507,Martinis,PhysRevX.7.011016,PhysRevA.101.012328}.

\item \emph{Mitigation of non-Markovian noise.} Second, 
by imposing $E'_\text{phys}=0$, the equation above tells us that the pseudo-environment $E_\text{PM}$ is the one which exactly counteracts all the effects of the original $E_\text{phys}$, resulting in the possibility to \emph{mitigate the effects non-Markovian noise} generated by the physical environment $E_\text{phys}$.
As a consequence, in this case the simulation of such a pseudo-environment can  be interpreted as an extrapolation/error-mitigation procedure on the lines of \cite{PhysRevX.7.021050,PhysRevLett.119.180509,PhysRevX.8.031027,PhysRevA.99.012334,9259940,PhysRevA.103.012611,PhysRevE.104.035309,ZhenyuCai,PRXQuantum.2.040330,Cai, PhysRevResearch.4.043140,Youngseok, Qin2023,PRXQuantum.4.040329}. The generality of these mitigation techniques usually rely on an underlying error model. For example, in \cite{PhysRevA.103.012611}, non-Markovian noise is analyzed in terms of  Lindblad equations with negative rates, whose explicit derivation can be involved unless pertubative assumptions are made. In this context, the technique presented here can be interpreted as a non-Markovian mitigation protocol whose underlying error model is defined by the pseudomode mapping. On one hand, this has the advantage of being an explicit non-perturbative model but its  applicability is, on the other hand, limited to noise originating from Gaussian Bosonic baths.

\item \emph{General bath restructuring.} Third, more generally, when $E'_\text{phys}$ is a modified version of $E_\text{phys}$, the formalism can be interpreted  as a way to  \emph{restructure}  some environmental properties without assuming them to be experimentally accessible. 
As an example, $E'_\text{phys}$ could specify a new bath having the very same properties as the original $E'_\text{phys}$ apart from a different temperature. This restructuring can be realized when the pseudo-environment defined by Eq.~(\ref{eq:1}) is coupled to the system alongside the original bath $E_\text{phys}$.
\end{itemize}

\begin{figure}[t!]
\centering
\includegraphics[width=\linewidth]{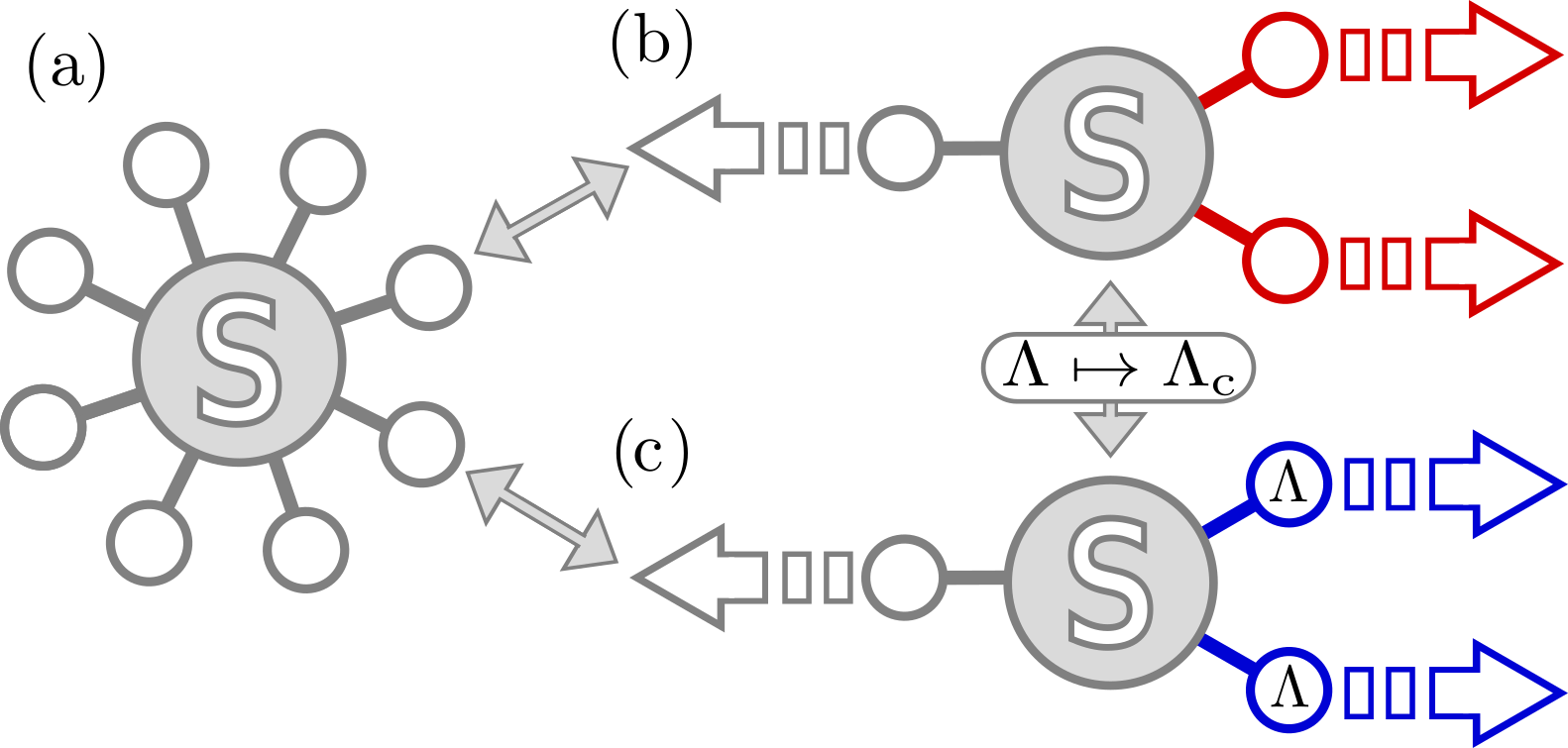}
 \caption{\label{fig:2}Graphical representation of the protocol to simulate unphysical pseudo-environments. In $(a)$, a system $S$ interacts with a continuum of bosonic degrees of freedom. In $(b)$, the reduced dynamics of the system is approximated by the one in which the system interacts with a discrete set of pseudomodes or stochastic fields, see Eq.~(\ref{eq:PM_mimic}). Some of these modes/fields can be unphysical (represented by the color red). In $(c)$, an ensemble of \emph{physical} pseudomodes/fields (represented by the color blue) leads to the same reduced system dynamics as the original pseudomode model in $(b)$ and the original open quantum system in $(a)$ once the parameter $\Lambda$ parametrizing the ensemble is analytically continued to the critical value $\Lambda_{\text{c}}$. This prescription for the physical implementation of unconstrained pseudomode model opens an avaneue to the applications presented in Fig.~\ref{fig:1}}
\end{figure}

The opening of the route towards these applications (for the simulation, mitigation, or reshaping of the properties of non-Markovian noise)
is conditioned on the possibility to physically reproduce the unconstrained pseudo-environment. \emph{In this article we describe a protocol to enable the experimental realization of such unphysical pseudomode models.}
This is achieved by analytical continuation of  measurement results over an ensemble  of physical systems \cite{Iblisdir}, see Fig.~\ref{fig:2}. 
By using bounds provided by the theory of polynomial approximation and extrapolation, we describe the limitations of this technique in terms of an interplay between bias and stability errors. Mainly, while the order of the extrapolating polynomial allows one to increase the predictive power of the method under perfect conditions (reducing its bias), it also increases the sensitivity to imprecision in the initial data (observable outcomes over a physical ensemble). This  leads to a limitation in the complexity of the effects of the environment on the system which can be recovered using this method.

This article is organized as follows. In section \ref{sec:OQS} we start by reviewing open quantum systems to introduce the pseudomode model in section \ref{sec:PM_main}. In section \ref{sec:ensemble}, we describe the main result of this article, i.e., a protocol to simulate the pseudomode model using an ensemble of physical systems. In section \ref{sec:applications}, we present an overview of the possible use of these results to simulate, mitigate, and restructure non-Markovian noise. In section \ref{sec:numerical}, we provide specific numerical examples for each of these use-cases for an environment described by a underdamped Brownian spectral density.
In section \ref{sec:stability}, we further provide an error analysis of the  algorithm. We finish presenting our conclusions in section \ref{sec:conclusions}.

\section{Open Quantum Systems}
\label{sec:OQS}
In this article, we focus on  Gaussian bosonic environments, whose effects on a system can be fully characterized by the correlation of the operator through which they couple to the system. In this section, tildes are used over some of the parameters to explicitly distinguish them from conceptually related ones used to describe the effective baths introduced in the next sections.

We consider a system $S$ interacting with a physical bosonic environment $E$ with a total Hamiltonian ($\hbar=1$)
\begin{equation}
\label{eq:original}
H = H_\text{S} + \hat{s}X +H_\text{E}\;,
\end{equation}
where $H_\text{S}$ is the Hamiltonian of the system and $H_\text{E}=\sum_{\tilde{k}}\omega_{\tilde{k}} \tilde{a}^\dagger_{\tilde{k}} \tilde{a}_{\tilde{k}}$ is the free Hamiltonian of the environment with $\tilde{a}_{\tilde{k}}$ the annihilation operator for the environmental mode ${\tilde{k}}$ with frequency $\omega_{\tilde{k}}$. The interaction term is a function of a generic system operator $\hat{s}$ and it has a linear dependency on the environment, i.e., 
\begin{equation}\label{real}
X=\sum_{\tilde{k}} \tilde{\lambda}_{\tilde{k}}(\tilde{a}_{\tilde{k}}+\tilde{a}_{\tilde{k}}^\dagger)
\end{equation}
where we introduced the coupling strengths  $\tilde{\lambda}_{\tilde{k}}\in \mathbb{R}$. The reduced dynamics of the system is 
\begin{equation}
\label{eq:rhoS}
    \tilde{\rho}_\text{S}(t)=\text{Tr}_\text{E}[\rho_{\text{S+E}}(t)]\;,
\end{equation}
where $\rho_{\text{S+E}}$ is the density matrix of the full system+environment.  In general, the reduced dynamics of the system depends on all the $n$-point free correlation functions involving the interaction operator $X(t)$ (where the time dependence indicates the Heisenberg picture). However, for Gaussian environments, all this information is encoded, through Wick's theorem, in the two-point correlation function
\begin{equation}
    C_\text{E}(t_2,t_1)=\text{Tr}_\text{E}[X(t_2)X(t_1)\rho_\text{E}^{\text{eq}}]\;,
\end{equation}
where the average is taken over the equilibrium distribution $\rho_\text{E}^{\text{eq}}$ of the environment. When the environment is initially in a thermal equilibrium at inverse temperature $\beta$, the correlation function is translational invariant and takes the form $C_\text{E}(t_2,t_1)=C_\text{E}(t_2-t_1)$, where
\begin{equation}
\label{eq:correlation}
C_\text{E}(t)=\int_0^\infty d\omega\frac{ J(\omega)}{\pi}\left[\coth\left(\frac{\beta\omega}{2}\right)\cos(\omega t)-i\sin(\omega t)\right],
\end{equation}
in terms of the spectral density function 
\begin{equation}
\label{eq:sp_d}
    J(\omega)=\pi\sum_{\tilde{k}} {\tilde{\lambda}^2_{\tilde{k}}}\delta(\omega-\omega_{\tilde{k}})\;.
\end{equation}
In the next section, we introduce the pseudomode model as a map between the reduced dynamics for the open quantum system described here and the one computed by averaging over the effects of a discrete set of ancillary harmonic quantum modes and classical stochastic fields.

\section{Pseudomode model}
\label{sec:PM_main}
{The pseudomode method consists in replacing the original continuum of environmental modes with a discrete set of dissipative harmonic modes and stochastic driving fields. The main purpose of these ancillary degrees of freedom is to reproduce the correlation function characterizing the original Gaussian environment and, ultimately, to reproduce the original reduced system dynamics in Eq.~(\ref{eq:rhoS}).} We refer to \cite{Tamascelli,Lambert,LuoSi} and  to Appendix \ref{sec:PM} and \ref{app:PM_DT} for more details.

The term \emph{``pseudomode model''}, $\text{PM}$, will be used to indicate a, possibly stochastic,  operator $\mathcal{M}$ acting linearly on the density matrix $\rho_{\text{S-PM}}$ in a space composed by the system and $N_\text{PM}$ ancillary Fock spaces. 
The influence of $\mathcal{M}$ can thereby be interpreted as the effect of an artificial environment made out of $N_\text{PM}$ dissipative harmonic modes (pseudomodes) and an optional stochastic driving field acting on the system. Perhaps the most important feature of the model is the existence of a non-pertubative relation between  these \emph{``pseudo-environments''} and  physical, Gaussian bosonic environments as described in section \ref{sec:OQS}. This  \emph{``pseudomode mapping''} defines the dependency of $\mathcal{M}$ on a set of parameters $G_\text{PM}$ which directly depend on the spectral density in Eq.~(\ref{eq:sp_d}), hence on the properties of the physical bath. 
This mapping is defined by matching the coupling statistics of the pseudomode model and the original bath [encoded in the correlation in Eq.~(\ref{eq:correlation})].

These definitions take a practical form in specifying the dynamics of the density matrix $\rho_{\text{S-PM}}$ of the pseudomode model as
\begin{equation}
\label{eq:dyn_S_PM}
    \dot{\rho}_{\text{S-PM}}(t;G_\text{PM})=\mathcal{M}(G_\text{PM})[\rho_{\text{S-PM}}(t;G_\text{PM})]\;.
\end{equation}
The parameters in $G_\text{PM}$ are optimized so that the reduced system dynamics 
\begin{equation}
   \rho_{\text{S}}(t;G_\text{PM})= \mathbb{E}[{\rho}_{\text{S}}(t;G_\text{PM})]\;,
\end{equation}
is equivalent to the one in the original open quantum system, i.e., that
\begin{equation}
\label{eq:PM_mimic}
    \tilde{\rho}_{\text{S}}(t)={\rho}_{\text{S}}(t;G_\text{PM})\;.
\end{equation}
Here, $\mathbb{E}$ indicates the expected value over the potentially, but not necessarily, non-trivial, stochastic properties of $\mathcal{M}$.

It is useful to stop for a moment in order to introduce a few examples describing how to qualitatively different environmental effects can be modeled using the formalism above.
\subsection{Reproducing environmental effects}
The pseudomode model is designed to exactly reproduce all the effects of a Gassian bosonic bath linearly coupled to a system. For concreteness, here we analyze how to model specific effects. 

\emph{Markovian effects.} Markovian effects arise as a consequence of memory-less contributions $C_\delta(t)=\Gamma\delta(t)$  to the correlation in Eq.~(\ref{eq:correlation}), so that $\Gamma$ can be interpreted as a decay rate. In this limit, the operator $\mathcal{M}$ takes a Lindblad form, i.e., it can be written as the sum of dissipators
\begin{equation}
\mathcal{M}\mapsto\mathcal{M}_\delta(\{\Gamma\})=\Gamma D_{\hat{s}}\;,
\end{equation}
where $D_{\hat{s}}[\cdot]=2\hat{s}\cdot \hat{s}^\dagger-\hat{s}^\dagger\hat{s}\cdot -\cdot \hat{s}^\dagger\hat{s}$, for a system operator $\hat{s}$.

\emph{Non-Markovian Classical effects.} This case, which includes the previous one, amounts in modeling  effects which can be ascribed to a symmetric contribution  $C_\text{class}(t)=C_\text{class}(|t|)$ to the correlation in Eq.~(\ref{eq:correlation}). The corresponding model can be written in terms of a stochastic drive
\begin{equation}
    \mathcal{M}\mapsto\mathcal{M}_\text{class}(\{c_n\})=-i\xi(t)[\hat{s},\cdot]\;,
\end{equation}
where the coefficients $\{c_n\}$, $n=1,\cdots,N_\xi$, define the spectral representation of  a Gaussian field with zero bias and correlation matching  $C_\text{class}(t)$, i.e., 
\begin{equation}
\label{eq:Cclass}
    C_\text{class}(t)=c_0+2\sum_{n=1}^{N_\xi}c_n \cos[n\pi t/T]\;,
\end{equation}
which can always be achieved by increasing the cut-off parameter $N_\xi$.

\emph{Non-Markovian Quantum effects.} This case, which includes both previous ones, amounts in modeling general effects related to a non-classical contribution  $C_\text{Q}(t)$ to the correlation in Eq.~(\ref{eq:correlation}). The corresponding model can be written in terms of a sum of terms taking the form
\begin{equation}
\label{eq:CQ}
\begin{array}{lll}
    \mathcal{M}\mapsto\mathcal{M}_\text{Q}(\{\Omega,\lambda,\Gamma,n\})&=&-i[\Omega a^\dagger a+\lambda \hat{s}(a+a^\dagger),\cdot]\\
    &&+\Gamma (n+1) D_{a}+\Gamma n D_{a^\dagger}\;,
    \end{array}
\end{equation}
written in terms of a pseudomode operator $a$ and dependent on its frequency $\Omega$, the coupling strength $\lambda$, the decay rate $\Gamma$, and the distribution $n$. This form implies the effects due to $N_\text{PM}$ pseuomodes to be encoded in a correlation contribution written as 
\begin{equation}
    C_\text{Q}(t)=\sum_{j=1}^{N_\text{PM}} C^j_\text{Q}(t)\;,
\end{equation}
where 
\begin{equation}
C^j_Q(t)=\lambda_j^2 [(n_j+1)e^{-i\Omega_j t}+n_j e^{i\Omega_j t}]\exp{[-\Gamma_j|t|]}\;.
\end{equation}

The pseudomode mapping consists in finding the optimal value for these parameters such that, in general,
\begin{equation}
\label{eq:dec}
    C_\text{E}(t)=C_\delta + C_\text{class} + C_\text{Q}(t)\;.
\end{equation}
By considering all these different effects together, the canonical pseudomode model associated with the environment $E$ is determined by the collection
\begin{equation}
\label{eq:G_main}
G_{{\text{PM}}}=\{\Omega_k,\lambda^2_k,\Gamma_{k},n_k,c_n\}\;,
\end{equation}
where $k=1,\cdots,N_\text{PM}$, $n=0,\cdots,N_\xi$, characterizing the environmental correlation in Eq.~(\ref{eq:dec}) in terms of pseudomodes and fields (where we omitted the Markovian component as its effects can always be included in the other degrees of freedom).

In \cite{Lambert}, it was shown that, the correlation of environments characterized by a Brownian spectral density in the underdamped regime, can be well approximated by a sum of purely quantum contributions, even in the zero-temperature case. In this case, some of the modes were shown to require imaginary couplings $\lambda$ to the system. In  \cite{LuoSi}, it was further shown that such an environment can always be modeled by a single, physical, quantum contribution with $n=0$ alongside a classical one representing a single, imaginary, stochastic field, for any temperature of the original bath. 
These results imply that, in order to correctly reproduce the reduced system dynamics, the artificial environment representing the pseudomode model might require to have ``unphysical'' parameters.

 While preventing  a direct physical interpretation, this enlarged parameter domain also implies the possibility for a more optimized description of the original environment. At the same time, it is worth to explicitly point out that, since the form of the dynamical equation \emph{does not} depend on the physicality of the parameters, its unphysical solutions can be interpreted as the \emph{analytical continuation of the physical solutions} (considered as functions of the variables in the set $G_{\text{PM}}$). For example, non-Hermitian contributions to the unitary dynamics  are not compensated by taking the Hermitian conjugate on the Hamiltonian when acting on the right of the density matrix \cite{Lambert}. This makes the procedure qualitatively different than  the orthodox concept of non-Hermitian quantum mechanics \cite{Bender,ju2022emergent}. \\

In order to give a more formal terminology, we define a parameter  $\eta\in G_\text{PM}$  as \emph{physical} if $\eta\in\mathbb{R}^+$ and \emph{unphysical} otherwise.
This language sets the basis for the generalizations which we are going to analyze in this article.

\subsection{Beyond the canonical pseuodomode model}
The pseudomodel described above is labeled \emph{``canonical''} because, despite its possible unphysicality, all the parameters are constrained to mimic the effects of \emph{physical environments} on the system through Eq.~(\ref{eq:PM_mimic}). Here, we are interested in a more general case whose parameters are \emph{unconstrained} to lie outside the canonical model, thereby corresponding to \emph{unphysical environments}. Given a  set of parameters $G_\text{PM}$ of an unconstrained pseudomode model, we define $E_{G_\text{PM}}$ as its corresponding pseudo-environment
\begin{equation}
\label{eq:EPM}
    E_{G_\text{PM}}\leftrightarrow G_\text{PM}\;,
\end{equation}
which is made out of the (potentially unphysical) modes and fields characterized by the parameters in $G_\text{PM}$, through Eq.~(\ref{eq:G_main}).
We will show that, while unphysical, these environments can be used to effectively change, or ``restructure'' the properties of a physical bath.

Given this general setting, our goal  is \emph{to describe how to reproduce the effects of such an unconstrained pseudomode model  from measurement results over physical ensembles}. We will achieve this by defining an analytical continuation protocol on the unphysical parameters to allow observable outcomes in a physical ensemble to reproduce the effects of any general unphysical pseudomode model. In turn, this will lead to a physical representation of models associated to physical environments (thereby defining a tool for their simulation) and models corresponding to more general effects, such as noise mitigation or the restructuring of some environmental properties such as temperature.

\section{Unphysical pseudomodes with physical ensembles}
\label{sec:ensemble}
In this section, we analyze a physical characterization of the unconstrained  pseudomode model. The possibility to do so stems from the fact that the analytical continuation considered here can intuitively be interpreted as a Wick's rotation (used to map time to an imaginary temperature-like quantity) on the parameter $\Lambda$. 
In this context, \cite{Iblisdir} (see also \cite{self2022estimating}) have shown that observable results on an ensemble of quantum systems can be used to perform a Wick rotation thereby opening an avenue to analyze classical statistical systems from quantum measurements and vice-versa.   

Here, we  implement a similar strategy to analytically continue  observables  extracted from a physical pseudomode model into observables of the corresponding unphysical pseudomode model. To start, we denote the unphysical/physical parameters as $\eta_j^\text{unphys}$/$\eta_j^\text{phys}$, where $k=1,\dots,N^\text{phys}$ and $j=1,\dots,N^\text{unphys}$, specifying their total number. These definitions can be used to write the set $G_{\text{PM}}$ characterizing the model as 
 \begin{equation}
     G_{{\text{PM}}}\equiv \left\{\eta_j^\text{unphys},\eta_k^\text{phys}\right\}\;.
 \end{equation}
We then consider a set of $N^\text{unphys}$ functions $\Xi_{j}:\mathbb{C}\rightarrow\mathbb{C}$ such that 
\begin{equation}
\label{eq:an_cont}
\begin{array}{lllll}
\Xi_{j}({\Lambda})&\in& \mathbb{R}^+&\text{for}&{\Lambda}\in D^\text{phys}\;,\\
\Xi_{j}({\Lambda}_c)&=&\eta_{j}^{\text{unphys}}&\text{for}&\Lambda_{\text{c}}\in\mathbb{C},~\Lambda_{\text{c}}\not\in D^\text{phys}\;.
\end{array}
\end{equation}
Here, the first set of conditions allow to interpret the values $\Xi_{j}({\Lambda})$  as physical parameters on the domain $D^\text{phys}$ which, here, is arbitrarily chosen to be $D^\text{phys}=[-1,1]$. The second condition  requires that the unphysical parameters $\eta_j^\text{unphys}$ can be recovered by analytical continuation of the functions $\Xi_j(\Lambda)$ for $\Lambda\mapsto\Lambda_{\text{c}}\in\mathbb{C}$. While not necessary, it is convenient to further impose 
\begin{equation}
\label{eq:physical_bound}
\Xi_j(-1)=0\;,\;\; \Xi_j(1)=1\;,
\end{equation}
to reflect bounds on accessible physical regimes.
 The parameter $\Lambda$ is the one we will, eventually analytically continue.
 We now define a ``physical regularization'' of the set $G_\text{PM}$ as
\begin{equation}
\label{eq:phys_vers}
G^\text{phys}_{{\text{PM}}}({\Lambda})\equiv \left\{\eta_j^\text{unphys}\mapsto\Xi_{j}({\Lambda}),\eta_k^\text{phys}\right\}\;,
\end{equation}
which uses $\Lambda$ to parametrize an ensemble of  pseudomodes models which are physical in the domain  $\Lambda\in [-1,1]$. 
Using the second condition in Eq.~(\ref{eq:an_cont}), it is possible to directly verify  that
\begin{equation}
\label{eq:G=G}
G^\text{phys}_{{\text{PM}}}({\Lambda_{\text{c}}})=G_{{\text{PM}}}\;.
\end{equation}
This shows that, indeed, \emph{the analytical continuation of the physically regularized model recovers the unconstrained pseuodomode model.}
For concreteness, it is possible to define the function $\Xi[\Lambda]$ as
\begin{equation}
\label{eq:Xi}
\Xi[\Lambda]=(1+\Lambda)/2\;,
\end{equation}
which, together with
\begin{equation}
\Lambda_{\text{c}}= -1+2i\;,
\end{equation}
fulfills the constraints in Eq.~(\ref{eq:an_cont}) and Eq.~(\ref{eq:physical_bound}) through the definition 
\begin{equation}
\label{eq:xiXixi}
    \Xi_j(\Lambda)=F_\Lambda(\eta_j^\mathcal{R})+\Xi(\Lambda)F_\Lambda(\eta_j^\mathcal{I})\;,
\end{equation}
where $\eta_j^\mathcal{R}\equiv\text{Re}[\eta_j^\text{unphys}]$, $\eta_j^\mathcal{I}\equiv\text{Im}[\eta_j^\text{unphys}]$, and where 
\begin{equation}
    F_\Lambda(x)\equiv\theta(x)x+\Xi^2(\Lambda)\theta(-x)|x|\;,
\end{equation}
is defined in terms of the step function $\theta(x)$=1 for $x\geq1$ and zero otherwise.
While other choices are possible, this will be the default one used throughout this article.\\

In summary, we formally defined physical versions of the pseudomodes parametrized by a \emph{single} parameter $\Lambda$ whose analytical continuation to $\Lambda\mapsto\Lambda_{\text{c}}$ reproduces the unphysical pseudomodes model as shown in Eq.~(\ref{eq:G=G}). Our attention now shifts towards the implementation of this analytical continuation using measurement results over the physical ensemble $G_\text{PM}^{\text{phys}}(\Lambda)$ corresponding to sweeping over different values of $\Lambda$.

\subsection{Analytical continuation}
\label{sec:an_cont_main}In the previous sections we showed that, given a pseudomode model defined by a set $G_\text{PM}(\Lambda)$ of unconstrained parameters, there exists a regularized 
version $G^\text{phys}_\text{PM}(\Lambda)$ whose parameters are physical for $\Lambda\in D^\text{phys}$. This implies that the corresponding models $\mathcal{M}^\text{phys}_\text{PM}(\Lambda)$ can be interpreted as a physical  ensemble parametrized by $\Lambda\in D^\text{phys}$.
Here,  the word ``physical'' takes an operative meaning \cite{Bridgman}, as each of these models can, 
in principle, be experimentally realized by coupling the system to a set of ancillary resonators and by driving it with a classical stochastic field. This means that, for such $\Lambda$s, and for a given time $t$, we can always perform a sequence of measurements to reconstruct the density matrix 
\begin{equation}
   {\rho}_{\text{S}}(t;\Lambda)\equiv{\rho}_{\text{S}}(t; \{\eta_j^\text{unphys}\mapsto\Xi_{j}({\Lambda}),\eta_k^\text{phys}\}) \;,
\end{equation}
which simply corresponds to the dynamics of the model as in Eq.~(\ref{eq:dyn_S_PM}). The functional dependence   of the density matrix in the parameter $\Lambda$ can then be analytical continued over to the value $\Lambda_{\text{c}}$ to finally achieve our goal, i.e., the simulation of the unphysical pseudomdode model corresponding to the density matrix
\begin{equation}
\label{eq:lkj}
       {\rho}_{\text{S}}(t;\Lambda_{\text{c}})\equiv{\rho}_{\text{S}}(t; \{\eta_j^\text{unphys}\mapsto\Xi_{j}({\Lambda_{\text{c}}}),\eta_k^\text{phys}\}) \;.
\end{equation}
This result can also be rephrased in terms of equivalent pseudo-environments. In fact, generalizing Eq.~(\ref{eq:EPM}), we can define $E^\text{phys}_\text{PM}(\Lambda)$ as the physical pseudo-environment corresponding to the regularized set $G^\text{phys}_\text{PM}(\Lambda)$, i.e., 
\begin{equation}
\label{eq:EPM2}
    E^\text{phys}_{G_\text{PM}}(\Lambda)\leftrightarrow G^{\text{phys}}_\text{PM}\;.
\end{equation}
In this way, Eq.~(\ref{eq:lkj}) can be interpreted as an analytical continuation of the corresponding pseudo-environments, i.e., 
\begin{equation}
\label{eq:EPM3}
    E^\text{phys}_\text{PM}(\Lambda_{\text{c}})=E_\text{PM}\;.
\end{equation}
While this can be considered just as an equivalent way to interpret Eq.~(\ref{eq:lkj}), it also allows us to grasp its meaning from a different point of view. In fact, Eq.~(\ref{eq:EPM3}) can be intuitive interpreted as manifesting the \emph{possibility to define a parametrization for physical environments such that, upon analytical continuation, the effects on the system are equivalent to those generated by a generalized, unphysical pseudomode model}. This different prospective paves the way for the applications which will be analyzed in section \ref{sec:applications}.

In practice, we can only assume the physical ensemble $E^\text{phys}_\text{PM}(\Lambda)$ to be realized for a discrete grid of points, which is the case we are going to analyze in the next section.

\subsection{Polynomial extrapolation}
\label{sec:pol_extr}
In this section, we analyze the analytical continuation of the expectation values of a system observable 
\begin{equation}
    f_n\equiv\langle\hat{O}_{\text{S}}\rangle(t;\Lambda)\equiv\text{Tr}_\text{S}[\hat{O}_{\text{S}}\rho_{\text{S}}(t;\Lambda)]\;,
\end{equation}
over the physical ensemble described by a set of $N_\text{exp}$ points $\Lambda_n\in\Lambda$ for $n=1,\cdots N_\text{exp}$. We note that this is a weaker version of the case considered in the previous section. In fact, the analytical continuation of the full density matrix can be considered as the limiting, and more resource-consuming, task where we analytically continue not just one, but a complete set of observables.

We now introduce the following practical procedure for analytical continuation. We first define the function $p_M(\Lambda)$ as the order $M$ polynomial which minimizes the distance
\begin{equation}
d^2=\sum_{n=0}^{N_\text{exp}}|p_M(\Lambda_n)-f_n|^2\;.
\end{equation} 
This least square fitting can be used to reconstruct the observable as
\begin{equation}
\langle \hat{O}_{\text{S}} \rangle^\text{reconstructed}\equiv p_M(\Lambda_{\text{c}})\;.
\end{equation}
The ability to reconstruct the full pseudomode model from an ensemble of pseudomodes constrained to physical dynamics opens up the possibility of several interesting applications that would be, otherwise, physically inaccessible. In the next section, we are going to analyze these applications, in terms of simulation,  restructuring, and mitigation of non-Markovian quantum noise.

\section{Applications}
\label{sec:applications}

In this section we describe different possible applications of our protocol. Mainly, we show that, by driving a system with classical noise and by coupling it to  ancillary harmonic modes, it is possible to analogically simulate,  mitigate, or restructure the effects of a Gaussian Bosonic environment. 

\subsection{Simulation}
We start by defining a procedure to simulate a physical environment  $E'$, which we assume to be well described by a pseudomode model characterized by a parameter set $G'_\text{PM}$. We note that this corresponds to imposing $E_\text{phys}=0$ in Eq.~(\ref{eq:1}), which then translates to
\begin{equation}
    E_\text{PM}=E'\;.
\end{equation}
In other words, a simulation of the open quantum system $S+E'$ can be achieved by coupling the (closed) system to  the pseudo-environment $E_\text{PM}$ corresponding to the pseudomode model with parameters $G'_\text{PM}$. As noted in our previous discussion, the pseudo-environment $E_\text{PM}$ is, in general, not directly physically realizable. To circumvent this problem,  we follow the analysis in section \ref{sec:ensemble}, and define a set of regularized parameters ${G}^\text{phys}_\text{PM}(\Lambda)$ corresponding to a collection of \emph{physical} pseudo-environments ${E}^{\text{phys}}_\text{PM}(\Lambda)$. 
By measuring observables in this ensemble, it is then possible to proceed with the analytical continuation protocol illustrated in section \ref{sec:ensemble}, i.e.,
\begin{equation}
    {G}^\text{phys}_\text{PM}(\Lambda_{\text{c}})=G_\text{PM}\;,~~~{E}^{\text{phys}}_\text{PM}(\Lambda_{\text{c}})=E_\text{PM}\;,
\end{equation}
which corresponds to  simulating the effects of the original environment $E'$ on the system.

\subsection{Mitigation}
\label{sec:anti_main}
Another interesting application is the use of the pseudomode model to mitigate the effects of a non-Markovian environment $E$,  which we assume to be well described by a pseudomode model characterized by a parameter set $G_\text{PM}=\{\Omega_k,g^2_k,\Gamma_{k},n_k,c_n\}$. The mitigation procedure correspond to imposing $E'_\text{phys}=0$ in Eq.~(\ref{eq:1}), which brings it to the form
\begin{equation}
\label{eq:over}
    E+E_{\overline{\text{PM}}}=0\;.
\end{equation}
This equation describes the (unphysical) pseudo-environment $E_{\overline{\text{PM}}}$  which, once coupled to the system, completely cancels all effects of the environment $E$. Since we assume the knowledge of the pseudomode model describing $E$, the parameters defining this ``antimode'' environment $E_{\overline{\text{PM}}}$ can be explicitly written as
\begin{equation}
\label{eq:PMbar_main}
G_{\overline{\text{PM}}}=\{\Omega_k,-g^2_k,\Gamma_{k},n_k,-c_n\}\;.
\end{equation}
To derive this equation we simply noted that, if the original set $G_\text{PM}$ describes a correlation $C_E(t)$, then the set $G_{\overline{\text{PM}}}$ must correspond to
 \begin{equation}
 \label{eq:CminusC}
 C_{\overline{\text{PM}}}(t)=-C_\text{PM}(t)\;,
 \end{equation}
 since Eq.~(\ref{eq:C_class}) and Eq.~(\ref{eq:CQ}) are linear in $c_n$ and $g_k^2$, respectively.
In other words, the compound effect of the pseudomode model and its ``antimode'' version does not have any influence on the system since the two correlations sum up to zero: A system simultaneously in contact with a Gaussian environment and its unphysical ``mirror'' or anti-environment should evolve as if no environment were present at all, i.e., noise-free. Because of its effective noise-cancelling action, the antimode environment is necessarily unphysical. However, following section \ref{sec:ensemble}, we can regularize its parameters to define the corresponding physical ensemble $E^\text{phys}_{\overline{\text{PM}}}(\Lambda)$ parametrized by $\Lambda$, from which gathering all the information needed for the analytical continuation
\begin{equation}
E^\text{phys}_{\overline{\text{PM}}}(\Lambda_{\text{c}})=E_{\overline{\text{PM}}}=-E\;,
\end{equation}
to achieve mitigation using measurements on a physical ensemble.

\subsection{Restructuring}
In its most general form, Eq.~(\ref{eq:1}) reads
\begin{equation}
\label{eq:rest}
E^{\text{restructuring}}_\text{PM}+E_{\text{phys}}  =E'_{\text{phys}}\;,
\end{equation}
and can, for example, be interpreted as a  ``restructuring'' of a given physical environment $E_\text{phys}$ (which we assume associated to a pseudo-environment $E_\text{PM}=E_\text{phys}$) into a version $E'_\text{phys}$ (associated to the pseudo-environment $E_{\text{PM}'}=E'_\text{phys}$) characterized by different physical properties (such as temperature or system-bath coupling). This is achieved by coupling the system to a pseudo-environment $E^{\text{restructuring}}_\text{PM}$ whose parameters must be a function of those in $E_\text{PM}$ and $E_{\text{PM}'}$ encoding the specific ``changes'' to be imposed on the bath. For the most interesting cases, the resulting environment is going to be unphysical so that, as done in the previous cases, we will need to resort to the analytical continuation procedure outlined in section \ref{sec:ensemble} for the physical implementation of its effects. 

Here, we note that, while more optimized versions might be possible (as shown in the example in section \ref{sec:restr}), in the worse case, the restructuring  can always be defined by using both the simulation and mitigation techniques presented in the previous sections. In fact, thanks to our assumptions and the definition in Eq.~(\ref{eq:over}) we can directly check that 
\begin{equation}
E^{\text{restructuring}}_\text{PM}=E_{\text{PM}'}+E_{\overline{\text{PM}}} \;,
\end{equation}
satisfies Eq.~(\ref{eq:rest}). In other words, in order to restructure a bath $E_\text{phys}$ into $E'_\text{phys}$ it is always possible to couple the system to both the anti-environment 
 relative to  $E_\text{phys}$ and an additional one simulating $E'_\text{phys}$. As mentioned, this worse case scenario can be optimized depending on the specific requirements of the restructuring as we will show in the example in section \ref{sec:restr}.

In the following, we are going to present explicit numerical examples for each of these applications.

\section{Numerical examples}
\label{sec:numerical}
To show a practical numerical implementation \cite{Qutip1,Qutip2,PhysRevA.98.063815,Li2022pulselevelnoisy} of the applications  presented in the previous section, here we present three examples: the mitigation and simulation of the effects of an environment characterized by a underdamped Brownian spectral density at zero and finite temperature and the  ``restructuring'' of the finite temperature case into the zero temperature one. We note that, for exposition clarity, the order in which these examples are presented is different with respect to the previously reported one.

For concreteness, we will focus on  a Gaussian Bosonic environment characterized by the spectral density function
\begin{equation}
\label{eq:spectral_density_main}
J^\text{B}(\omega)=\frac{\gamma\lambda^2\omega}{(\omega^2-\omega_0^2)^2+\gamma^2\omega^2}\;,
\end{equation}
written in terms of a resonance frequency $\omega_0$, a frequency-width $\gamma$, and a (frequency)$^{3/2}$ strength $\lambda$. It describes a Ohmic behavior at low frequency, i.e.,
\begin{equation}
    J^\text{B}(\omega)\sim\alpha\omega~\text{for}~ \omega\ll\omega_0\;,
\end{equation}
 and it has a polynomial cut-off at high-frequencies, i.e., 
 \begin{equation}
   J^\text{B}(\omega)\sim\alpha\omega_0^4/\omega^3 ~\text{for}~ \omega\gg\omega_0 \;, 
 \end{equation}
 in terms of the adimensional scale $\alpha=\lambda^2\gamma/\omega_0^4$. We further restrict to the underdamped regime which requires 
\begin{equation}
    \omega_0^2 - \gamma^2/4>0\;.
\end{equation}
By inserting the spectral density $J_B(t)$ in Eq.~(\ref{eq:correlation}) the correlation function can be computed  as \cite{LuoSi}
\begin{equation}
    C^\text{B}(t;\beta)=C^\text{B}_\text{class}(t;\beta)+C^\text{B}_\text{Q}(t)\;,
\end{equation}
where
\begin{equation}
\renewcommand{\arraystretch}{2.2}
\begin{array}{lll}
    C^\text{B}_\text{class}(t;\beta)  &=&\displaystyle \frac{\lambda^2}{4\Omega}\coth{(\beta(\Omega+i\Gamma)/2)}e^{i\Omega |t|} e^{-\Gamma|t|}\\
   &&\displaystyle -\frac{\lambda^2}{4\Omega}\coth{(\beta(-\Omega+i\Gamma)/2)}e^{-i\Omega |t|} e^{-\Gamma|t|}\\
   &&-\displaystyle\frac{\lambda^2}{4\Omega}\left(e^{-i\Omega t}+e^{i\Omega t}\right)e^{-\Gamma |t|}\\
   &&+\displaystyle\frac{2i}{\beta}\sum_{k>0}J(\omega^\text{M}_{k})e^{-|\omega^\text{M}_{k}||t|}\;.
    \end{array}
\end{equation}
This expression is written in terms of the Matsubara frequencies $\omega^\text{M}_k=2\pi k i/\beta$ ($k=1,\cdots,\infty$),
is the symmetric, or ``classical'' contribution which contains all temperature effects which can, thereby, be modeled within the statistics of a single temperature dependent, classical field $\xi_\beta(t)$, defined explicitly in appendix \ref{sec:PM}, and such that
\begin{equation}
\label{eq:Cxx}
    C^\text{B}_\text{class}(t;\beta)=\mathbb{E}[\xi_\beta(t_2)\xi_\beta(t_1)]\;,
\end{equation}
where $t=t_2-t_1$. We can label the remaining contribution 
\begin{equation}
\begin{array}{lll}
    C^\text{B}_\text{Q}(t) &=&\displaystyle\frac{\lambda^2}{2\Omega}\exp{[-i\Omega t-\Gamma |t|]}\;.
    \end{array}
\end{equation}
as ``quantum'', since it does not have specific symmetries, but it has the form of Eq.~(\ref{eq:C_PM}), i.e., it can be reproduced using a single harmonic mode initially at zero temperature. 

To be specific, we will consider the open system to be made out of a two-level system  coupled to the Brownian environment $B$ described above. 
Specifically, given a full density matrix $\rho_{\text{S+B}}(t)$ in the system+environment space whose dynamics is described by a full Hamiltonian $H_{S+B}$ as
\begin{equation}
\label{eq:OQS}
    \dot{\rho}_{\text{S+B}}(t)=-i[H_{S+B},\rho_{\text{S+B}}(t)]\;,
\end{equation}
we want to compute the reduced dynamics encoded in
\begin{equation}
\rho_{\text{S}}(t)=    \text{Tr}_\text{B}[\rho_{\text{S+B}}(t)]\;.
\end{equation}
By  defining the system Hamiltonian as
\begin{equation}
    H_{\text{S}}=\frac{\omega_s}{2} \sigma_z + \frac{\Delta}{2}  \sigma_x\;,
\end{equation}
and assuming that the operator
$\hat{s}=\sigma_x$ mediates the interaction to the environment, the reduced dynamics can be computed by using the pseudomode model, i.e., by computing
\begin{equation}
\rho_{\text{S}}(t)= \text{Tr}_{\text{B}}[\rho_{\text{S+PM}}(t)]\;,
\end{equation}
which requires to solve the differential equation
\begin{equation}
\label{eq:Tzerodyn}
    \dot{\rho}_{\text{S+PM}}=-i[H_{\text{S}}+\xi_\beta(t)\hat{s},\rho_{\text{S+PM}}]+L_{a_\text{res}}[\lambda_\text{res};\rho_{\text{S+PM}}]\;.
\end{equation}
Here, $\rho_{\text{S+PM}}$ is the density matrix in the system+pseudomode space. This dynamics is driven by a single  field $\xi_\beta(t)$ whose statistics depends on the inverse temperature $\beta$ to reproduce the Matsubara contribution to the correlation in Eq.~(\ref{eq:Mats}) and a single pseudomode $a_\text{res}$ to describe the resonant properties of the spectral density in Eq.~(\ref{eq:spectral_density_main}). Following the standard pseudomode mapping \cite{Lambert,Tamascelli,LuoSi}, we have
\begin{equation}
    L_{a_\text{res}}[\lambda_\text{res};\cdot]=-i[H_\text{res}(\lambda_\text{res}),\cdot]+ \Gamma_\text{res} D^{T=0}_{a_\text{res}}[\cdot]
\end{equation}
in terms of the Hamiltonian 
\begin{equation}
    H_\text{res}(\lambda_\text{res})=\Omega_\text{res} a_\text{res}^\dagger a_\text{res}+\lambda_\text{res} \hat{s} (a_\text{res}+a^\dagger_\text{res})\;,
\end{equation}
the zero-temperature dissipator
\begin{equation}
    D^{T=0}_{a}[\rho]=2a\rho a^\dagger+a^\dagger a\rho +\rho a^\dagger a\;,
\end{equation}
written for a generic operator $a$\;, and the coefficients 
\begin{equation}
\lambda_\text{res}=\sqrt{\lambda^2/2\Omega}\;,~\Omega^2_\text{res}=\omega_0^2-\Gamma^2_\text{res}\;,~\Gamma_\text{res} =\gamma/2\;.
\end{equation}
The initial state is $\rho_{\text{S+PM}}(0)=\rho_{\text{S}}(0)\otimes \ketbra{0}{0}$ in terms of the vacuum $\ket{0}$ annihilated by the harmonic mode $a$.

An interesting feature of this hybrid model associated to the spectral density in Eq.~(\ref{eq:spectral_density_main}) is that \cite{LuoSi}:
\begin{itemize}
\item  all the parameters associated with the resonant mode $a_\text{res}$ are physical, i.e., $\lambda_\text{res}, \Omega_\text{res}, \Gamma_\text{res}$ are real and positive.
\item all temperature effects are encoded in the statistics of the field $\xi_\beta(t)$; i.e., in its autocorrelation function.
\end{itemize}
Given this specific environment, we are now going to analyze three specific applications in terms of mitigation, restructuring, and analog simulation. 
\begin{figure}[t!]
\includegraphics[width = \columnwidth]{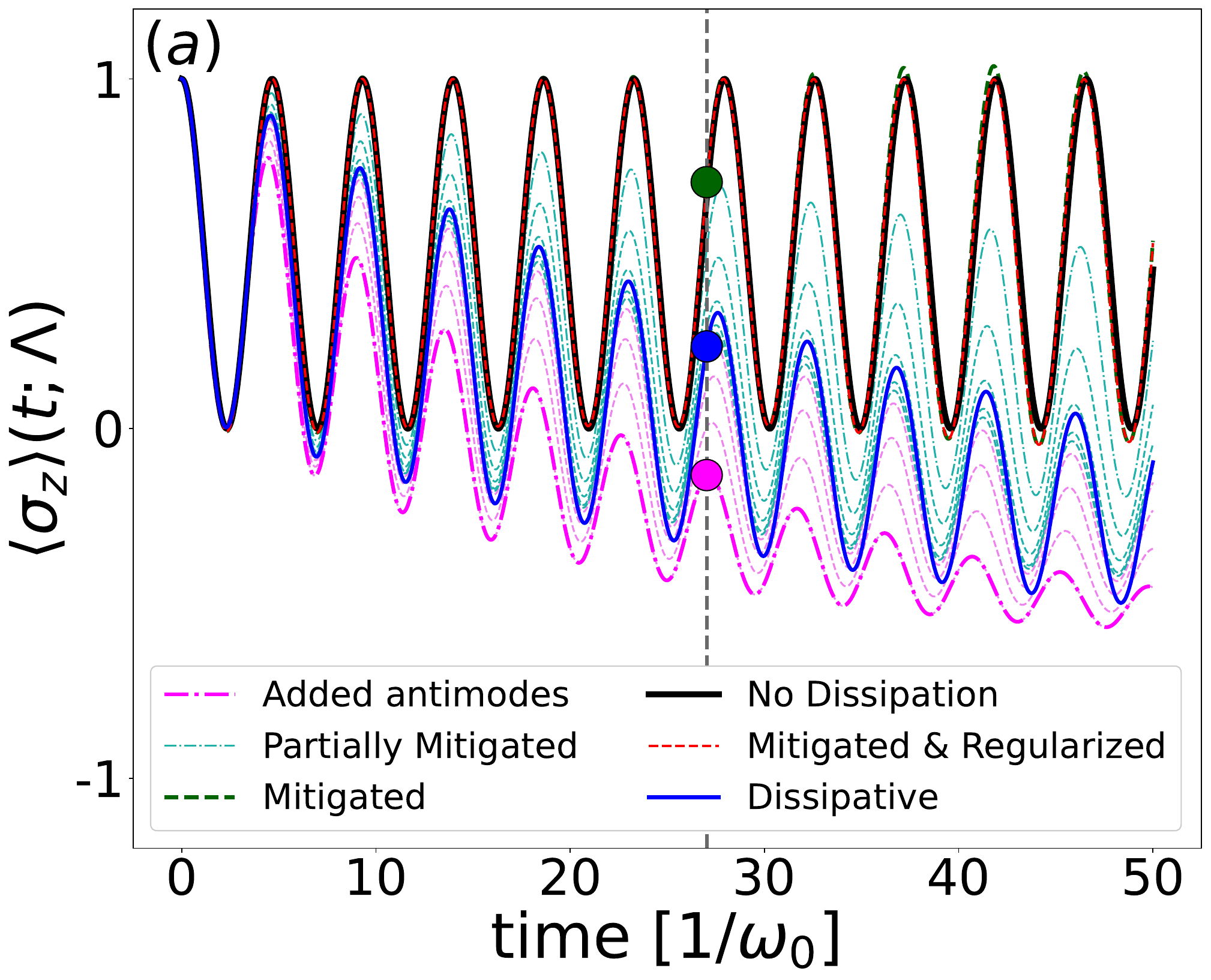} 
\includegraphics[width = .9\columnwidth]{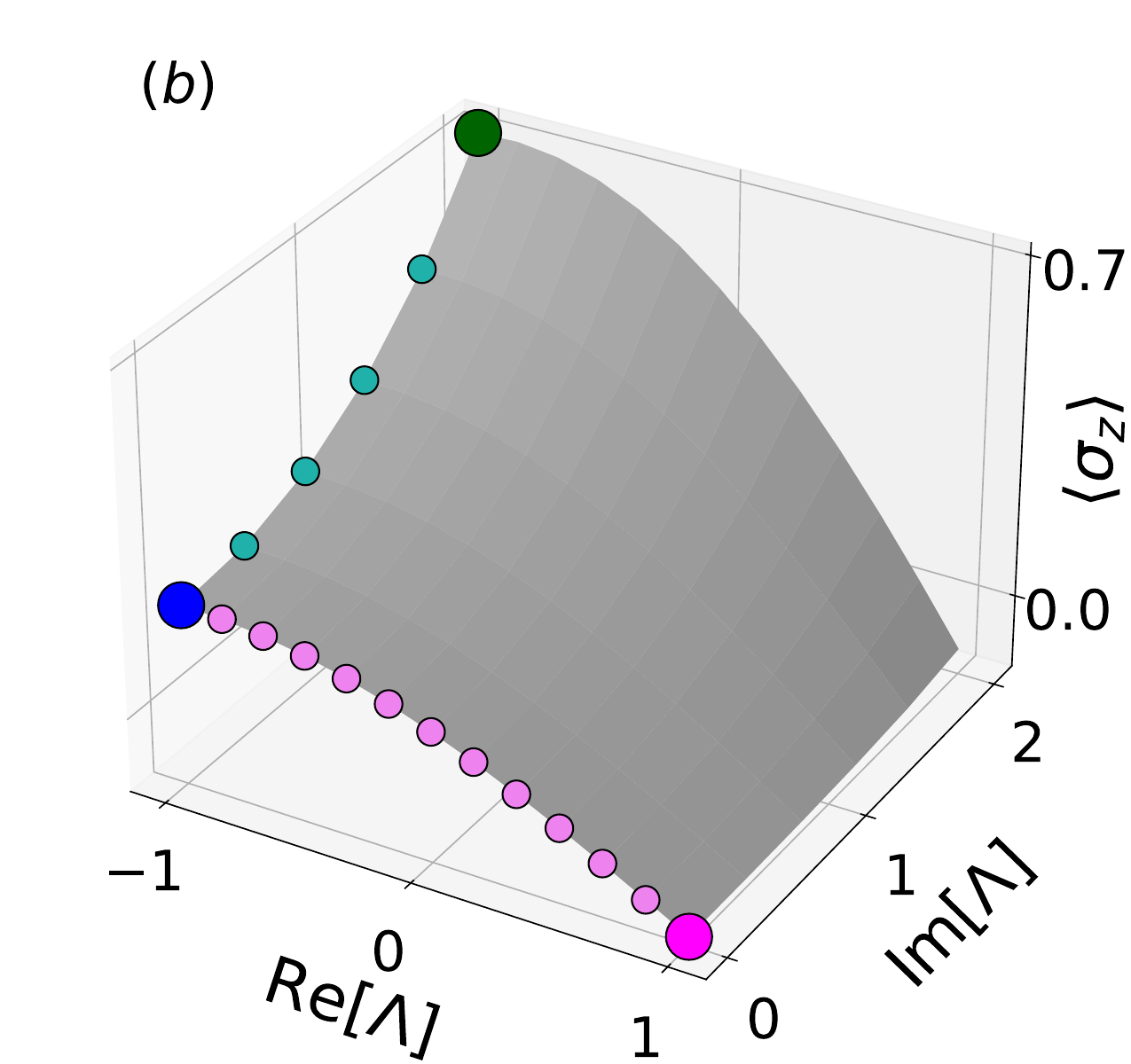} 
\caption{\label{fig:brownian_dynamics_T0} Mitigation of an underdamped Brownian environment at zero temperature. $(a)$ Dynamics of the observable $\sigma_z$. In blue, the original dissipative dynamics, solving Eq.~(\ref{eq:Tzerodyn}). A physical version of the antimode model is added, to generate the different violet curves corresponding to different $\Lambda\in[-1,1]$ in Eq.~(\ref{eq:PMantiPM}). In green, partial and full mitigation as an analytical continuation on the function $\sigma_z(t;\Lambda)$ for $\Lambda\in[-1,-1-2i]$. In red, a regularization is applied to the analytically continuation of the full density matrix. $(b)$ Analytical continuation at the specific time described by the grey line in $(a)$. Violet points corresponds to the measured observables for different real $\Lambda$. The green points correspond to the partially and fully implemented analytical continuation. In blue, the original value in the solely presence of the bath $B$. The overall grey surface is the polynomial $p_M(t;\Lambda)$ used to define the analytical continuation as an extrapolation. The specific parameters used are: $\alpha=0.02$, $\Gamma=0.3\omega_0$, $\omega_s=\Delta=\omega_0$, $M=10$, $N_\text{exp}=12$, $N_\xi=100$.}
\end{figure}
\subsection{Noise mitigation}
In this section, we are going to analyze the mitigation of the effects of the environment specified by the spectral density in Eq.~(\ref{eq:spectral_density_main}) both at zero and finite temperature to ultimately compute observables corresponding to the noise-free dynamics
\begin{equation}
\label{eq:free}
    \dot{\rho}_\text{free}=-i[H_{\text{S}},\rho_\text{free}]\;.
\end{equation}
To do this, we add stochastic driving and coupling to ancillary quantum modes to the original $S+B$ open quantum system such that, after analytical continuation of a single parameter, their correlation is exactly the opposite of the original one $C^\text{B}(t)$. We can achieve this using the formalism described in section \ref{sec:anti_main}. 

First, we want to define the unphysical ``antimode'' model $\overline{\text{PM}}$ whose correlation satisfies  
\begin{equation}
\label{eq:CbarmC}
    C_{\overline{\text{PM}}}=-C^\text{B}(t)\;.
\end{equation} 
Using Eq.~(\ref{eq:PMbar_main}) this can be done by introducing a  resonant ``anti-mode'' $\bar{a}_\text{res}$ and an ``anti-field'' $\bar{\xi}_{\beta}$, whose parameters are the same as those associated with $a_\text{res}$ and ${\xi}_{\beta}$ in Eq.~(\ref{eq:Tzerodyn}), except for an additional complex-rotation in the interaction to the system. Specifically,
\begin{equation}
\label{eq:lxi}
    \begin{array}{lll}
\bar{\lambda}_\text{res}\equiv i\lambda_\text{res}\;,~
\bar{\xi}_{\beta}\equiv i {\xi}_{\beta} ~(\text{or}~ \bar{c}_n=-c_n)\;,
    \end{array}
\end{equation}
where the field coefficients $c_n$ are explicitly defined in Eq.~\ref{eq:xi_cn}.

 To continue, we need to define a physical ensemble whose effects are the same as the antimode model after analytical continuation. Before doing this, it is worth differentiating  between the zero and finite temperature case.
 \begin{figure}[t!]
\includegraphics[width = \columnwidth]{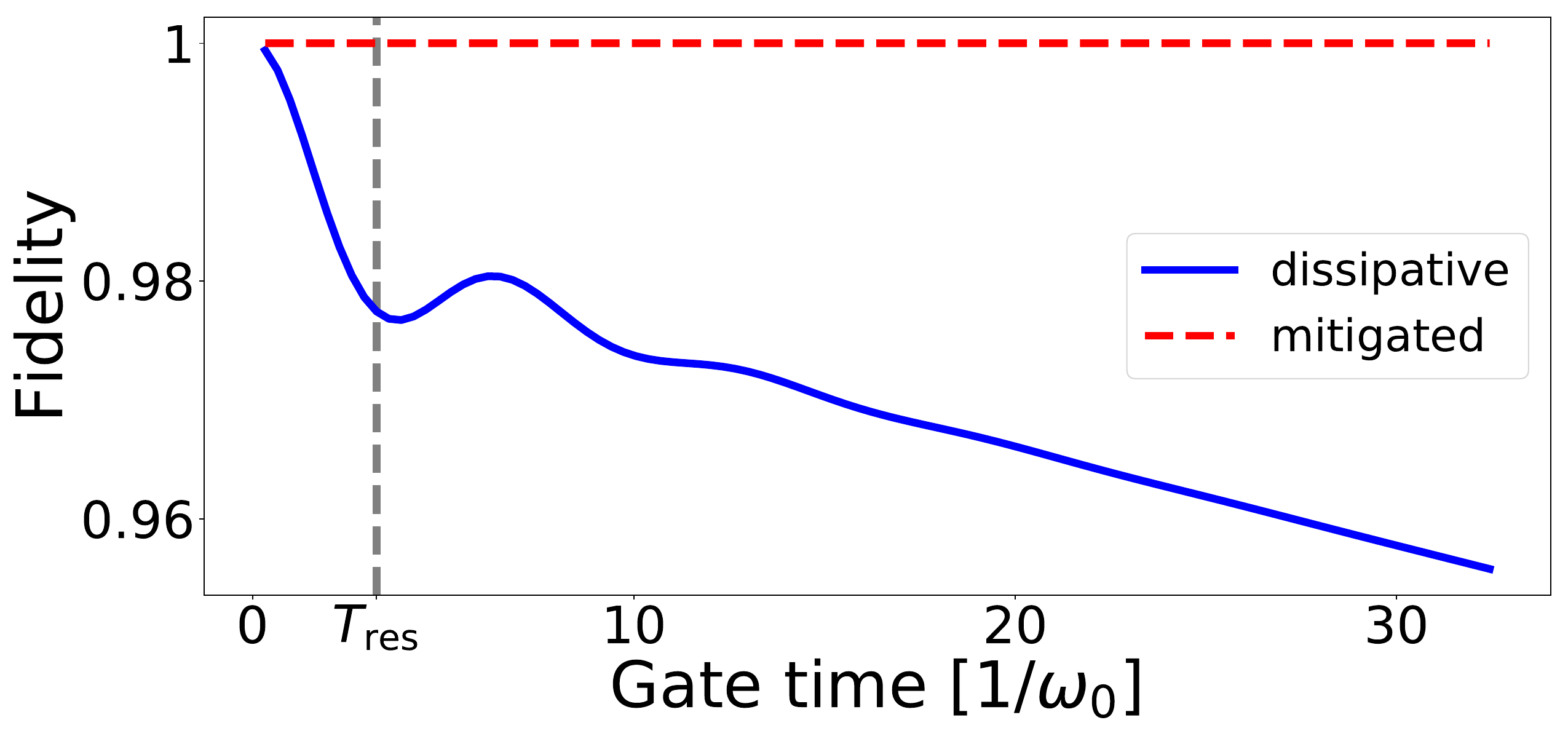} 
\caption{\label{fig:gate} Gate-fidelity against gate-time for a noisy $\theta=\pi$ rotation around the $z$-axis on a two-level system before and after analytical continuation. Here, noise is modeled through the coupling to an underdamped Brownian environment at zero temperature. The dashed-grey vertical line represents the time $T_\text{res}$ where the two-level system is on resonance with the environment; i.e., when $\omega_s=\theta/T_\text{res}=\Omega$. The specific parameters used are: $\lambda=0.2 \omega_0^{3/2}$, $\Gamma=0.5\omega_0$, $\omega_s=\Delta=0$, $M=7$, $N_\text{exp}=12$, and $N_\xi=100$.}
\end{figure}
\subsubsection{Zero Temperature}
In the zero-temperature limit, it is interesting to note that a simplification occurs in the classical expression for the correlation function as \cite{Lambert,LuoSi}:
\begin{equation}
\label{eq:Mats}
\begin{array}{lll}
    C^\text{B}_\text{class}(t;\beta=\infty)&=&\displaystyle\frac{i}{\pi}\int_0^\infty dx J^\text{B}(ix)e^{-x|t|}\;.
      \end{array}
\end{equation}
This zero-temperature ``Matsubara'' contribution is negative at zero time, hinting to the fact that imaginary fields are necessary to ensure the correct sign is reproduced in Eq.~(\ref{eq:Cxx}).  This is, in fact, the case, and the field $\xi_{\beta=\infty}(t)$ is purely imaginary; which, using Eq.~(\ref{eq:lxi}), corresponds to a real anti-field $\bar{\xi}_{\beta=\infty}(t)$. Therefore, the stochastic part of the model does not require  introducing any additional analytical continuation procedure [see Eq.~(\ref{eq:xiXixi})].  Thus, the analytical continuation needs only to be performed upon a single unphysical parameter (the coupling $\overline{\lambda}_\text{res}$ between the system and the resonant mode). 

Given this construction of the antimode model, we can now define its physically-regularized version, i.e., we want to couple the system to a physical ensemble which can be analytically continued to the antimode model above. Specifically, given the Hamiltonian $H_{S+B}$ in the original system+bath space, we couple the system to a physical regularized-antimode $\bar{a}^\text{phys}_\text{res}$ and drive it by a physical field $\bar{\xi}^\text{phys}_{\beta=\infty}(t)$, such that the dynamics is described by
\begin{equation}
\label{eq:SBanti}
    \dot{\rho}_{\text{S+B}}=-i[H_{S+B}+\bar{\xi}^\text{phys}_{\beta=\infty}\hat{s},\rho_{\text{S+B}}]+{L}_{\bar{a}^\text{phys}_\text{res}}[\Xi(\Lambda)\lambda_\text{res};\rho_{\text{S+B}}].
\end{equation}
Here, the physicality of the field $\bar{\xi}^\text{phys}_{\beta=\infty}(t)=\bar{\xi}^\mathcal{I}_{\beta=\infty}(t)$ explicitly corresponds to having an autocorrelation function of the form
\begin{equation}
\label{eq:barC1}
\begin{array}{lll}
    \mathbb{E}[\bar{\xi}^\text{phys}_{\beta=\infty}(t_2)\bar{\xi}^\text{phys}_{\beta=\infty}(t_1)]&=&\displaystyle -\frac{i}{\pi}\int_0^\infty dx J^\text{B}(ix)e^{-x|t|}\\
    &=& -C^\text{B}_\text{class}(t;\beta=\infty)>0~~\forall t\;.
    \end{array}
\end{equation}
In parallel, the parameters characterizing the physically-regularized antimode $\bar{a}^\text{phys}_\text{res}$  are the same as those for the antimode $\bar{a}_\text{res}$ in Eq.~(\ref{eq:Tzerodyn}), except for the presence of the additional parameter $\Lambda$ [introduced through the function $\Xi$ defined in Eq.~(\ref{eq:Xi})], renormalizing the coupling to the system into the physical domain. We assume such a coupling to be constrained in the range $[0,\lambda_\text{res}]$, corresponding to $\Lambda\in[-1,1]$.
The correlation function of this mode is then given by
\begin{equation}
\label{eq:barC2}
    \bar{C}_\text{res}^B(t)=\Xi^2(\Lambda) C_\text{res}^B(t)\;,
\end{equation}
which, by construction, gives rise to a minus sign for $\Lambda=\Lambda_{\text{c}}=-1+2i$ using Eq.~(\ref{eq:Xi}). In this way, Eq.~(\ref{eq:barC1}) and Eq.~(\ref{eq:barC2}) fulfill the  identity in Eq.~(\ref{eq:CbarmC}) defining the antimode model. In other words, by performing the analytical continuation $\Lambda\mapsto\Lambda_{\text{c}}$, the model in Eq.~(\ref{eq:SBanti}) adds a correlation $-C_\text{res}^\text{B}(t)$ to the open quantum system in the $(S+B)$ space in Eq.~(\ref{eq:OQS}), i.e., it completely counteracts, in principle, all environmental noise ultimately leading to the free dynamics in Eq.~(\ref{eq:free}).

We give a graphical exemplification of this proceudure in Fig.~\ref{fig:brownian_dynamics_T0}. It is important to note that, in order to compute the dynamics plotted in this figure, we did not solve the differential equation in Eq.~(\ref{eq:SBanti}) which involves the original environmental continuum. Instead, the physically-regularized antimode model is introduced on top of the deterministic pseudomode model characterizing such a  continuum. Specifically, we considered
\begin{equation}
\label{eq:PMantiPM}
    \begin{array}{lll}
\dot{\rho}_{\text{S+PM}+\overline{\text{PM}}}&=&-i[H_{\text{S}}+H_{\text{PM}}+\overline{\xi}_{\beta=\infty}\hat{s},\rho_{\text{S+PM}+\overline{\text{PM}}}]\\
&&+\displaystyle\sum_{j=1}^3 L_{a_j}[\lambda_j;\rho_{\text{S+PM}+\overline{\text{PM}}}]\\
&&+L_{\bar{a}^\text{reg}_\text{res}}[\Xi(\Lambda)\lambda_\text{res};\rho_{\text{S+PM}+\overline{\text{PM}}}]\;,
\end{array}
\end{equation}
which corresponds to using the antimode model alongside the deterministic pseudomode model in \cite{Lambert} [used to describe the continuum $B$ in Eq.~(\ref{eq:SBanti})]. We refer to Appendix \ref{sec:UB} for a brief overview.
The solutions of this differential equation for different values of $\Lambda$ in $\Lambda_n\in[-1,1]$ are used to compute the observable $\langle\hat{O}_{\text{S}}\rangle^\text{phys}(t;\Lambda)$.
This is done by interpolating the values $\langle\hat{O}_{\text{S}}\rangle^\text{phys}(t;\Lambda_n)$ with an order $M$ polynomial $p_M(t;\Lambda)$ which is then analytically continued to $\Lambda\mapsto\Lambda_{\text{c}}$ to compute $\langle\hat{O}_{\text{S}}\rangle(\Lambda_{\text{c}};t)=\langle\hat{O}_{\text{S}}\rangle^\text{free}(t)$ corresponding to Eq.~(\ref{eq:free}). 
To improve the accuracy of the reconstruction, we also perform a regularization procedure by analytically continuing a complete set of observables (in this case, the three Pauli matrices) and imposing physicality on the extrapolated result. Specifically, the regularization we considered here corresponds to writing
\begin{equation}
\label{eq:phys_reg}
    \rho^\text{reg}_S(t)=\frac{1}{2}\left(1+\sum_{i=x,y,z} \frac{\text{Re}\left[\langle\sigma_i \rangle(t;\Lambda_{\text{c}})\right]}{Z}\right)\;,
\end{equation}
where 
\begin{equation}
    Z=\text{max}\left[1,\sum_{i=x,y,z}\left\{\text{Re}\left[\langle\sigma_i \rangle(t;\Lambda_{\text{c}})\right]\right\}^2\right]\;.
\end{equation}
In Fig.~\ref{fig:brownian_dynamics_T0}(a) we show the original dissipative dynamics alongside the ones determined by coupling additional noise (i.e., the regularized antimode environment) to the system. The values of these observables are then used to recover the noise-free dynamics. In  Fig.~\ref{fig:brownian_dynamics_T0}(b) the underlying analytical continuation procedure is shown explicitly in the complex plane for a specific time of the dynamics. As shown in  Fig.~\ref{fig:brownian_dynamics_T0}(a), the effect of the further regularization in Eq.~(\ref{eq:phys_reg}) helps in reproducing the correct values for the noise-free dynamics.

We present a further exemplification in Fig.~\ref{fig:gate}, where we plot the fidelity of a simple single-qubit gate, consisting of a $\pi/2$ rotation around the $y$-axis in the Bloch sphere. The qubit is in contact with a zero-temperature bath which lowers the quality of the gate as the gate-time is increased. We show that the effects of the bath on this single-qubit operation can be mitigated through analytical continuation.   
\begin{figure}[t!]
\includegraphics[width = \columnwidth]{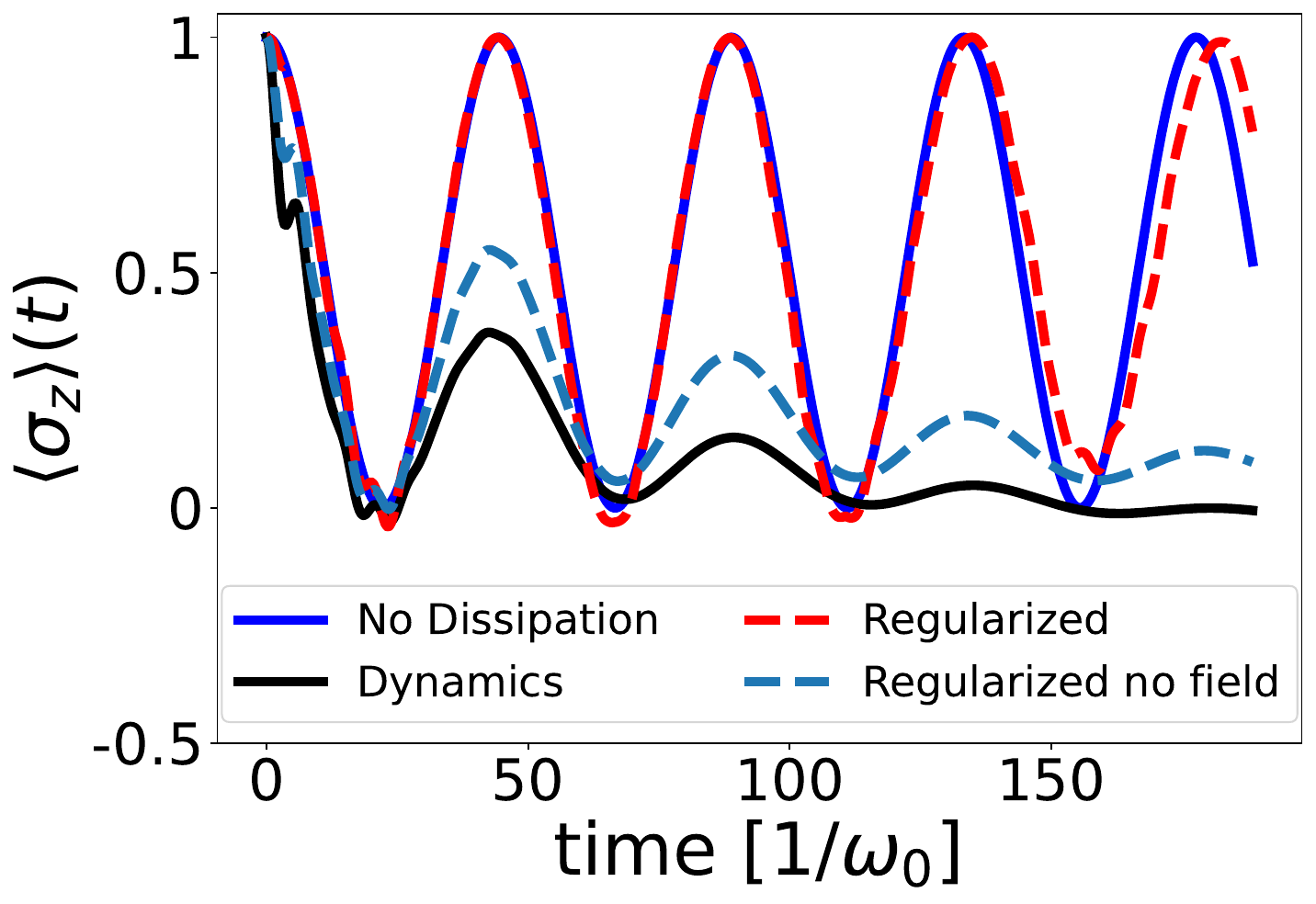} 
\caption{\label{fig:mitigation_finiteT} Mitigation of an underdamped Brownian environment at finite temperature. The black/blue curves show the dissipative/free dynamics. The dashed red/blue curves show the mitigation with/without a stochastic field. The original system-bath is here modeled with the hierarchical equations of motion.
The specific parameters used here are: $\alpha=8\times 10^{-3}$, $\Gamma=0.1\omega_0$, $\omega_s=\Delta=0.1\omega_0$, $M=10$, $N_\text{exp}=12$, $N_\xi=5000$, and $\beta=1/\omega_0$.}
\end{figure}
\subsubsection{Finite Temperature}
As mentioned in the introduction to this section, the effects of a bath with spectral density given by Eq.~(\ref{eq:spectral_density_main}) in the underdamped regime can be modeled using a single resonant harmonic mode and a classical stochastic field even at finite temperature. In fact, \emph{all temperature effects of the bath are encoded in the statistics of this driving field}. As a consequence, in order to mitigate temperature effects, we only need to update the way we handle this classical stochastic process with respect to the (zero temperature) analysis in the previous section. Such a case was, in fact, rather special as the stochastic driving field in the antimode model was real, thereby not requiring regularization and analytical continuation. This is no longer the case at finite temperature where, in order to satisfy its defining Eq.~(\ref{eq:Cxx}), the field $\xi_\beta(t)$ is no longer restricted to imaginary values but must, in general, be written as
\begin{equation}
    \xi_\beta(t)=\bar{\xi}^\mathcal{I}_\beta(t)-i \bar{\xi}^\mathcal{R}_\beta(t)\;,
\end{equation}
in terms of real and imaginary parts. Here, the notation simply follows from using the definition $\bar{\xi}_\beta(t)=\bar{\xi}^\mathcal{R}_\beta(t)+i \bar{\xi}^\mathcal{I}_\beta(t)$ for the  antifield in Eq.~(\ref{eq:lxi}). The mitigation of the imaginary part follows the same procedure as in the zero-temperature case, i.e., it only requires the introduction of a corresponding real anti-field. However, the mitigation of the real part now requires to be regularized and analytically continued, similarly to what was done before for the system-pseudomode coupling. In turn, this leads to the following modification to the regularized antimode model in Eq.~(\ref{eq:SBanti}) as
\begin{equation}
\label{eq:SBantiT}
\begin{array}{lll}
    \dot{\rho}_{\text{S+B}}&=&-i[H_\text{S+B}+\bar{\xi}^\text{reg}_\beta(t;\Lambda)\hat{s}(t),\rho_{\text{S+B}}]\\
    &&+{L}_{\bar{a}_\text{res}^\text{reg}}[\Xi(\Lambda)\lambda_\text{res};\rho_{\text{S+B}}]\;,
    \end{array}
\end{equation}
in terms of the physically regularized version of the antifield which, following Eq.~(\ref{eq:xiXixi}), reads
\begin{equation}  
\label{eq:field_anti}\bar{\xi}^\text{reg}_\beta(t;\Lambda)=\bar{\xi}^\mathcal{R}_\beta+\Xi(\Lambda)\bar{\xi}^\mathcal{I}_\beta\;.
\end{equation}
These equations  define the ensemble whose physical observables can be used to mitigate the noise of the environment $B$ at finite temperature. In Fig.~\ref{fig:mitigation_finiteT}, we analyze a specific example to highlight this model.
\subsection{Restructuring the environment}
\label{sec:restr}
\begin{figure}[t!]
\includegraphics[width = \columnwidth]{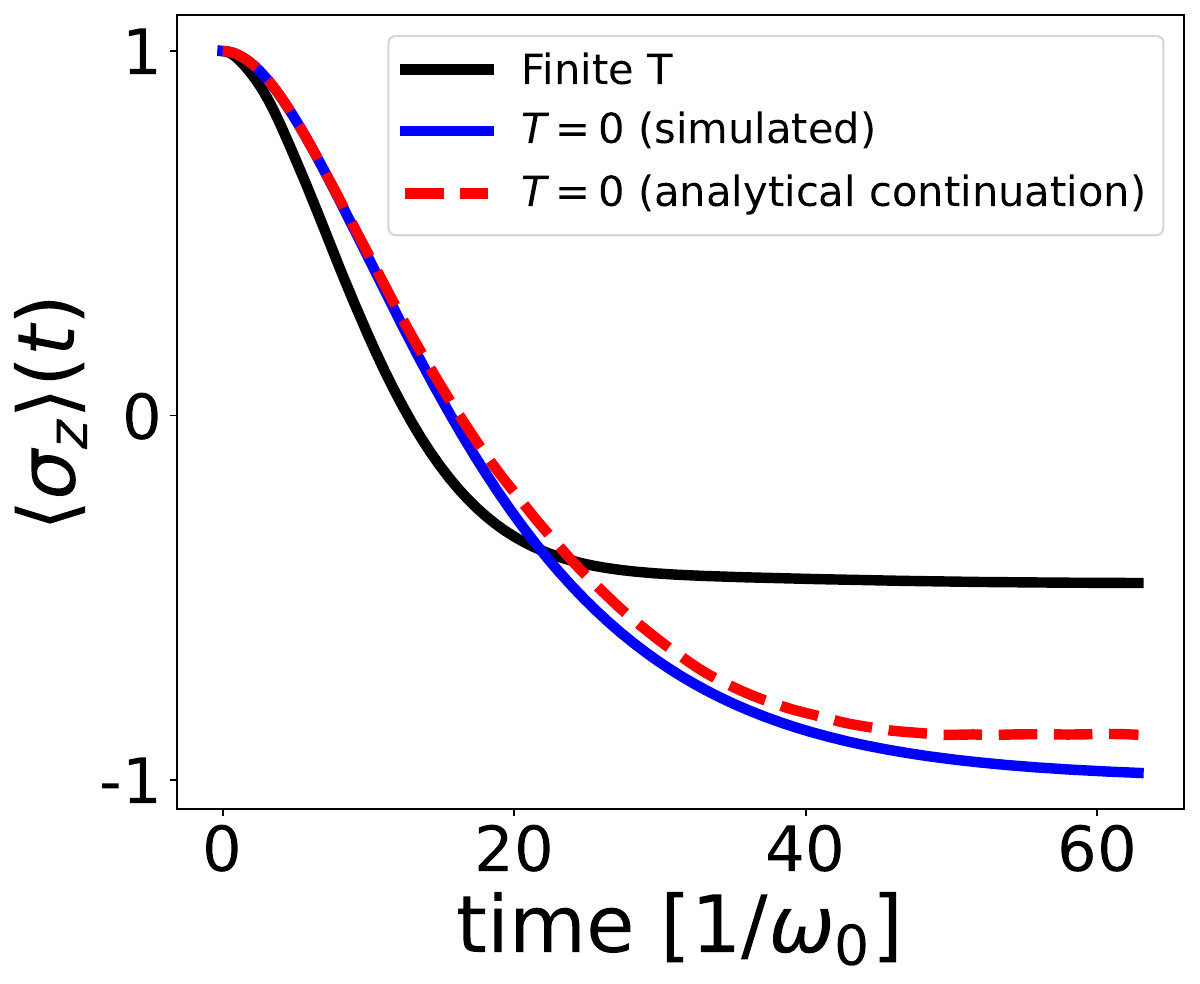} 
\caption{\label{fig:reshape} Restructuring of a finite temperature environment into a zero temperature one. The black/blue curves correspond to the dissipative dynamics when the system is in contact with a finite/zero temperature environment. For the dashed red curve, a stochastic field is added to effectively approximate the zero-temperature dynamics after analytical continuation. The specific parameters used here are:  $\omega_z=\omega_0$, $\Delta=0$, $\lambda=0.1 \omega_0^{3/2}$, $\gamma=0.3 \omega_0$, and $\beta=1/\omega_0$.}
\end{figure}
In this section, we consider a scenario in which one might wish to modify, i.e., \emph{restructure}, some properties of an environment without assuming them to be directly accessible. We specifically focus on the case in which the temperature of an environment is algorithmically reduced by using the analytical continuation procedure presented in the previous sections.

Interestingly, in the underdamped Brownian bath considered here, the modeling of this situation does not require the introduction of any additional \emph{quantum} degree of freedom. Indeed, it is possible to algorithmically modify the temperature of the original bath by simply performing analytical continuation over the intensity of a classical stochastic drive.  In fact, we can explicitly write the finite temperature stochastic pseudomdomode model as
\begin{equation}
\label{eq:SBantiT}
\begin{array}{lll}
    \dot{\rho}^{\beta}_{\text{S}+\text{PM}}&=&-i[H_\text{S}+\xi_\beta(t)\hat{s},\rho^{\beta}_{\text{S}+\text{PM}}]+{L}_{a_\text{res}}[\lambda_\text{res};\rho^{\beta}_{\text{S}+\text{PM}}]\;,
    \end{array}
\end{equation}
whose field satisfies Eq.~(\ref{eq:Cxx}). Therefore, the differential equation for a pseudomode model at a different temperature $\beta'$ can be obtained by simply adding an extra field $\xi_{\Delta\beta}(t)$ to the previous equation with
\begin{equation}
\mathbb{E}[\xi_{\Delta\beta}(t)\xi_{\Delta\beta}(0)]=
    C^\text{B}_\text{class}(t;\beta')-C^\text{B}_\text{class}(t;\beta)\;.
\end{equation}
Given these considerations, the temperature $\beta$ of a bath $B$ acting on a system $S$ can be algorithmically modified to a new temperature $\beta'$ by driving the system with a regularized stochastic field 
\begin{equation}
    \xi_{\Delta\beta}(t;\Lambda)=\xi^\mathcal{R}_{\Delta\beta}(t)+\Xi(\Lambda)\xi^\mathcal{I}_{\Delta\beta}(t)\;.
\end{equation}
To be more explicit, the dynamics in the full $(S+B)$ space becomes
\begin{equation}
\begin{array}{lll}
    \dot{\rho}_{\text{S+B}}&=&-i[H_{\text{S+B}}+\xi_{\Delta\beta}(t;\Lambda)\hat{s},\rho_{\text{S+B}}]\;,
    \end{array}
\end{equation}
so that, assuming the original dynamics to be well described by the model in Eq.~(\ref{eq:SBantiT}), the analytical continuation to $\Lambda_{\text{c}}$ effectively changes the classical correlation function to $C^\text{B}_\text{class}(t;\beta')$; thereby achieving the mentioned algorithmic change in temperature. We give an example of this procedure in Fig.~\ref{fig:reshape}.

\subsection{Simulation}
\begin{figure}[t!]
\includegraphics[width = \columnwidth]{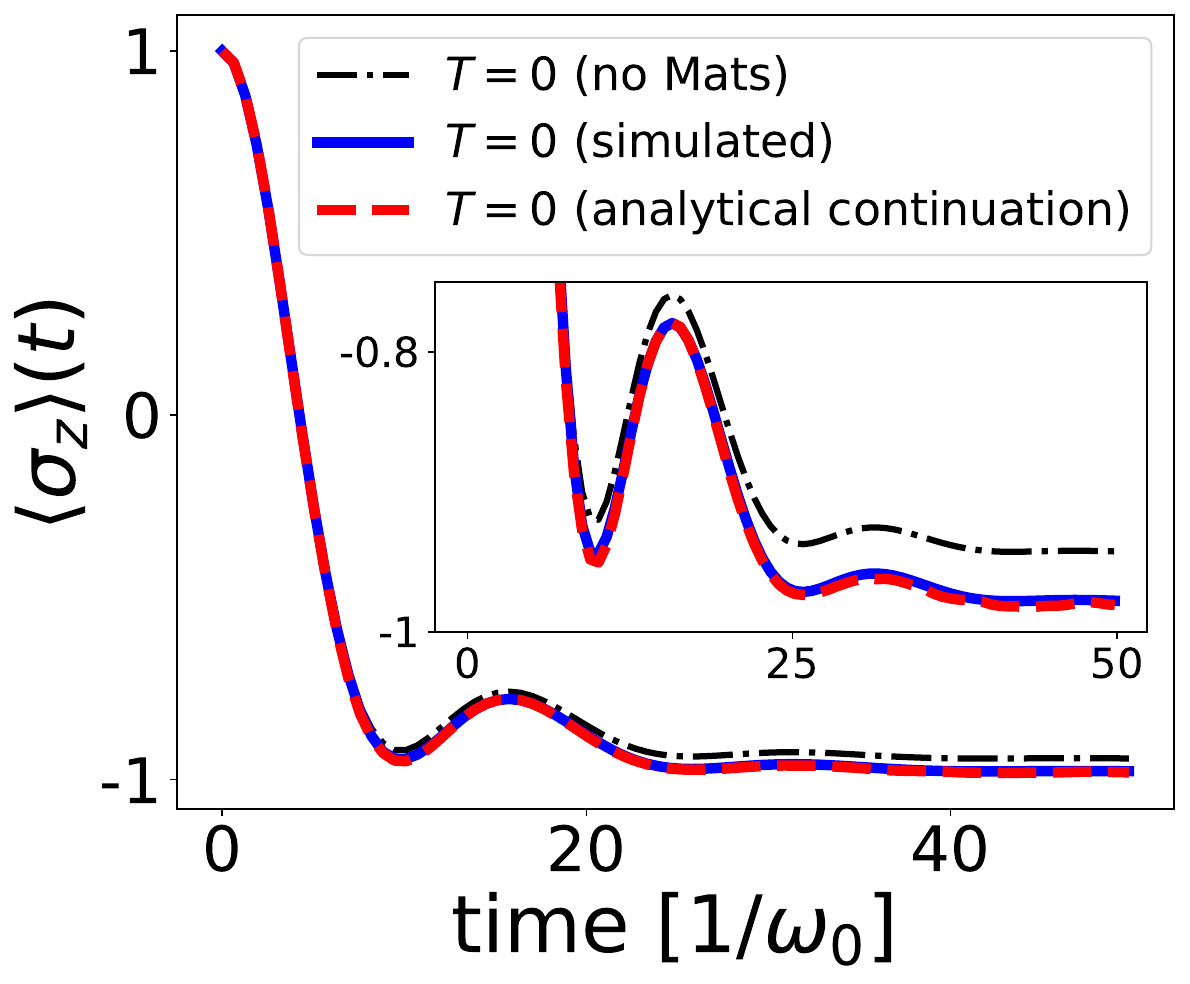} 
\caption{\label{fig:simulation} Analog simulation of a  zero temperature environment. The blue curve shows the true dissipative dynamics, in dashed-red the simulated one obtained by coupling the system to a harmonic mode and by analytically continuing a classical stochastic drive. The dashed black curve shows, the simulation done without the additional field, which is necessary to predict the correct hybridization to the environment, when the system is in contact with a zero temperature environment. The dashed red curve shows the dynamics after analytical continuation is performed upon an additional stochastic field. In the inset, 
 a restricted scale of the main plot is shown to better highlight the effects of the simulation. The specific parameters used here are: $\omega_s=\omega_0$, $\Delta=0$, $\lambda=0.3 \omega_0^{3/2}$, $\gamma=0.3 \omega_0$,  $N_\text{exp}=12$, and $N_\xi=10^3$.
}
\end{figure}
\begin{figure}[t!]
\includegraphics[width = \columnwidth]{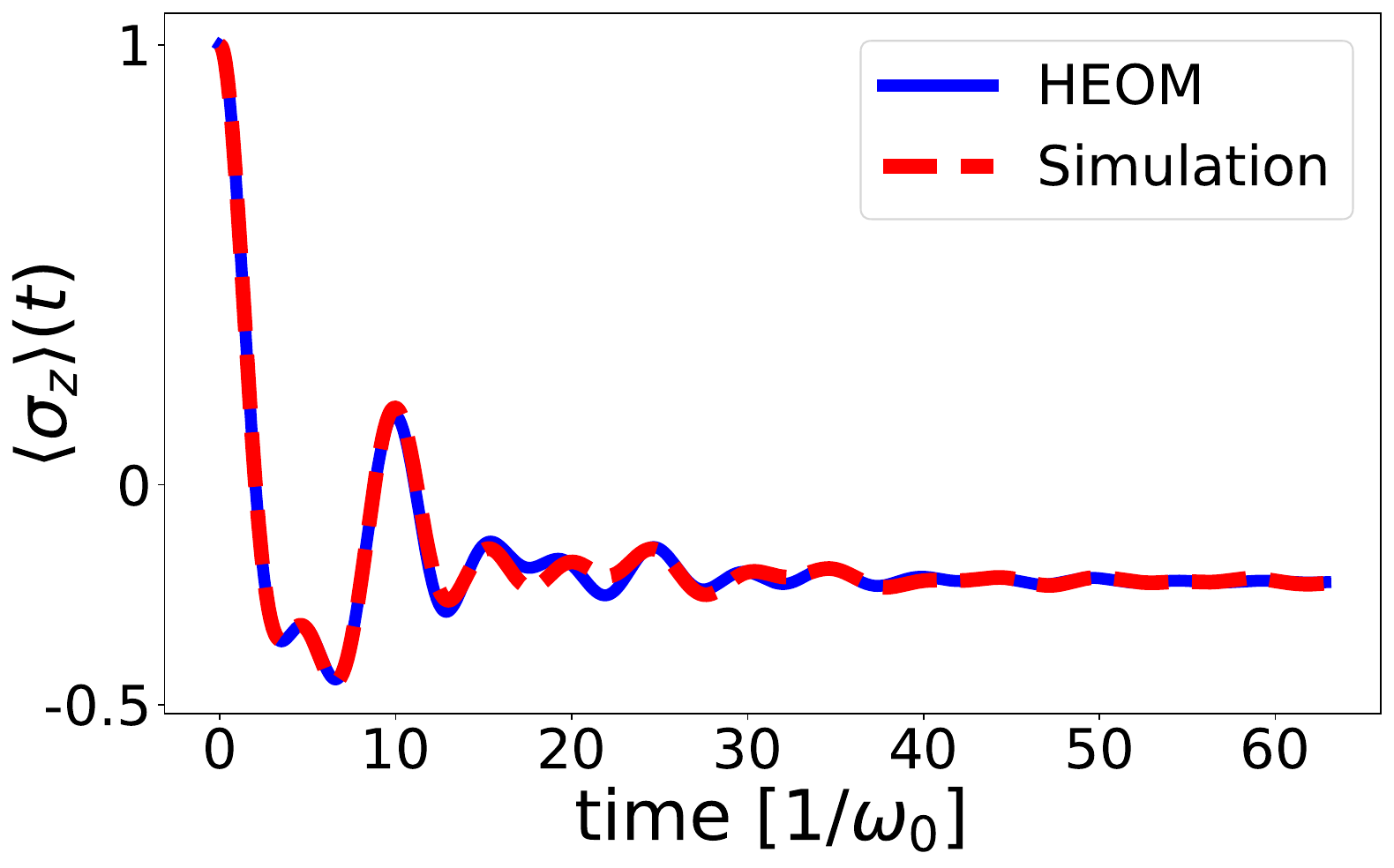} 
\caption{\label{fig:Energy_transfer} Dynamics of a two-level system corresponding to a model for excitonic energy transfer dynamics in a molecular dimer system \cite{Smith} at room temperature. The blue curve shows the results simulated by the HEOM and the dashed red curve shows the ones arising from a quantum simulation protocol in which the system is coupled to a single harmonic mode at zero temperature and a classical field. The high-temperature regime allows this field to be physical, thereby not requiring any analytical continuation procedure. 
The parameters were chosen to be on the same order as the ones used in Fig.~5 in \cite{Smith} and they explicitly are: $\omega_0=220 \;\text{cm}^{-1}$, $\omega_s=100 \;\text{cm}^{-1}$, $\Delta=100 \;\text{cm}^{-1}$, $\gamma=20 \;\text{cm}^{-1}$, $\alpha=2\lambda^2/\omega_0^2=80 \;\text{cm}^{-1}$, $T=300 \;\text{K}$ [corresponding to $1/\beta=k_B T=(k_B T) \lambda_0/(2\pi\hbar c) \hbar\omega_0=0.94\hbar\omega_0$, in terms of the Boltzmann constant $k_B$, the speed of light $c$, the wavelength $\lambda_0=(1/220)\;\text{cm}$] in units such that $\hbar=2\pi c = 1$. For consistency, in this simulation, the coupling operator to the environment is set to $\hat{s}=\sigma_z$.}
\end{figure}

Perhaps the most immediate application of the formalism presented here is in the analog simulation of the non-Makovian effects of an environment. This can be implemented in a rather direct way since the resonant pseudomode is physical. Given a closed system with Hamiltonian $H_{\text{S}}$, the effects of a Brownian environment at inverse temperature $\beta$ can be reconstructed from the physical ensemble obtained from Eq.~(\ref{eq:SBantiT}) with the replacement
\begin{equation}
\label{eq:field_sim}
    \xi_\beta(t)\mapsto \xi^\mathcal{R}_\beta(t)+\Xi(\Lambda)\xi^\mathcal{I}_\beta(t)\;.
\end{equation}
\begin{figure}[t!]
\includegraphics[width = \columnwidth]{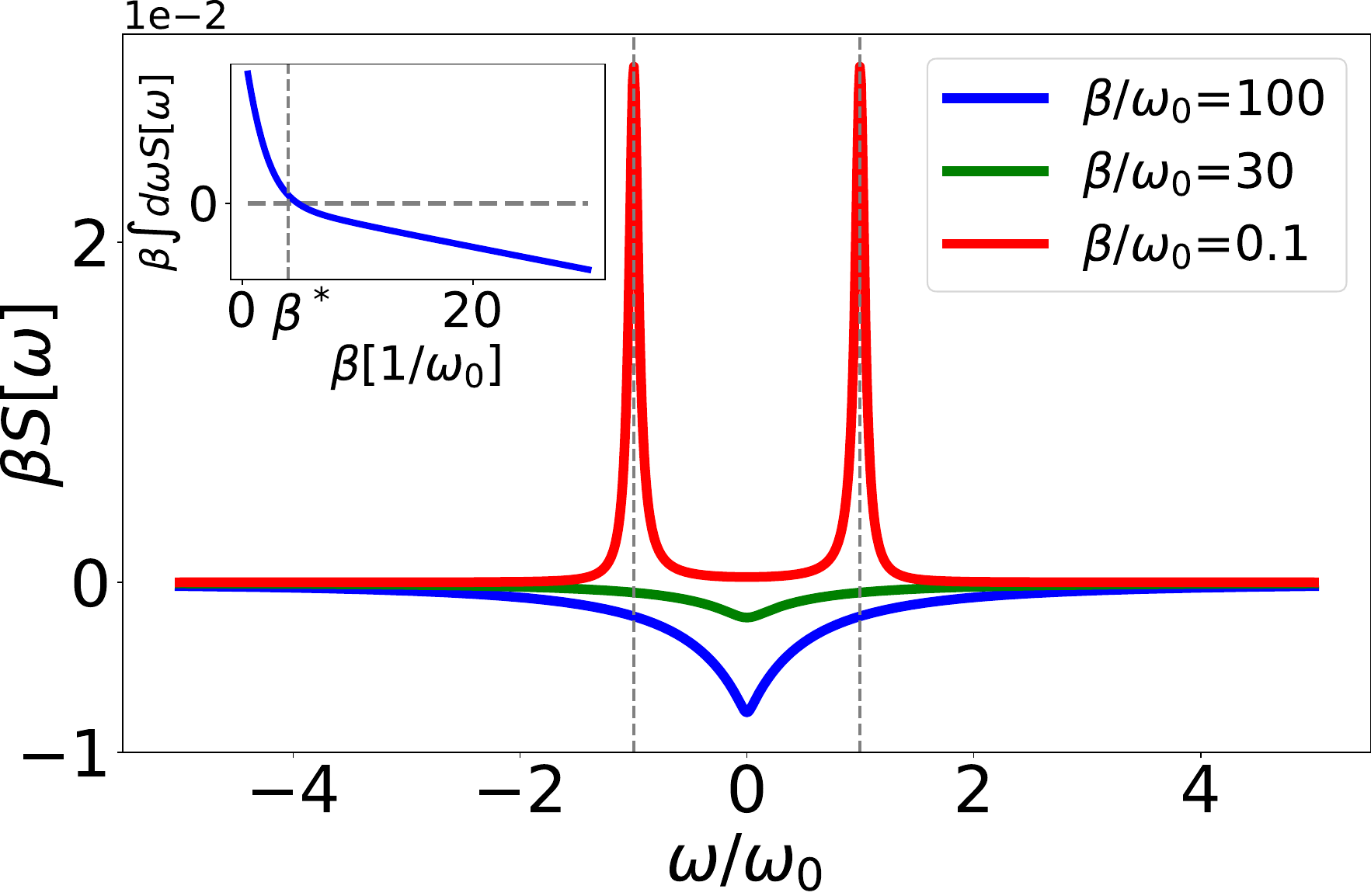} 
\caption{\label{fig:fourier} Classical spectrum corresponding to an underdamped Brownian spectral density as a function of frequency. At low temperatures, the correlation is characterized by the Matsubara contribution leading to a negative spectrum. In this regime, imaginary fields are required for the simulation of this enviornment. At higher temperatures, resonant terms (vertical lines highlight $\omega=\pm\omega_0$) dominate, leading to a positive spectrum. In this regime, the classical contribution to the correlation can be simulated using physical (real) fields. In the inset, we show, in arbitrary units,  the integral of the spectrum as a function of inverse temperature $\beta$ to highlight the cross-over (as the curve intersects zero) between the two regimes. Other parameters used here are: $\gamma = 0.1 \omega_0$, and $\lambda=0.1/\sqrt{2\pi} \omega_0^{3/2}$.}
\end{figure}
It is interesting to note that this expression is the ``dual'' of Eq.~(\ref{eq:field_anti}). In fact, while Eq.~(\ref{eq:field_anti}) is used to \emph{mitigate} the classical effects of the environment, Eq.~(\ref{eq:field_sim}) is used to \emph{simulate} them. We present a specific example of this analog procedure in Fig.~\ref{fig:simulation}, where we simulate the effects of a zero temperature environment.
 As mentioned above, the only obstacle preventing from a direct simulation of an environment is the imaginary component of the field $\xi(t)$. It is then worth asking whether there are regimes where such component is negligible. To do this, we can recall that the stochastic fields depend on the square-root of the parameters $c_n$ defining the spectral decomposition of the classical contribution to the bath correlation function, see Eq.~(\ref{eq:xi_spectral_representation_2_main}). Assuming the time dynamics to be the largest time-scale in the model, these coefficients become 
\begin{equation}
    c_n\rightarrow S_\text{class}[\omega]\; d\omega\;,
\end{equation}
which can be interpreted as a continuum version of Eq.~(\ref{eq:xi_cn}) and written in terms of the ``classical'' spectrum of the bath
\begin{equation}
    S_\text{class}[\omega]=\frac{1}{2\pi}\int_{-\infty}^\infty d\tau \;C_\text{class}(\tau)e^{i\omega \tau}\;.
\end{equation}
In the limit, the question about the possibility to simulate a bath using physical fields then becomes equivalent to check whether this quantity is positive, i.e., $S_\text{class}[\omega]\in\mathbb{R}_+$ for all $\omega$.
For the Brownian spectral density considered here, the expression for the classical contribution to the correlation is given explicit in Eq.~(\ref{eq:CQCc}) in terms of a sum of decaying exponentials, corresponding to a spectrum represented as an infinite sum of Lorentzians functions. In Fig.~\ref{fig:fourier}, we plot the spectrum for a specific set of parameters. From this figure we can recover the previously analyzed case: at low temperatures, the classical correlation is determined by the Matsubara contribution to the correlation which is negative, corresponding to imaginary fields. However, at higher temperatures, resonant terms in the correlation become dominant leading to a positive spectrum. This argument can be used (see Appendix \ref{sec:UB}) to derive an estimate for the inverse temperature locating the cross-over between the two regimes as
\begin{equation}
    \beta^*\omega_0=2\sqrt{4-(\gamma/\omega_0)^2}\;.
\end{equation}
For $\beta<\beta^*$, it is possible to produce a direct analogical simulation of non-Markovian effects at higher temperatures, such as in the case of the excitonic energy transfer  in molecular dimer systems \cite{Ishizaki_2,doi:10.1073/pnas.0908989106,doi:10.1146/annurev-conmatphys-020911-125126,Lambert13,Smith}. In Fig.~\ref{fig:Energy_transfer}, we simulate one of the environments analyzed in \cite{Smith}.

\section{Error Analysis}
\label{sec:stability}
The limitations of the method we presented rely on the interplay between imperfections in the experimental data and the complexity of the environmental effects to be mitigated.
\begin{figure}[t!]
\includegraphics[width = \columnwidth]{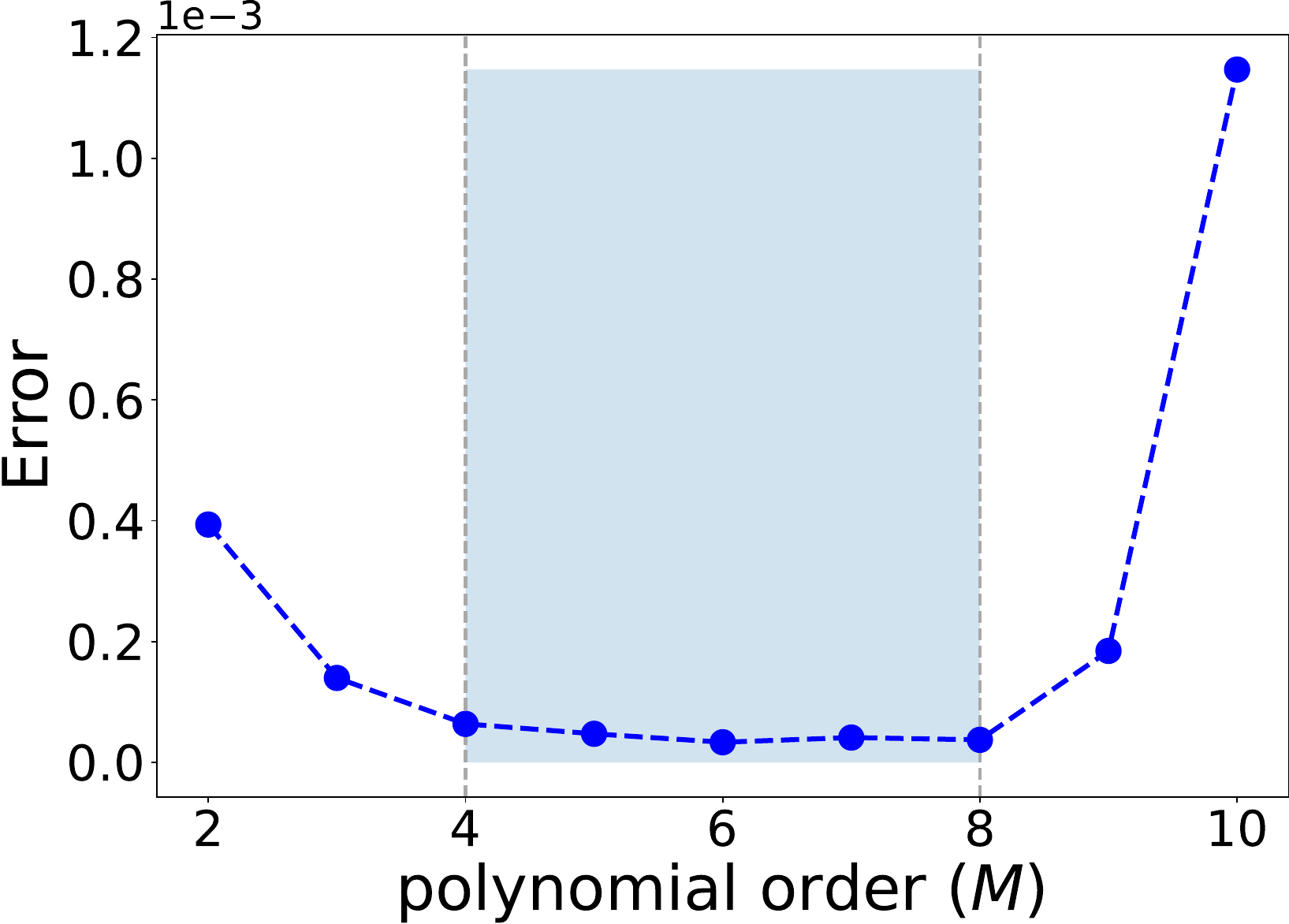} 
\caption{\label{fig:brownian_dynamics_T0_error} Reconstruction error as a function of the order $M$ of the interpolating polynomial. The error refers to the open quantum system analyzed in Fig.~\ref{fig:brownian_dynamics_T0}. More precisely, we defined $\text{Err}=|\langle\sigma_z(t)\rangle^\text{free}-\langle\sigma_z\rangle^\text{reg}(\Lambda_{\text{c}};t)|$ where the averages are taken with respect to $\rho^\text{free}(t)$ and $\rho^\text{reg}(t)$, respectively, and the time corresponds to the grey vertical line in Fig.~\ref{fig:brownian_dynamics_T0}. For small values of $M$ the error is biased, i.e., the polynomial is not able to encode the complexity of the environmental effects we want to mitigate. For high values of $M$, the algorithm is unstable with respect to errors in the initial data. In this example, the artificially injected error is modeled as a normal variable with zero mean and standard deviation $\sigma=10^{-5}$. As the variance of the error decreases, the optimal parameter range of the algorithm (shaded blue) extends towards the right, allowing to model more subtle features.}
\end{figure}

It is possible to make this intuitive consideration more precise by estimating the error in the computed observables after analytical continuation. In Appendix \ref{app:err}, we follow \cite{Demanet} and use the properties of Chebyshev polynomials (see, for example, \cite{Trefethen,Mason,Pachon}) to present precise upper bounds on the expected error. In general, the expected difference between the true observable following a noise-free dynamics and the one reconstructed through analytical continuation can be written as
\begin{equation}
\text{Err}=\text{Err}_\text{bias}+\text{Err}_\text{stability}\;.
\end{equation}
This expression depends on the sum of two qualitatively different terms characterizing the stability and the bias of the algorithm. In the case of perfect knowledge on the measured observables, the stability term is zero. In this case, the error is only due to the bias, i.e., by how precisely the $M$-order fitting polynomial is able to reconstruct the analytical continuation. In general, the bias becomes exponentially accurate by increasing $M$. Unfortunately, sensitivity to imperfections in the initial data does also increase exponentially with the order of the polynomial. This is encoded in the term $\text{Err}_\text{stability}$, which is also inversely proportional to the square-root of the number $N_\text{exp}$ of data available. 

This interplay between stability and bias is a common feature to extrapolation algorithm and it can be interpreted as the balance between underfitting and overfitting. For example, in Fig.~\ref{fig:brownian_dynamics_T0_error} we computed the error as a function of the degree of the interpolating polynomial showing the origin of an optimal range for the degree $M$.

\section{Conclusions}
\label{sec:conclusions}
We presented an extrapolation technique to mitigate, restructure, and simulate the effects of Gaussian non-Markovian environments on a quantum system. The method relies on the interaction of the system with ancillary leaky modes and a stochastic driving field to define a physical ensemble parameterized by a single parameter. 
Measurements results over this ensemble can be used to define an analytical continuation procedure, allowing us to use these ancilla modes to perform one of the following tasks:
\begin{itemize}
    \item Simulating the effects of a non-Markovian environment.
    \item Mitigating non-Markovian noise affecting the system.
    \item Restructuring some of the properties of a given environment (already interacting with the system) without directly accessing it.
\end{itemize}
We presented the details and proof of the general formalism, as well as several numerical examples to showcase the flexibility of the algorithm to adapt to different environments (such as zero and finite temperature ones, and a physically motivated example related to simulating excitonic energy transfer in a molecular dimer systems). These examples also demonstrated how the above range of applications can be used in practise (such as noise mitigation for the dynamics of quantum observables or quantum gate operations, simulation of zero-temperature environments, and restructuring of a finite temperature bath to a zero-temperature one).
The algorithm used in all these tasks is based on polynomial extrapolation and it is thereby ultimately limited by  instability against imperfections in experimental data.

As an outlook, the protocol presented here could also be adapted to specify or optimize digital quantum algorithms for the mitigation of non-Markovian noise in quantum computing tasks, or for the  simulation of zero-temperature environments for ground-state engineering, see \cite{lambert2023fixing}. In this context, it would  be relevant to  relax any requirement on prior knowledge about the environment by preceding the presented protocols with noise characterization techniques such as in \cite{PRXQuantum.4.040329}. We further note that the regularization procedure used in here could be extended to bigger system sizes by analytically continuing a reconstructed version of the full state using classical shadow tomography \cite{Aaronson1,Aaronson2,Huang2020}. In addition, the possibility of restructuring a given environment could, possibly, be helpful to push experimental setups past the limits imposed by the presence of physical baths elusive of direct manipulation.
\section{Acknowledgments}
 M.C. acknowledges support from NSFC (Grant No.~11935012) and NSAF (Grant No.~U2330401). F.N. is supported in part by: Nippon Telegraph and Telephone Corporation (NTT) Research, the Japan Science and Technology Agency (JST) [via the Quantum Leap Flagship Program (Q-LEAP), and the Moonshot R\&D Grant Number JPMJMS2061], and the Asian Office of Aerospace Research and Development (AOARD) (via Grant No. FA2386-20-1-4069).

 \nocite{apsrev41Control}
\bibliographystyle{apsrev4-2}
\bibliography{bib}
\newpage
\appendix

\section{The pseudomode model: main ideas}
\label{sec:PM}
The pseudomode method consists in replacing the original continuum of environmental modes with a discrete set of dissipative harmonic modes and stochastic driving fields. 
The main purpose of these ancillary degrees of freedom is to reproduce the correlation function characterizing the original Gaussian environment and, ultimately, to reproduce the original reduced system dynamics in Eq.~(\ref{eq:rhoS}).

 Explicitly, the pseudomode model consists in the following linear differential equation 
 \begin{equation}
\label{eq:Lindblad}
\dot{\rho}_{\text{S-PM}-\xi}(t)=L_{\text{S-PM}}[\rho_{\text{S-PM}}(t)]\;,
\end{equation}
for a (possibly stochastic) density matrix $\rho_{\text{S-PM}}(t)$ whose Hilbert space includes the system and $N_\text{PM}$ pseudomodes (PM). The Lindblad superoperator in the full system+pseudomodes space is written as the sum of three parts
\begin{equation}
\label{eq:LPM_main}
    L_{\text{S-PM}-\xi}=L_{\text{S}}+\sum_{k=1}^{N_\text{PM}}L^k_{\text{S-PM}}[\cdot]+L^\xi_S\;,
\end{equation}
where $L_{\text{S}}=-i[H_{\text{S}},\cdot]$ describes the free dynamics of the system and depends on its Hamiltonian $H_{\text{S}}$. The second term describes the dynamics of the quantum degrees of freedom as
\begin{equation}
\label{eq:superop}
L^k_{\text{S-PM}}[\cdot]=-i[H^k_{\text{PM}},\cdot] +  D_k[\cdot]\;.
\end{equation}
Here, $H^k_\text{PM}=\Omega_k a^\dagger_k a_k+\hat{s}X^k_{\text{PM}}$,  in terms of the annihilation operator $a_k$ associated with the $k$th pseudomode with frequency $\Omega_k\in\mathbb{C}$, $k=1,\dots,N_\text{PM}$. The coupling to the system is described by the interaction operator 
\begin{equation}
\label{eq:XPM}
X^k_{\text{PM}}=g_k (a_k+a_k^\dagger)\;,
\end{equation}
where $g_k\in\mathbb{C}$ constitutes an important conceptual difference from \eqref{real}, where the realness of the $\tilde{\lambda}_{\tilde{k}}$ couplings is required for Hermiticity. The dissipative properties of the $k$th pseudomode are independently characterized by the superoperator
\begin{equation}
\label{eq:dissipator}
\begin{array}{lll}
D_k[\cdot]&=&(n_k + 1)\Gamma_k \left(2a_k[\cdot] a_k^\dagger- a^\dagger_k a_k [\cdot]-[\cdot] a^\dagger_k a_k\right)\\
&&+n_k\Gamma_k \left(2a^\dagger_k[\cdot] a_k- a_k a^\dagger_k [\cdot]-[\cdot] a_k a^\dagger_k\right)\;,
\end{array}
\end{equation}
where $\Gamma_k, n_k\in\mathbb{C}$. We set the initial condition for Eq.~(\ref{eq:Lindblad}) to be $\rho_{\text{S}}(0)\otimes\rho_\text{PM}(0)$, where
\begin{equation}
\rho_\text{PM}(0)=\rho^\text{eq}_\text{PM}=\prod_{k=0}^{N_\text{PM}} \exp{[-\beta_k\Omega_k a^\dagger_k a_k]}/Z_k\;,
\end{equation}
where $Z_k$ are the coefficients needed to impose unity trace and where we assume the consistency condition $2n_k+1=\coth{(\beta_k\Omega_k/2)}$.
The presence of dissipation in this pseudo-environment does not break the Gaussianity of the model whose effect on the system are fully characterized by the two-point correlation
\begin{equation}
    C_\text{Q}(t)=\text{Tr}_{\text{PM}}[X_{\text{PM}}(t)X_{\text{PM}}(0)\rho^{\text{eq}}_{\text{PM}}]\;,
\end{equation}
with 
\begin{equation}
    X_{\text{PM}}(t)=\exp{\{L_{\text{S-PM}}^\dagger[\cdot] t\}}X_{\text{PM}}\;,
\end{equation}
in terms of $X_\text{PM}=\sum_k X^j_\text{PM} $ and $L_{\text{S-PM}}^\dagger$, the adjoint of $L_{\text{S-PM}}=\sum_{k=1}^{N_\text{PM}}L^k_{\text{S-PM}}$. Explicitly, each pseudomode independently contributes to the correlation function above as 
\begin{equation}
\label{eq:corr_lind_to_prove}
C_{\text{Q}}(t)=\sum_{k=0}^{N_\text{PM}} C_{\text{PM}}^k(t)\;,
\end{equation}
where
\begin{equation}
\label{eq:C_PM}
C_\text{PM}^k(t)=g^2_k\left[(n_k+1)e^{-i\Omega_k t}+n_ke^{i\Omega_k t}\right]\exp{[-\Gamma_k |t|]}
\end{equation}
characterizes the effect of the $k$th pseudomode, see \cite{Tamascelli,Lambert,LuoSi}.

The last term in Eq.~(\ref{eq:LPM_main}) is optional and its non-trivial presence defines a stochastic hybrid \cite{LuoSi} version of the fully deterministic pseudomode model above. In fact, this contribution can be interpreted as a stochastic driving of the system, i.e.
\begin{equation}
\label{eq:Lxi}
    L_{\text{S}}^\xi=-i\xi(t)[\hat{s},\cdot]\;,
\end{equation}
where the field $\xi(t)$ is assumed to be a Gaussian stochastic process with zero mean. Furthermore, we assume its auto-correlation function 
\begin{equation}
\label{eq:C_class}
C_\text{class}(t)=    \mathbb{E}[\xi(t)\xi(0)]\;,
\end{equation}
to be stationary (translationally invariant in time) so that it can be used to model classical properties of the original environment. In the spirit of this method, this driving field is allowed to  explore complex values; so that, in general, $\text{Im}[\xi(t)]\neq 0$. As shown in \cite{LuoSi}, it is possible to write this field in terms of the following spectral decomposition
\begin{equation}
\label{eq:xi_spectral_representation_2_main}
    \xi(t)=\sqrt{c_0} \xi_0+\sum_{n=1}^{N_\xi}\sqrt{2c_n}\left[\xi_n \cos(\omega_n t)+\xi_{-n}\sin(\omega_n t)\right],
\end{equation}
with  $\omega_n = n\pi/T$, and where $\xi_n$, $n=-N_\xi,\cdots,N_\xi$ ($N_\xi\in\mathbb{N}$) are Gaussian random variables with zero mean and unit variance.
 The coefficients 
\begin{equation}
\label{eq:xi_cn}
    c_n=\frac{1}{2T}\int_{-T}^Td\tau\cos(n\pi \tau/T)\;C_\text{class}(\tau)\;,
\end{equation}
define the spectral decomposition by ensuring that
\begin{equation}
\label{L2_expansion}
    C_\text{class}(t)=c_0+2\sum_{n=1}^{N_\xi}c_n \cos(\omega_n t)\;.
\end{equation}
 In the following, we will denote the average over the stochastic realization of the field $\xi(t)$ as $\mathbb{E}[\cdot]$.

It is important to stop for a moment and note that, within this formalism,  the reduced density matrix
\begin{equation}
\label{eq:rho_tilde}
    {\rho}_\text{S}(t)=\mathbb{E}\left[\underset{\text{PM}}{{\text{Tr}}}[\rho_{\text{S-PM}}(t)]\right]\;.
\end{equation}
 is as a function of the parameter set
\begin{equation}
\label{eq:G}
G_{\text{PM}}\equiv\{\Omega_k,g^2_k,\Gamma_{k},n_k,c_n\}\;,
\end{equation}
parametrizing the model. \emph{The pseudomode model consists in the characterization of these parameters; i.e., it can be interpreted as a map from the original Gaussian open quantum system to}  $G_{\text{PM}}$.
For Gaussian environments, a sufficient condition for 
\begin{equation}
\label{eq:main}
\rho_\text{S}(t)=\tilde{\rho}_\text{S}(t)\;,
\end{equation}
to hold,  is that 
\begin{equation}
\label{eq:correlationModel}
C_\text{E}(t)=C_{\text{PM}}(t)\equiv C_\text{Q}(t)+C_\text{class}(t)\;,
\end{equation}
which can be interpreted as \emph{the optimization equation defining the paramaters of the pseudomode model in Eq.~(\ref{eq:G}).}

We note that the second term in Eq.~(\ref{eq:correlationModel}) is the auto-correlation function of the stochastic process $\xi(t)$ defined in Eq.~(\ref{eq:C_class}) which is symmetric, i.e. classical, by construction. For generic spectral densities, the correlation in Eq.~(\ref{eq:correlation}) cannot be exactly reproduced by a finite number of psedudomodes; while this is possible, on average, within the stochastic hybrid approach \cite{LuoSi}. 
At the same time, it is important to stress that, while adding stochastic resources to the model might lead to a more efficient simulation of the open quantum system, it is not necessary. In other words, by imposing $\xi(t)=0$, the model above reduces to the fully deterministic pseudomode model.
However, its inclusion leads to some advantages, like a reduction of the Hilbert space dimension and the possibility to include all temperature effects in the noise statistics. In practice, this allows to initialize all the pseudomodes at zero temperature, \emph{for any temperature of the original bath}, i.e., to impose $\rho_\text{PM}(0)=\prod_{k=0}^{N_\text{PM}} \ketbra{0}{0}$. This is going to be the preferred choice throughout this article.
\section{Pseudomode model: additional details}
\label{app:PM_DT}
In this section we present more details about the deterministic and hybrid pseudomode model based on the work done in  \cite{Tamascelli,Lambert,LuoSi}.

\subsection{The deterministic pseudomode model}
We consider a Gaussian open quantum system $S+E$ made out of a system $S$ and its environment $E$ such that the interaction operator has the free correlation function $C_{\text{E}}(t)$ given in Eq.~(\ref{eq:correlation}). Following the strategy developed in \cite{Tamascelli}, we then proceed with the following steps
\begin{itemize}
\item[a.] Introduce an open quantum system $S+E'$ in terms of an alternative environment $E'$, such that the interaction operator with the system has a correlation given by $C_{\text{E}'}(t)=\sum_{k=0}^N C_{\text{PM}}^k(t)$, with the $C_{\text{PM}}^k(t)$ defined in Eq.~(\ref{eq:C_PM}). The environment $E'$ is made by $(N_{\text{PM}}+1)$ pseudomodes and $(N+1)$ bosonic baths for each of them. The reduced system dynamics of this model is equivalent to that of the original one, provided $C_{\text{E}}(t)=C_{\text{E}'}(t)$.
\item[b.] Show that reduced dynamics in the System+Pseudomode space of the open quantum system $S+E'$ above is equivalent to the one originating from the Lindblad Eq.~(\ref{eq:Lindblad}).
\item[c.] As a consistency check, we show that this Lindblad model has correlations given by Eq.~(\ref{eq:corr_lind_to_prove}).
\end{itemize}
The next three sections, labelled $a,b,c$ will prove the three points itemized above.
\subsubsection{Pseudomodes as an open quantum system}
Let us consider an open quantum system in which a subsystem of interest $S$ interacts with an alternative environment $E'$ made out of a set of $(N+1)$ pseudomodes $a_k$, each coupled to its own bosonic bath (whose modes are denoted as $b_{k\alpha}$) as
\begin{equation}
H = H_{\text{S}}+H_{\text{B}} + \hat{s}X_{\text{PM}}\;,
\end{equation} 
where the Hamiltonian of the environment $H_{\text{B}}$ is
\begin{equation}
\label{eq:H_B}
H_{\text{PM}}=\sum_k H_{\text{PM}}^{k}\;,
\end{equation}
with 
\begin{equation}
H_{\text{PM}}^{k}=\Omega_k a_k^\dagger a_k +\sum_{\alpha}\left[ \frac{ig_{k\alpha}}{\sqrt{2\omega_\alpha}}(b^\dagger_{k\alpha} a_k-a_k^\dagger b_{k\alpha})+\omega_{k\alpha}b^\dagger_{k\alpha}b_{k\alpha}\right].
\end{equation}
We also defined the interaction operator 
\begin{equation}
X_{\text{PM}}=\sum_{k=0}^{N_{\text{PM}}} X^k_{\text{PM}}=\sum\lambda_k/\sqrt{2\Omega_k} (a_k+a_k^\dagger)\;.
\end{equation}
We assume that the environment of each pseudomode is populated by modes with both positive and negative frequencies, i.e., $\omega_{k\alpha}\in(-\infty,+\infty)$. We further suppose that the spectral density associated with the bath of the $k$th pseudomode is constant, i.e., 
\begin{equation}
\label{eq:spectral_density}
J_k(\omega)=\pi\sum_\alpha\frac{g^2_{k,\alpha}}{2\omega_{k,\alpha}}\delta(\omega-\omega_{k,\alpha})=\Gamma_k\;,
\end{equation}
and that each of the pseudomodes $a_{k}$ and the corresponding environmental modes $b_{k\alpha}$ are in an initial state $ \rho^{0}_k=\exp{[-\beta_{k}\Omega_{k}a^\dagger_{k}a_{k}]}/Z_k$ and $\rho^0_{k\alpha}=\exp{[-\beta_{k\alpha}\omega_{k\alpha}b^\dagger_{k\alpha}b_{k\alpha}]}/Z_{k\alpha}$ respectively (with $Z_k$ and $Z_{k\alpha}$ imposing unit trace), together with the consistency condition 
\begin{equation}
\label{eq:unorthodox}
\beta_{k\alpha}\omega_{k\alpha}=\beta_k\Omega_k\;,
\end{equation}
which, in the continuum limit, reads
\begin{equation}
\label{eq:unorthodox_2}
\beta_{k}(\omega)\omega=\beta_k\Omega_k\;.
\end{equation}
We explicitly highlight the slight abuse in notation as $\beta_k(\omega)$ denotes the inverse temperature associated with environmental mode at frequency $\omega$ for the pseudomode $k$, while $\beta_k$ denotes the inverse temperature associated with the pseudomode $k$).
We note that these unorthodox conditions define a state for the full environment 
\begin{equation}
\begin{array}{lll}
\rho_{\text{E'}}&=&\displaystyle\prod_{k}\rho^0_{k}\prod_{k,\alpha}\rho^0_{k,\alpha}=\exp{[F_0]}/Z_0\;,
\end{array}
\end{equation}
which is not a  thermal state but it is the closest quantum idealization of classical white noise (see \cite{Gardiner}, pag.~164). Here, we defined 
\begin{equation}
F_0=\sum_k\beta_k\Omega_k a^\dagger_k a_k+\sum_{k\alpha}\beta_{k\alpha}\omega_{k\alpha} b^\dagger_{k\alpha} b_{k\alpha}\;,
\end{equation}
and $Z_0$ as a constant to impose unit trace.
The free correlation function of the interaction operator can be obtained as
\begin{equation}
C_{\text{PM}}(t_1,t_2)=\text{Tr}_{\text{E'}}\left[X_{\text{PM}}(t_2)X_{\text{PM}}(t_1)\rho_{\text{E'}}\right]\;,
\end{equation}
where $X_{\text{PM}}(t)=e^{i H_B t}X_{\text{PM}}e^{-i H_B t}$.
The unorthodox definition of the equilibrium state of the bath is designed to allow (see Appendix \ref{app:proof_commutation})
\begin{equation}
\label{eq:to_prove}
[H_B,F_0]=0\;,
\end{equation}
which, in turn, makes the correlation translational invariant in time since
\begin{equation}
\label{eq:corr_transl_inv}
\begin{array}{lll}
C_{\text{PM}}(t_1,t_2)&=&\text{Tr}_{\text{E'}}\left[e^{i H_B (t_2-t_1)}X_{\text{PM}}e^{-i H_B (t_2-t_1)}X_{\text{PM}}\right.\\
&&~~~~~~\times \left.e^{-i H_B t_1}\rho_{\text{E'}}e^{i H_B t_1}\right]\\
&=&\text{Tr}_{\text{E'}}\left[X_{\text{PM}}(t_2-t_1)X_{\text{PM}}\rho_{\text{E'}}\right]\\
&\equiv&C_{\text{PM}}(t_2-t_1)\;.
\end{array}
\end{equation}
Note that this is non-trivial because the Hamiltonian $H_B$ involves a Jaynes-Cummings interaction between the pseudomodes and the modes of their baths. Because of this, the previous relation does not hold in the presence of a true thermal equilibrium \cite{Gardiner}. 
In order to make progress evaluating $C_{\text{PM}}(t)$, we can first compute the formal solution for the Heisemberg equation of motion $\dot{b}_{k\alpha}=i[H_{\text{B}},b_{k\alpha}]$, and use it in the Heisemberg equation of motion for the pseudomodes $\dot{a}_{k}=i[H_{\text{B}},a_{k}]$. This leads to the following result for the Laplace transforms $\bar{x}_k, \bar{p}_k$ of the quadratures $x_k=a^\dagger_k+a_k$ and $p_k=i(a^\dagger_k-a_k)$
\begin{equation}
\begin{array}{lll}
s\bar{x}_k&=&\displaystyle x_k(0)+\left[\Omega_k-\int^\infty_{-\infty}d\omega\frac{J_k(\omega)\omega}{\pi(s^2+\omega^2)}\right]\bar{p}_k\\
&&\displaystyle-s\int_{-\infty}^\infty ~d\omega\frac{J_k(\omega)}{\pi(s^2+\omega^2)}\bar{x}_k-x^\text{in}_k\\
s\bar{p}_k&=&\displaystyle p_k(0)+\left[\Omega_k-\int^\infty_{-\infty}d\omega\frac{J_k(\omega)\omega}{\pi(s^2+\omega^2)}\right]\bar{x}_k\\
&&\displaystyle-s\int_{-\infty}^\infty ~d\omega\frac{J_k(\omega)}{\pi(s^2+\omega^2)}\bar{p}_k-p^\text{in}_k\;,
\end{array}
\end{equation}
where 
\begin{equation}
\begin{array}{lll}
x^\text{in}_k&=&\displaystyle\sum_{k,\alpha}\frac{g_{k,\alpha}}{\sqrt{2\omega_{k,\alpha}}}\left(\frac{b^\dagger_{k,\alpha}(0)}{s-i\omega_{k,\alpha}}+\frac{b_{k,\alpha}(0)}{s+i\omega_{k,\alpha}}\right)\\
p^\text{in}_k&=&\displaystyle i\sum_{k,\alpha}\frac{g_{k,\alpha}}{\sqrt{2\omega_{k,\alpha}}}\left(\frac{b^\dagger_{k,\alpha}(0)}{s-i\omega_{k,\alpha}}-\frac{b_{k,\alpha}(0)}{s+i\omega_{k,\alpha}}\right)\;.
\end{array}
\end{equation}
Now, using the expression for the spectral density in Eq.~(\ref{eq:spectral_density}) we find
\begin{equation}
\begin{array}{lll}
s\bar{x}_k&=&x_k(0)+\Omega_k\bar{p}_k-\Gamma_k \bar{x}_k-x^\text{in}_k\\
s\bar{p}_k&=&p_k(0)-\Omega_k\bar{x}_k-\Gamma_k \bar{p}_k-p^\text{in}_k\;,
\end{array}
\end{equation}
which leads to
\begin{equation}
[(s+\Gamma_k)^2+\Omega_k^2]\bar{x}_k=(s+\Gamma_k)[x_k(0)-x^\text{in}]+\Omega_k[p_k(0)-p^\text{in}_k]\;.
\end{equation}
We can now use this result in the  expression for the translational invariant correlation Eq.~(\ref{eq:corr_transl_inv}) to obtain
\begin{equation}
\begin{array}{lll}
C(t)&=&\displaystyle\sum_k\frac{\lambda_k^2}{2\Omega_k}\mathcal{L}_t^{-1}\left\{\text{Tr}_{\text{E'}}[\bar{x}_kx_k(0)\rho]\right\}\\
&=&\displaystyle\sum_k\frac{\lambda_k^2}{2\Omega_k}\frac{1}{2\pi i}\int ds ~e^{st}\left\{\frac{[s+\Gamma_k]\langle x_k(0)x_k(0)\rangle}{(s+\Gamma_k^2)+\Omega_k^2}\right.\\
&&+\displaystyle\left.\frac{\Omega_k\langle p_k(0)x_k(0)\rangle}{(s+\Gamma_k^2)^2+\Omega_k^2}\right\}\;,
\end{array}
\end{equation}
where we defined $\langle\cdot\rangle\equiv\text{Tr}_{\text{PM}}(\cdot\prod_k\rho_k)$, the trace being over the pseudomodes space. We note that, in the above derivation, translational invariance in time (derived thanks to the condition in Eq.~(\ref{eq:unorthodox})) was essential as it allowed a great simplification through the identity $\langle x^\text{in}x(0)\rangle=0$. Using $\langle x_k(0)x_k(0)\rangle=2 n_k+1$ (with $2n_k+1=\coth(\beta_k\Omega_k/2)$), and $\langle p_k(0)x_k(0)\rangle=-i$, we obtain
\begin{equation}
\label{eq:CCC}
C_{\text{PM}}(t)=\sum_{k=0}^{N_{\text{PM}}}\frac{\lambda_k^2}{2\Omega_k}\left(n_k e^{i\Omega_k t}+(1+n_k) e^{-i\Omega_k t}\right)e^{-\Gamma_k t}\;,
\end{equation}
as in Eq.~(\ref{eq:C_PM}) in the main text upon the definition $g_k=\lambda_k^2/2\Omega_k$.
Since, by hypothesis, this is a Gaussian open quantum system, the reduced dynamics of the system is fully determined by the functional form of $C_{\text{PM}}(t)$ through Dyson equation. As a consequence, this model reproduces the same dynamics as the original model as long as $C_{\text{E}}(t)=C_{\text{PM}}(t)$.
\subsubsection{Dissipative pseudomodes}
\label{eq:secsec}
The dynamics in the System+Pseudomodes space can be explicitly written in terms of the following influence functional expression
\begin{equation}
\label{eq:Inf_func}
\rho_{\text{S-PM}}=\hat{T}\exp{\{\hat{F}_t[\cdot]\}}\rho_{\text{S-PM}}(0)\;,
\end{equation} 
where $\hat{T}$ is the time-ordering operator and where the influence superoperator is
\begin{equation}
\label{eq:Inf_func_2}
\hat{F}_t[\cdot]=-\int_0^t dt_2\int_0^{t_2}dt_1 G(t_1,t_2)[\cdot]\;,
\end{equation}
where (using the shorthand $G\equiv G(t_1,t_2)[\cdot]$)
\begin{equation}
\begin{array}{lll}
G&=&\displaystyle\sum_k\langle{B_k^\dagger(t_2) B_k(t_1)}\rangle [a_k(t_2) a_k^\dagger(t_1)[\cdot]-a_k^\dagger(t_1) [\cdot] a_k(t_2)]\\
&&\displaystyle\sum_k\langle{B_k(t_2) B^\dagger_k(t_1)}\rangle [a_k^\dagger(t_2) a_k(t_1)[\cdot]-a_k(t_2)[\cdot]a_k^\dagger(t_1)]\\
&&\displaystyle\sum_k\langle{B_k(t_1) B^\dagger_k(t_2)}\rangle [[\cdot]a_k^\dagger(t_1) a_k(t_2) -a_k(t_2)[\cdot]a_k^\dagger(t_1)]\\
&&\displaystyle\sum_k \langle{B^\dagger_k(t_1) B_k(t_2)}\rangle[[\cdot]a_k(t_1) a_k^\dagger(t_2)-a_k^\dagger(t_2) [\cdot] a_k(t_1)],
\end{array}
\end{equation}
where we defined 
\begin{equation}
    B_k(t_i)=\sum_{\alpha}{g_{k,\alpha}}/{\sqrt{2\omega_{k,\alpha}}}b_{k,\alpha}(t_i)\;,
\end{equation}
for $i = 1,2$ to characterize the interaction operator between the pseudomodes and their environment [see Eq.~(\ref{eq:H_B})]. Notice that the interaction picture used in the previous expression implies a change of frame defined by $U=\exp[i (H_{\text{S}}+H_{\text{PM}})t]$, where $H_{\text{S}}+H_\text{PM}=H_{\text{S}}+\sum_k H^k_{\text{PM}}$ is the Hamiltonian in the System+Pseudomode space.
Explicitly, 
\begin{equation}
\label{eq:BB}
\renewcommand{\arraystretch}{2.2}
\begin{array}{lll}
\langle B_k^\dagger(t_2) B_k(t_1) \rangle&=&\displaystyle\frac{1}{\pi}\int_{-\infty}^\infty d\omega J_k(\omega) n_k(\omega) e^{-i\omega(t_1-t_2)}\\
&=&\displaystyle 2\Gamma_k n_k\delta(t_2-t_1)\\
\langle B_k(t_2) B_k^\dagger(t_1) \rangle&=&\displaystyle\frac{1}{\pi}\int_{-\infty}^\infty d\omega  J_k(\omega) [1+n_k(\omega)] e^{-i\omega(t_1-t_2)}\\
&=&\displaystyle 2\Gamma_k [1+n_k]\delta(t_2-t_1)\;,
\end{array}
\end{equation}
where $2n_k(\omega)+1=\coth{\beta_{k}(\omega)\omega/2}$ and, importantly, we used the condition in Eq.~(\ref{eq:unorthodox}) to obtain $2n_k(\omega)+1=\coth{{\beta_{k}\Omega_k}/{2}}$, effectively implying
\begin{equation}
n_k(\omega)\mapsto n_k(\Omega_k)\equiv n_k\;.
\end{equation}
Using Eq.~(\ref{eq:BB}) into Eq.~(\ref{eq:Inf_func_2}), we get
\begin{equation}
\begin{array}{lll}
F_t[\cdot]&=&-\displaystyle\left\{\sum_k (1+n_k)\Gamma_k\left[a_k^\dagger(t) a_k(t)[\cdot]+[\cdot]a_k^\dagger(t) a_k(t)\right.\right.\\
&&-\displaystyle\left. 2a_k(t)[\cdot]a_k^\dagger(t)\right]t\\
&&+\displaystyle\sum_k n_k\Gamma_k\left[a_k(t) a_k^\dagger(t)[\cdot]+[\cdot]a_k(t) a_k^\dagger(t)\right.\\
&&-\displaystyle\left.\left. 2a_k^\dagger(t)[\cdot]a_k(t)\right]t\right\}\;,
\end{array}
\end{equation}
where we used $\int_0^t dt' \delta(t-t')=1/2$ (see Eq. 5.3.12 in \cite{Gardiner}). Going back to the Shroedinger picture, using Eq.~(\ref{eq:Inf_func}), and taking a time derivative, we get 
\begin{equation}
\label{eq:S-PM_app}
\dot{\rho}_{\text{S-PM}}=L_{\text{S-PM}}[\rho_{\text{S-PM}}]\;,
\end{equation}
where 
\begin{equation}
\label{eq:superop}
L_{\text{S-PM}}[\cdot]=-i[H_{\text{S}}+H_{\text{PM}},\cdot] +\sum_{k=0}^{N_{\text{PM}}} D_k[\cdot]\;,
\end{equation}
and 
\begin{equation}
\begin{array}{lll}
D_k[\cdot]&=&(n_k + 1)\Gamma_k \left(2a_k[\cdot] a_k^\dagger- a^\dagger_k a_k [\cdot]-[\cdot] a^\dagger_k a_k\right)\\
&&+n_k\Gamma_k \left(2a^\dagger_k[\cdot] a_k- a_k a^\dagger_k [\cdot]-[\cdot] a_k a^\dagger_k\right)\;,
\end{array}
\end{equation}
This is Eq.~(\ref{eq:Lindblad}) in the main text.
\subsubsection{Correlations for the dissipative pseudomodes}
As a consistency check, we now show that the correlation function computed in Eq.~(\ref{eq:CCC}) can, equivalently, also be computed using the Lindblad model developed in the previous section. Specifically, we want to show that
\begin{equation}
\label{eq:C'=C}
C'_{\text{PM}}(t)=C_{\text{PM}}(t)\;,
\end{equation}
where $C'_{\text{PM}}(t)=\text{Tr}_{\text{PM}}[X_{\text{PM}}(t)X_{\text{PM}}(0)]$, with $X_{\text{PM}}(t)=\exp{\{L_{PM}^\dagger[\cdot] t\}}X_{\text{PM}}$, where 
\begin{equation}
L^\dagger_\text{PM}=i[H_{\text{S}}+H_{\text{PM}},\cdot]+\sum_k D^\dagger_k[\cdot]\;,
\end{equation}
 with
\begin{equation}
\begin{array}{lll}
D^\dagger_k[\cdot]&=&(n_k + 1)\Gamma_k \left(2a^\dagger_k[\cdot] a_k-  [\cdot]a^\dagger_k a_k-a^\dagger_k a_k[\cdot]\right)\\
&&+n_k\Gamma_k \left(2a_k[\cdot] a^\dagger_k-  [\cdot]a_k a^\dagger_k-a_k a^\dagger_k[\cdot] \right)\;,
\end{array}
\end{equation}
We then have
\begin{equation}
\begin{array}{lllll}
L^\dagger_\text{PM}[a_k]&=&i[\Omega_k a^\dagger_k a_k,a_k]+D^\dagger_k[a_k]&=&-i\Omega_k a_k-\Gamma_k a_k\\
L^\dagger_\text{PM}[a^\dagger_k]&=&i[\Omega_k a^\dagger_k a_k,a^\dagger_k]+D^\dagger_k[a^\dagger_k]&=&i\Omega_k a_k-\Gamma_k a_k\;,
\end{array}
\end{equation}
so that
\begin{equation}
\begin{array}{lll}
C'_{\text{PM}}(t)&=&\displaystyle\sum_{k=0}^{N_{\text{PM}}}\frac{\lambda_k^2}{2\Omega_k}\langle(a_k e^{-i\Omega_k}+a^\dagger_k e^{-i\Omega_k})(a_k+a^\dagger_k)\rangle e^{-\Gamma_k t}\\
&=&\displaystyle\sum_{k=0}^{N_{\text{PM}}}\frac{\lambda_k^2}{2\Omega_k}\left(n_k e^{i\Omega_k t}+(1+n_k) e^{-i\Omega_k t}\right)e^{-\Gamma_k t}\;,
\end{array}
\end{equation}
which implies Eq.~(\ref{eq:C'=C}).
\subsubsection{Proof of Eq.~(\ref{eq:to_prove})}
\label{app:proof_commutation}
Here, we prove that
\begin{equation}
[H_B,F_0]=0\;,
\end{equation}
where
\begin{equation}
\begin{array}{lll}
F_0&=&\displaystyle\sum_k\beta_k\Omega_k a^\dagger_k a_k+\sum_{k,\alpha}\beta_{k,\alpha}\omega_{k,\alpha} b^\dagger_{k,\alpha} b_{k,\alpha}\\
H_B&=&\displaystyle\sum_k \left[\Omega_k a_k^\dagger a_k +i\sum_{\alpha} \frac{g_{k,\alpha}}{\sqrt{2\omega_\alpha}}(b^\dagger_{k\alpha} a_k-a_k^\dagger b_{k\alpha})\right.\\
&&\displaystyle\left.+\sum_{\alpha}\omega_{k\alpha}b^\dagger_{k,\alpha}b_{k,\alpha}\right]\;.
\end{array}
\end{equation}
In fact, we have
\begin{equation}
\begin{array}{lll}
[H_B,F_0]&=&\displaystyle i\sum_{k,\alpha} \frac{g_{k,\alpha}}{\sqrt{2\omega_\alpha}}[(b^\dagger_{k\alpha} a_k-a_k^\dagger b_{k\alpha}), F^{\text{eq}}]\\
&=&\displaystyle i\sum_{k,\alpha} \frac{g_{k,\alpha}}{\sqrt{2\omega_\alpha}}(b^\dagger_{k\alpha} a_k+a_k^\dagger b_{k\alpha})(\beta_k\Omega_k-\beta_{k,\alpha}\omega_{k,\alpha})\\
&=&0\;,
\end{array}
\end{equation}
where, in the last step, we used Eq.~(\ref{eq:unorthodox}).
\subsection{A hybrid pseudomode model}
As shown in \cite{LuoSi}, it is possible replace some of the quantum degrees of freedom present in the fully deterministic pseudomode model presented in the previous section with a classical stochastic colored noise $\xi(t)$ which is stationary, Gaussian, and with zero mean. In fact, the addition of a driving term $\xi(t)\hat{s}$ in the Hamiltonian in Eq.~(\ref{eq:H_B}) effectively adds, after averaging over the noise, a term
\begin{equation}
    C_\text{class}(t)=\mathbb{E}[\xi(t_2)\xi(t_1)]\;,
\end{equation}
where $t_2-t_1=t$ and which accounts for the statistics of the field. In turn, the dynamics of this hybrid model can be explicitly written by adding the stochastic noise in Eq.~(\ref{eq:S-PM_app}) to write the Lindblad operator as
\begin{equation}
    L_{\text{S-PM}}=L_{\text{S}}+\sum_{k=1}^{N_\text{PM}}L^k_{\text{S-PM}}[\cdot]-i[\xi(t)\hat{s},\cdot]\;.
\end{equation}
While we refer to \cite{LuoSi} for more details, here we note that the addition of noise in the model corresponds to a decomposition of the effects of the original environment into a classical and a quantum part, i.e., 
\begin{equation}
    C_\text{E}(t)=C_\text{Q}(t) + C_\text{class}(t)\;.
\end{equation}
in which the second term is symmetric under time reversal (so that it can be modeled using classical resources) while the first term is more general and requires ancillary quantum degrees of freedom.
Interestingly, this decomposition is \emph{not} unique allowing for further possibility of optimization in the model. For example, it is possible to chose the classical contribution $C_\text{class}(t)$ [hence the field $\xi(t)$] in such a way that all the pseudomodes modeling $C_\text{Q}(t)$ are, initially, at zero temperature. This is the choice which is used throughout this article.
\subsection{Underdamped Brownian Spectral Density}
\label{sec:UB}
In this section, we describe the explicit form of the pseudomode model for the spectral density $J_B(t)$ in Eq.~(\ref{eq:spectral_density_main}). At zero temperature, the deterministic pseudomode mode can be defined using a single resonant mode at $N_\text{mats}$ zero-frequency Matsubara modes. This follows from the decomposition
\begin{equation}
    C^B(t;\beta=\infty)=C_\text{res}(t)+M(t)\;,
\end{equation}
in terms of the resonant and Matsubara contributions
\begin{equation}
\label{Mzero}
\renewcommand{\arraystretch}{2.2}
    \begin{array}{lll}
C_\text{res}(t)&=&\displaystyle\frac{\lambda^2}{2\Omega}\exp{[-i\Omega t-\Gamma|t|]}\\
M(t)&=&\displaystyle-\frac{\lambda^2\gamma}{\pi}\int_0^\infty  \frac{dx ~x e^{-x|t|}}{[(\Omega+i\Gamma)^2+x^2][(\Omega-i\Gamma)^2+x^2]}\;.
    \end{array}
\end{equation}
It is possible to model these terms by introducing a resonant and $N_\text{Mats}$ Matsubara harmonic modes $a_\text{res}$, $a_k$ ($k=1,\cdots,N_\text{Mats}$) and define a pseudomode model through the Lindblad operator
\begin{equation}
    L^B_\text{PM}= L_\text{res}+L_\text{Mats}\;,
\end{equation}
where 
\begin{equation*}
   L_\text{res}[\cdot]=-i[H_\text{res},\cdot]+\Gamma_\text{res}[2a_\text{res}\cdot a^\dagger_\text{res}-a^\dagger_\text{res}a_\text{res}\cdot-\cdot a^\dagger_\text{res}a_\text{res}],
\end{equation*}
 with 
 \begin{equation}
     H_\text{res}=\Omega_\text{res}a^\dagger_\text{res}a_\text{res}+\lambda_\text{res}(a_\text{res}+a^\dagger_\text{res})\hat{s}\;.
 \end{equation} 
 These parameters are explicitly given by 
 \begin{equation}
     \begin{array}{lll}
\lambda_\text{res}&=&\lambda^2/2\Omega\\
\Omega_\text{res}&=&\Omega=\sqrt{\omega_0^2-\Gamma^2_\text{res}}\\
\Gamma_\text{res}&=&\Gamma=\gamma/2\;.
     \end{array}
 \end{equation}
 The remaining contribution to the Lindbladian can be written as 
 \begin{equation}
     L_\text{Mats}=\sum_{k=1}^{N_\text{Mats}}L^k_{\text{Mats}}\;,
 \end{equation}
 in terms of  
 \begin{equation*}
     L^k_\text{Mats}[\cdot]=-i[H^k_\text{Mats},\cdot]+\Gamma^k_\text{Mats}[2a_k\cdot a^\dagger_k-a^\dagger_ka_k\cdot-\cdot a^\dagger_k a_k]\;,
 \end{equation*}
 with 
 \begin{equation}
     H^k_\text{Mats}=\Omega_ka^\dagger_ka_k+\lambda_k(a_k+a^\dagger_k)\hat{s}\;.
 \end{equation}
 These parameters can be estimated by fitting the corresponding correlation 
\begin{equation}
\label{eq:CCC}
C_{\text{Mats}}(t)=\sum_{k=1}^{N_{\text{Mats}}}\frac{\lambda_k^2}{2\Omega_k}\left(n_k e^{i\Omega_k t}+(1+n_k) e^{-i\Omega_k t}\right)e^{-\Gamma_k t}\;,
\end{equation}
to $M(t)$, see \cite{Lambert}. However, because of the absence of oscillatory behavior in $M(t)$, it is possible to directly impose $\Omega_k=0$, $k=1,\cdots,N_\text{Mats}$. 

It is interesting to note that, in general, the number of Matsubara modes needed to reproduce the effects of $M(t)$ depends on the simulation time. In fact, we  note that the domain of integration in the integral in  Eq.~(\ref{Mzero}) is effectively restricted to values $x\ll 1/t$ by the presence of the exponential \cite{Loss}. Therefore,
\begin{equation}
\begin{array}{lll}
    M(t)&\overset{t\gg 1/\Omega,1/\Gamma}{\simeq}&-\displaystyle\frac{\lambda^2\gamma}{\pi}\frac{\partial}{\partial t}\int_0^\infty \frac{e^{-x|t|}}{\omega_0^4}=-\displaystyle\frac{\alpha}{\pi}\frac{1}{t^2}\;,
    \end{array}
\end{equation}
where 
\begin{equation}
    \alpha=\lambda^2\gamma/\omega_0^2\;.
\end{equation}
This shows that $M(t)$ is characterized by an asymptotic ($\sim1/t^2$) polynomial decay which contrasts with the exponential decay present in the  Matsubara correlation, see Eq.~(\ref{eq:CCC}).

At finite temperatures, the decomposition of the correlation becomes
\begin{equation}
    C^\text{B}(t;\beta)=C^\text{B}_\text{res}(t;\beta) +M(t;\beta)\;,
\end{equation}
where
\begin{equation}
\renewcommand{\arraystretch}{2.2}
\begin{array}{lll}
    C^\text{B}_\text{res}(t;\beta)    
    &=&\displaystyle \frac{\lambda^2}{4\Omega}\coth{(\beta(\Omega+i\Gamma)/2)}e^{i\Omega |t|} e^{-\Gamma|t|}\\
   &&\displaystyle -\frac{\lambda^2}{4\Omega}\coth{(\beta(-\Omega+i\Gamma)/2)}e^{-i\Omega |t|} e^{-\Gamma|t|}\\
   &&-\displaystyle\frac{\lambda^2}{4\Omega}\left(-e^{-i\Omega t}+e^{i\Omega t}\right)e^{-\Gamma |t|}\\
   
    M(t;\beta)&=&\displaystyle\frac{2i}{\beta}\sum_{k>0}J^B(\omega^\text{M}_{k})\exp{[-|\omega^\text{M}_{k}||t|]}\;,
    \end{array}
\end{equation}
in terms of $\omega^\text{M}_k=2\pi k i /\beta$ ($k=1,\cdots,\infty$). In order to reproduce it, we can introduce three ``resonant'' modes $a_{j;\text{res}}$, $j=1,2,3$ alongside $N_\text{mats}$ ``Matsubara'' modes $a_k$, $k=1,\cdots N_\text{mats}$ characterized by the Lindbladian
\begin{equation}
\label{eq:dynamics_B_app}
L^\beta_\text{PM}=L^\beta_\text{res}+L^\beta_\text{Mats}\;.
\end{equation}
The Matsuabara contribution is formally the same as in the zero temperature case, but its parameters are defined to fit
$M(t;\beta)$ instead of its zero-temperature limit in Eq.~(\ref{Mzero}).
However, the resonant part has to be updated and defined by
\begin{equation}
    L^\beta_\text{res}[\cdot]=-i[H^\beta_\text{res},\cdot]+ D^\beta_\text{res}[\cdot]\;,
\end{equation}
where
\begin{equation}
\begin{array}{lll}
H^\beta_\text{res}&=&\displaystyle \sum_{j=1}^{3}\lambda^j_\text{res}[a^j_\text{res}+(a^j_\text{res})^\dagger]\hat{s}+\Omega^j_\text{res} (a^j_\text{res})^\dagger a^j_\text{res}\\
D^\beta_\text{res}[\rho]&=&\displaystyle{\sum_{j=1}^{N_\text{mats}}}\Gamma_j[(n_j+1)(2 a_j\rho a_j^\dagger-a_j^\dagger a_j\rho-\rho a_j^\dagger a_j)\\
&&\displaystyle +n_j(2 a^\dagger_j\rho a_j-a_j a^\dagger_j\rho-\rho a_j a^\dagger_j)]\;,
\end{array}
\end{equation}
as a function of the parameters
\begin{equation}
\label{eq:underdamped_full_PM_param}
    \begin{array}{lllllllll}
    \lambda_1&=&\displaystyle\sqrt{\frac{\lambda^2}{2\Omega}}&\lambda_2&=&\displaystyle\sqrt{\frac{I_B\lambda^2}{4\Omega}}&\lambda_3&=&\displaystyle\sqrt{\frac{-I_B\lambda^2}{4\Omega}}\\
    \Omega_1&=&\Omega&\Omega_2&=&0&\Omega_3&=&0\\
    \Gamma_1&=&\Gamma&\Gamma_2&=&\Gamma-i\Omega&\Gamma_3&=&\Gamma+i\Omega\\
    n_1 &=&\displaystyle\frac{R_B-1}{2}&n_2&=&0&n_3&=&0\;,\\
    \end{array}
\end{equation}
characterizing the resonant modes through the definitions 
\begin{equation}
    \begin{array}{lll}
R_B&=&\text{Re}\{\coth[\beta(\Omega+i\Gamma)/2)\}\\
I_B&=&\text{Im}\{\coth[\beta(\Omega+i\Gamma)/2)\}\;.
    \end{array}
\end{equation}
 We note that other choices which only use two resonant modes initially prepared with complex-temperature values are possible \cite{Paul}.

The hybrid pseudomode model relies on writing the correlation function of the bath in the following alternative form
\begin{equation}
    C^\text{B}(t;\beta)=C^\text{B}_\text{class}(t;\beta) +C^\text{B}_\text{Q}(t) \;,
\end{equation}
where
\begin{equation}
\label{eq:CQCc}
\renewcommand{\arraystretch}{2.2}
\begin{array}{lll}
    C^\text{B}_\text{class}(t;\beta)    
    &=&\displaystyle \frac{\lambda^2}{4\Omega}\coth{(\beta(\Omega+i\Gamma)/2)}e^{i\Omega |t|-\Gamma|t|}\\
   &&\displaystyle -\frac{\lambda^2}{4\Omega}\coth{(\beta(-\Omega+i\Gamma)/2)}e^{-i\Omega |t|-\Gamma|t|}\\
   &&-\displaystyle\frac{\lambda^2}{4\Omega}\left(e^{-i\Omega t}+e^{i\Omega t}\right)e^{-\Gamma |t|}\\
   &&+\displaystyle\frac{2i}{\beta}\sum_{k>0}J^B(\omega^\text{M}_{k})\exp{[-|\omega^\text{M}_{k}||t|]}\\
    C^\text{B}_\text{Q}(t;\beta) &=&\displaystyle\frac{\lambda^2}{2\Omega}\exp{[-i\Omega t-\Gamma |t|]}\;,
    \end{array}
\end{equation}
Since the ``classical'' contribution $C^\text{B}_\text{class}(t)$ is symmetric under time reversal, it can be reproduced by a single classical stochastic process $\xi^B(t)$ with autocorrelation function 
\begin{equation}
    \mathbb{E}[\xi^B(t_2)\xi^B(t_1)]=C^\text{B}_\text{class}(t_2-t_1)\;.
\end{equation}
The remaining ``quantum'' contribution can be reproduced with a single ``resonant'' pseudomode characterized by the same $L_\text{res}$ Lindblad as the one defined in the zero temperature case.

Despite the presence of a series dependent on the Matsubara frequencies, the spectrum $S[\omega]$ can be computed in a closed form.
In fact, we can write
\begin{equation}
\renewcommand{\arraystretch}{2.2}
\begin{array}{lll}
    S[\omega]&=&\displaystyle\int_{-\infty}^\infty dt\; C(t) e^{-i\omega t}\\
    &=&\displaystyle\int_0^\infty \frac{d\bar{\omega}}{2\pi}J(\bar{\omega})\int_{-\infty}^\infty dt\left\{[\text{coth}(\beta\bar{\omega}/2)-1]e^{i(\bar{\omega}-\omega)t}\right.\\

    &&+\left.[\text{coth}(\beta\bar{\omega}/2)+1]e^{-i(\bar{\omega}+\omega)t}\right\}\\

    &=&\displaystyle \theta(\omega)J(\omega)[\text{coth}(\beta\omega/2)-1]\\
    &&+\displaystyle \theta(-\omega)J(-\omega)[\text{coth}(-\beta\omega/2)+1]\;,
\end{array}
\end{equation}
and
\begin{equation}
S[\omega]=S_\text{Q}[\omega]+S_\text{class}[\omega]\;,
\end{equation}
in terms of the Fourier transforms of the quantum and classical contributions to the correlation 
\begin{equation}
    S_{\text{Q}/\text{class}}[\omega]=\int_{-\infty}^\infty dt\; C_{\text{Q}/\text{class}}(t) e^{-i\omega t}\;.
\end{equation}
Assuming continuity and $J(0)=0$, the zero-frequency limit becomes
\begin{equation}
    S[0]=\lim_{\omega\rightarrow 0^+} J(\omega)\text{coth}(\beta\omega/2)=\frac{2}{\beta}J^*(0)
\end{equation}
where $J^*(\omega)\equiv J(\omega)/\omega$ which we assume to be defined at $\omega=0$ by continuity. For the Brownian spectral density above, we have
\begin{equation}
    S^\text{B}_\text{Q}[\omega]=\frac{\lambda^2\Gamma}{\Omega}\frac{1}{(\Omega+\omega)^2+\Gamma^2}\;,
\end{equation}
which allows to write the classical spectrum in a closed form as
\begin{equation}
\begin{array}{lll}
S^\text{B}_\text{class}&=&S^\text{B}[\omega]-S^\text{B}_\text{Q}[\omega]\;,
\end{array}
\end{equation}
where
\begin{equation}
\begin{array}{lll}
S^\text{B}[\omega]
    &=&\displaystyle \theta(\omega)J^\text{B}(\omega)[\text{coth}(\beta\omega/2)-1]\\
    &&+\displaystyle \theta(-\omega)J^\text{B}(-\omega)[\text{coth}(-\beta\omega/2)+1]\;,
\end{array}
\end{equation}
We can further write the zero-frequency limit in the following explicit form
\begin{equation}
    S^\text{B}[0]=\frac{2}{\beta}\frac{\gamma\lambda^2}{\omega_0^4}\;.
\end{equation}
We now present an intuitive estimation for the temperature above which this spectrum is positive. To do this, we note that, for low temperatures, the classical spectrum is negative (due to the fact that the classical correlation is dominated by the Matsubara contribution) and peaked at zero frequency while, at high temperature, the spectrum becomes positive over all frequencies. As a consequence of this consideration, there must be a cross-over temperature $\beta^*$ such that $S_\text{class}[0]=0$. Using the results above, this temperature can be computed explicitly. In fact, we have
\begin{equation}
    S^\text{B}_\text{class}[0]=S^\text{B}[0]-S^\text{B}_\text{Q}[0]=\frac{4\Gamma\lambda^2}{\beta\omega_0^4}=\frac{\lambda^2\Gamma}{\Omega\omega_0^2}\;.
\end{equation}
By imposing $S^\text{B}_\text{class}[0]=0$, the value $\beta^*$ can be written as
\begin{equation}
    \beta^*\omega_0=2\sqrt{4-(\gamma/\omega_0)^2}\;.
\end{equation}

\section{Error Analysis}
\label{app:err}
In this section, we present an analytical analysis of the estimation error of the polynomial extrapolation algorithm presented in section \ref{sec:pol_extr}. To start, we briefly review the main setting of the extrapolation protocol presented in the main article.

\subsection{General setting}
Following section \ref{sec:pol_extr}, we define $f(\Lambda)$ as the analytical continuation, in the parameter $\Lambda\in\mathbb{C}$, of the 
expectation value of the system variable $\hat{O}_S$ at time $t$. We note that, in this appendix, all time-dependences are omitted to simplify the notation.  Explicitly, the above definition amounts to the requirement that
\begin{equation}
    f(\Lambda)\equiv\langle \hat{O}_{\text{S}} \rangle{(\Lambda})\;,
\end{equation}
within the restricted  domain $\Lambda\in[-1,1]$, which parametrize the physical ensemble where $\hat{O}_S$ is measured. 
Despite this complete functional information not being practically available, we can assume to have access to   $N+1$ experimental values 
\begin{equation}
f^\text{exp}_n\equiv\langle \hat{O}_{\text{S}} \rangle{(\Lambda_n)}+\epsilon_n\equiv f^\text{true}_n+\epsilon_n\;,
\end{equation}
in the set $\Lambda_n\in[-1,1]$, $n=0,\dots,N$. Here, we further defined $f^\text{true}_n\equiv\langle \hat{O}_{\text{S}} \rangle{(\Lambda_n)}$ as the ``true'' value of the observable affected by noise through unbiased Gaussian random variables $\epsilon_n$ with zero mean and variance $\sigma^2_\text{exp}$. This data can now be used to introduce an order $M$ polynomial $p^\text{exp}_M(\Lambda)$ which minimizes the least-squares distance
\begin{equation}
\label{eq:least_d}
d^2=\sum_{n=0}^N|p^\text{exp}_M(\Lambda_n)-f^\text{exp}_n|^2\;.
\end{equation} 
In other words, $p^\text{exp}_M(\Lambda)$  constitutes our best estimate  for the analytical function $f(\Lambda)$ using the discrete set of available data $f^\text{exp}_n$. As explained in section \ref{sec:pol_extr}, we are ultimately interested in reconstructing the expectation  $\langle \hat{O}_{\text{S}} \rangle^\text{reconstructed}$ corresponding to an analytical continuation at a specific unphysical value denoted by $\Lambda_c\in\mathbb{C}$, i.e.,
\begin{equation}
\label{eq:OS_app}
\langle \hat{O}_{\text{S}} \rangle^\text{reconstructed}\equiv p^\text{exp}_M(\Lambda_{\text{c}})\;.
\end{equation}
In the following, we further limit the analysis to the case $N\geq M$, so that the number of data points available is bigger than the order of the extrapolating polynomial.

The goal of the following section is to provide an estimate for the extrapolation error of the quantity in Eq.~(\ref{eq:OS_app}), i.e., 
\begin{equation}
\label{eq:Errrror_app}
\begin{array}{lll}
\text{Err}&=&\displaystyle\mathbb{E}_{\text{exp}}[|\langle \hat{O}_{\text{S}} \rangle(\Lambda_{\text{c}})-\langle \hat{O}_{\text{S}} \rangle^\text{reconstructed}|]\\
&\equiv&\displaystyle \mathbb{E}_{\text{exp}}[|f(\Lambda_{\text{c}})-p^\text{exp}_M(\Lambda_{\text{c}})|]\;.
\end{array}
\end{equation}
where $\langle \hat{O}_{\text{S}} \rangle{(\Lambda_{\text{c}})}$ is the ``true'' analytically-continued value which we ultimately want to compute. Importantly, the average 
$\mathbb{E}_{\text{exp}}$ takes into account  the  uncertainties on the experimental data encoded in the random variables $\epsilon_n$.

To analytically approach the estimation of this error, in the next section we follow \cite{Demanet} and take advantage of an explicit extrapolation form, expressed in terms of Chebyshev polynomials. 

\subsection{ Chebyshev Polynomials}
\label{section:Chebyshev}
This subsection reviews the definitions and properties of the  Chebyshev polynomials of first kind, adapting the references \cite{Trefethen,Mason,Pachon}.

The Chebyshev polynomials of first kind are defined as
\begin{equation}
\label{eq:Cheb_def}
    T_m(x)=\cos{n\theta}\;,
\end{equation}
for $m\geq 0$ and for $x\in[-1,1]$, such that $x=\cos{\theta}$, which immediately implies the bound
\begin{equation}
\label{eq:bound_T_x}
|T_m(x)|\leq 1~~~~\text{for}~~~x\in[-1,1]\;.
\end{equation}
 While not immediately apparent from the definition, these functions satisfy the recurrence relation
\begin{equation}
    T_m(x)=2x T_{m-1}(x)-T_{m-1}(x)\;,
\end{equation}
which, together with the initial conditions $T_0(x)=1$ and $T_1(x)=x$ imply that each $T_m(x)$ is a  polynomial of order $m$. It is further possible to analytically continue these polynomials and to write them as
\begin{equation}
T_m(z)=\left[\left(z+\sqrt{z^2-1}\right)^m+\left(z-\sqrt{z^2-1}\right)^m\right]/2\;,
\end{equation}
where $z\in\mathbb{C}$.
Given a parameter $\rho>0$, it is possible to define the so-called Bernstein  ellipse 
\begin{equation}
E_\rho=\{z\in\mathbb{C}:z={(w+w^{-1})}/{2},w\in\mathbb{C},|w|=\rho>1\}\;,
\end{equation}
such that 
\begin{equation}
\begin{array}{lll}
|T_m(z\in\Gamma_\rho)|&=&|(\rho^m e^{im\theta}+\rho^{-m}e^{-im\theta})|/2\;.
\end{array}
\end{equation}
Given a point $z\in\mathbb{C}$ is very useful to define the following function
\begin{equation}
\label{eq:rhoz}
\rho_z=\left|z+\sqrt{z^2-1}\right|\;,
\end{equation}
i.e., such that $z\in\Gamma_{\rho_z}$ or, in other words, the Bernstein ellipse with parameter $\rho_z$ passes through the point $z$.

In the following, we will denote by $\tilde{E}_\rho$ the open Bernstein  ellipse, i.e., the region of the complex plane inside $E_\rho$ (i.e., the one containing the origin).
This constitutes a rather interesting identity because it implies that, on the Bernstein ellipse $E_\rho$ the Chebyshev polynomials satisfy the bounds
\begin{equation}
\label{eq:bound_T}
{[\rho^m-\rho^{-m}]}/{2}\leq |T_m(z\in\Gamma_\rho)|\leq{[\rho^m+\rho^{-m}]}/{2}\;,
\end{equation}
and, since $\rho>1$, the cleaner, but less strict one,
\begin{equation}
\label{eq:T_bound}
|T_m(z\in\Gamma_\rho)|\leq\rho^m\;.
\end{equation}
The Chebyshev polynomials $T(x)$ are a basis for functions in the interval $[-1,1]$. This means that every function $g(x)$  can be written \cite{Trefethen}, in this interval, in terms of the Chebyshev series
\begin{equation}
\label{eq:Cheb_series}
g(x)=\sum_{m=0}^\infty g_m T_m(x)=\sum_{m=0}^\infty g_m (z_\theta^m+z_\theta^{-m})/2\equiv G(z_\theta)\;.
\end{equation}
where $g_m\in\mathbb{C}$, and where we used Eq.~(\ref{eq:Cheb_def}) to define the function $G(z)$ on the unit circle $z_\theta=\exp[i\theta]$, $\theta\in[0,2\pi]$.
It is further possible to follow an elegant geometric construction \cite{Trefethen}, to show that an analytical continuation of $g(x)$ in $\tilde{E}_\rho$ corresponds to analytically continue $G(z)$ inside the annulus $\rho^{-1}\leq|z|\leq\rho$. Intuitively, this is a consequence of the fact that $G(z)$ can be interpreted as the pull-back of $g(x)$ on the unit circle by the Joukoasky map $J(z)=(z+z^{-1})/2$, $z\in\mathbb{C}$ (which projects the unit circle onto the real axis, i.e., $J(z)=\text{Re}[z]$ for $|z|=1$). In other words, $F(z)=g(x)$.
In turn, this allows to take advantage of the Laurent series in Eq.~(\ref{eq:Cheb_series}) to write \cite{Trefethen}
\begin{equation}
\frac{g_m}{2}=\frac{1}{2\pi i}\int_{|z|=\rho}dz~z^{-1+m}G(z)\;,
\end{equation}
where the factor $1/2$ should not appear on the left-hand side for $m=0$.  By further assuming $|g(z)|\leq Q_\rho$ for $x\in\tilde{E}_\rho$ the expression above implies
\begin{equation}
\label{eq:bound_an}
    |g_0|\leq Q_\rho\;,~\text{and}~|g_m|\leq 2 Q_\rho \rho^{-m}~\text{for}~m\geq 1\;,
\end{equation}
see theorem 8.1 in \cite{Trefethen}.
We can now use these bounds to analyze the error made by truncating the Chebyshev series in Eq.~(\ref{eq:Cheb_series}). To do this, we can define $g_M(x)$ as the truncated series
\begin{equation}
\label{eq:fm}
g_M(x)=\sum_{m=0}^M g_m T_m(x)\;.
\end{equation}
Using Eq.~(\ref{eq:bound_T_x}) and (\ref{eq:bound_an}), the error made by this truncation can be quantified as
\begin{equation}
\label{eq:ff_M_interval}
\begin{array}{lll}
\displaystyle\sup_{x\in[-1,1]}|g(x)-g_M(x)|&=&\displaystyle\sup_{x\in[-1,1]}\left|\sum_{m=M+1}^\infty g_m T_m(x)\right|\\
&\leq&\displaystyle 2Q_\rho\sum_{m=M+1}^\infty \rho^{-m}\\
&=&\displaystyle 2Q_\rho \frac{\rho^{-M}}{\rho-1}\;.
\end{array}
\end{equation}
We note that this bound is valid for all parameters $\rho>1$ defining an open Bernstein ellipse $\tilde{E}_\rho$ inside which $g(x)$ admits an analytical continuation. 

In the following subsection, we show how to use these results in order to provide an estimate for the error made in a polynomial extrapolation.
\subsection{Error analysis using Chebyshev Polynomials}
\label{app:err_Cheb}
In this subsection, we adapt the results elegantly presented in \cite{Demanet} to analyze in more detail the expression for the extrapolation error in Eq.~(\ref{eq:Errrror_app}) which averages the difference between the ``true'' extrapolation value $f(\Lambda_c)=\langle\hat{O}_S\rangle(\Lambda_c)$ and its $M$-order polynomial estimate $p^\text{exp}_M(\Lambda_c)$. Assuming that the function $f(\Lambda)$ admits analytical continuation, we can always write it in terms of a Chebyshev series as
\begin{equation}
\label{eq:full_series}
f(z)=\sum_{m=0}^\infty a_m T_m(z)\;
\end{equation}
with $a_m\in\mathbb{C}$. Similarly, the polynomial $p^\text{exp}_M(z)$ can be written as a truncated series
\begin{equation}
p_M(z)=\sum_{m=0}^M c_m T_m(z)\;,
\end{equation}
whose coefficients $c_n$ solve the least-squares optimization defined by the distance in Eq.~(\ref{eq:least_d}). Explicitly, we can write $c_m\equiv \vec{c}_m$, where
\begin{equation}
    {\vec{c}}=(T^\dagger T)^{-1}T^\dagger (\vec{f}+\vec{\epsilon})
\end{equation}
written in terms of the vector $\vec{\epsilon}=(\epsilon_0,\cdots,\epsilon_N)^T$ and
\begin{equation}
\label{eq:vector}
\vec{f}=\left(\begin{array}{c}f_0\\\vdots\\f_N\end{array}\right)=\left(\begin{array}{c}f(\Lambda_0)\\\vdots\\f(\Lambda_N)\end{array}\right)\equiv\left(\begin{array}{c}\langle\hat{O}\rangle_{\Lambda_0}\\\vdots\\\langle\hat{O}\rangle_{\Lambda_N}\end{array}\right),
\end{equation}
and
\begin{equation}
\label{eq:Matrix}
T=\left(\begin{array}{ccc}T_0(\lambda_0)&\cdots& T_M(\Lambda_0)\\\vdots&&\vdots\\T_0(\Lambda_N)&\cdots& T_M(\Lambda_N)\end{array}
\right)\;.\end{equation}
In order to introduce some of this definitions to analyze the full series in Eq.~(\ref{eq:full_series}), we can decompose it as
\begin{equation}
    f(z)=f_M(z)+\sum_{m=M+1}^\infty a_m T_m (z)
\end{equation}
in terms of the truncated series 
\begin{equation}
    f^M(z)\equiv\sum_{m=0}^M a_m T_m (z)\;.
\end{equation}
In this way, the coefficients $a_m$ ($m=0,\dots,M$) can be written explicitly as $a_m\equiv\vec{a}_m$, where
\begin{equation}
\label{eq:a_new}
\vec{a}=(T^\dagger T)^{-1}T^\dagger\vec{f}^M\;.
\end{equation}
Here, we introduced the vector \begin{equation}
\vec{f}^M=\left(\begin{array}{c}f^M(\Lambda_0)\\\vdots\\f^M(\Lambda_N)\end{array}\right)\;,
\end{equation}
which, as can be noted by comparison, is not equivalent to the one in Eq.~(\ref{eq:vector})  since it relies on the truncated series rather than the ``true'' value.
With this notation, we have
\begin{equation}
\begin{array}{lll}
\text{Err}&=&\mathbb{E}_\text{exp}\left[\displaystyle\left|\sum_{m=0}^\infty a_mT_m(z)-\sum_{m=0}^Mc_mT_m(z)\right|\right]\\
&\leq&\displaystyle \mathbb{E}_\text{exp}\left[\left|\sum_{m=0}^M (\vec{a}_m-\vec{c}_m)T_m(z)\right|\right]\\
&&\displaystyle +\left|\sum_{m=M+1}^\infty a_mT_m(z)\right|\\
&\leq&\displaystyle||(T^\dagger T)^{-1}T^\dagger||_\infty||(\vec{f}_M-\vec{f})||_\infty \sum_{m=0}^M|T_m(z)|\\
&&\displaystyle+\sum_{m=M+1}^\infty |a_m|\;|T_m(z)|\\
&&\displaystyle +\mathbb{E}_\text{exp}\left[||(T^\dagger T)^{-1}T^\dagger\vec{\epsilon}||_\infty\right] \sum_{m=0}^M|T_m(z)|\\
&\equiv&\displaystyle\text{Err}_1+\text{Err}_2\;,
\end{array}
\end{equation}
where we introduced the infinity norm and used Eq.~(\ref{eq:normdec}), see section \ref{sec:norms}  and defined
\begin{equation}
\label{eq:errs}
\begin{array}{lll}
\text{Err}_1&=&\displaystyle ||(T^\dagger T)^{-1}T^\dagger||_\infty||(\vec{f}_M-\vec{f})||_\infty \sum_{m=0}^M|T_m(z)|\\
&&\displaystyle+\sum_{m=M+1}^\infty |a_m|\;|T_m(z)|\\
\text{Err}_2&=&\displaystyle \mathbb{E}_\text{exp}\left[||(T^\dagger T)^{-1}T^\dagger\vec{\epsilon}||_\infty\right] \sum_{m=0}^M|T_m(z)|\;.
\end{array}
\end{equation}
Our goal is now to compute the quantities in the expressions above.
To begin, using Eq.~(\ref{eq:norm_inf_2}), we have
\begin{equation}
\label{eq:tmp1}
 ||(T^\dagger T)^{-1}T^\dagger||_\infty \leq\sqrt{N+1} ||(T^\dagger T)^{-1}T^\dagger||_2\;.
\end{equation}
Now, since $T$ is a $(N+1)\times(M+1)$ matrix, we can write it, in singular value decomposition, as 
\begin{equation}
    T=V_{(N+1)\times (N+1)}\Sigma_{(N+1)\times (M+1)} U_{(M+1)\times (M+1)}\;,
\end{equation}
where $U$ and $V$ are unitaries and $\Sigma$ has non-zero elements only on the diagonal. Omitting the size of the matrices we have
\begin{equation}
(T^\dagger T)^{-1}T^\dagger=U^{-1}(\Sigma^\dagger\Sigma)^{-1}\Sigma V^{-1}\;.
\end{equation}
Since $(\Sigma^\dagger\Sigma)^{-1}\Sigma$ has non-zero elements only on the diagonal, this constitutes a singular value decomposition for $(T^\dagger T)^{-1}T^\dagger$. Moreover, the diagonal elements of $(\Sigma^\dagger\Sigma)^{-1}\Sigma$ are the inverse of the diagonal elements of $\Sigma$. Using Eq.~(\ref{eq:sigmaA}), this implies
\begin{equation}
\label{eq:1divsigma}
||(T^\dagger T)^{-1}T^\dagger||_2=\frac{1}{\text{min}(\sigma_T)}\;.
\end{equation}
Using this result into Eq.~(\ref{eq:tmp1}), we obtain
\begin{equation}
\label{eq:tmp2}
 ||(T^\dagger T)^{-1}T^\dagger||_\infty \leq \frac{\sqrt{N+1}}{\text{min}(\sigma_T)}\;,
\end{equation}
where $\sigma_T$ are the singular values of the matrix $T$.
In order to compute $||\vec{f}-\vec{f}_M||_\infty$, we note that all evaluations of $f(z)$ and $f_M(z)$ inside its expression are within the interval $[-1,1]$. This implies that we can use Eq.~(\ref{eq:ff_M_interval}) to  deduce that
\begin{equation}
\label{eq:tmp3}
||\vec{f}-\vec{f}_M||_\infty\leq 2Q_\rho\frac{\rho^{-M}}{\rho-1}\;.
\end{equation}
The remaining two terms can be immediately bounded using Eq.~(\ref{eq:T_bound}), Eq.~(\ref{eq:bound_an}),  
and Eq.~(\ref{eq:rhoz})
as
\begin{equation}
\label{eq:tmp4}
\begin{array}{cll}
\displaystyle\sum_{m=0}^M|T_m(z)|&\leq&\displaystyle\sum_{m=0}^M\rho_z^{m}\\
&=&\displaystyle\frac{1-\rho_z^{M+1}}{1-\rho_z}\\
\displaystyle\sum_{m=M+1}^\infty |a_m|\;|T_m(z)|&\leq&\displaystyle 2 Q_\rho\sum_{m=M+1}^\infty \left(\frac{\rho_z}{\rho}\right)^m\\
&=&\displaystyle 2 Q_\rho \frac{(\rho_z/\rho)^{M+1}}{(1-\rho_z/\rho)}\;.
\end{array}
\end{equation}
Using Eqs.~(\ref{eq:tmp2},\ref{eq:tmp3},\ref{eq:tmp4}) into the first line of Eq.~(\ref{eq:errs}), we get
\begin{equation}
\label{eq:ErrErrErr1}
\text{Err}_1=\displaystyle 2 Q_\rho\left[\frac{\sqrt{N+1}}{\text{min}(\sigma_T)}\frac{\rho^{-M}(1-\rho_z^{M+1})}{(\rho-1)(1-\rho_z)}+ \frac{(\rho_z/\rho)^{M+1}}{(1-\rho_z/\rho)}\right].
\end{equation}
Note that a more conservative bound can be found in place of the first bound in Eq.~(\ref{eq:tmp4}) as
\begin{equation}
\label{eq:alt_bound}
\begin{array}{lll}
\sum_{m=0}^M|T_m(z)|&\leq&\displaystyle\sum_{m=0}^M\rho_z^{m}\leq\displaystyle (M+1)\rho_z^M\;,
\end{array}
\end{equation}
which would result in the less tight bound
\begin{equation}
\label{eq:less_tight_bound}
\text{Err}_1=\displaystyle2 Q_\rho \left(\frac{\rho_z}{\rho}\right)^M\left[\frac{(M+1)\sqrt{N+1}}{\text{min}(\sigma_T)}+\displaystyle  \frac{(\rho_z/\rho)}{(1-\rho_z/\rho)}\right],
\end{equation}
instead of Eq.~(\ref{eq:ErrErrErr1}), see \cite{Demanet}, theorem $6$. 
Reference \cite{Demanet} further proves a lower bound for $\text{min}(\sigma_T)$ in the case of equispaced points $\Lambda_r=-1+2r/N$,$r=0,\dots,N$ and the oversampling condition $N\geq 2M$. Using Eq.~(20) of \cite{Demanet} into the second equation of theorem 4 of \cite{Demanet}, the bound reads
\begin{equation}
\text{min}(\sigma_T)^2\geq\frac{1}{25}\left(\frac{N-M^2/2}{2M+1}-\frac{27\sqrt{N}}{32\pi}\right)\;,
\end{equation}
which can be inserted in Eq.~(\ref{eq:less_tight_bound}). The tightness of this bound can be further relaxed (using the second equation of theorem 3 into the second equation of theorem 4 of \cite{Demanet}) to
\begin{equation}
\text{min}(\sigma_T)^2\geq\left(\frac{2N}{125(2M+1)}\right)\;.
\end{equation}
We now compute
\begin{equation}
\label{eq:TTTERR}
\text{Err}_2=\mathbb{E}_\text{exp}\left[\left|\sum_{m=0}^M [(T^\dagger T)^{-1}T^\dagger\vec{\epsilon}]_mT_m(z)\right|\right]\;.
\end{equation}
First, we give a tight estimation of this term which can be used for numerical analysis. To do this, we just note that
\begin{equation}
\begin{array}{lll}
\text{Err}_2^2&=& \displaystyle\mathbb{E}_\text{exp}^2\left[\left|\sum_{m=0}^M [(T^\dagger T)^{-1}T^\dagger\vec{\epsilon}]_mT_m(z)\right|\right]\\
&\leq&\displaystyle\mathbb{E}_\text{exp}^2\left[\left|\sum_{j=0}^N\epsilon_j \sum_{m=0}^M [(T^\dagger T)^{-1}T^\dagger]_{m j}T_m(z)\right|\right]\\
&\leq&\displaystyle\mathbb{E}_\text{exp}\left[\left|\sum_{j=0}^N\epsilon_j \sum_{m=0}^M [(T^\dagger T)^{-1}T^\dagger]_{m j}T_m(z)\right|^2\right]\\
&\leq&\displaystyle\mathbb{E}_\text{exp}\left[\sum_{j=0}^N\epsilon^2_j \sum_{m=0}^M |[(T^\dagger T)^{-1}T^\dagger]_{m j}|^2|T_m(z)|^2\right]\\
&\leq&\displaystyle\sigma^2\sum_{j=0}^N\sum_{m=0}^M |[(T^\dagger T)^{-1}T^\dagger]_{m j}|^2|T_m(z)|^2\;,
\end{array}
\end{equation}
which leads to
\begin{equation}
\label{eq:stab_stab}
\text{Err}_2^2\leq\sigma_\text{exp}^2\sum_{j=0}^N\sum_{m=0}^M |[(T^\dagger T)^{-1}T^\dagger]_{m j}|^2|T_m(z)|^2\;.
\end{equation}
This relatively tight bound can now be directly numerically evaluated. However, we now  go back to Eq.~(\ref{eq:TTTERR}), i.e., 
\begin{equation}
\label{eq:Err2Err2}
\text{Err}_2\leq\displaystyle \mathbb{E}_\text{exp}\left[||(T^\dagger T)^{-1}T^\dagger\vec{\epsilon}||_\infty \right]\sum_{m=0}^M|T_m(z)|\;,
\end{equation}
with the intention to find an analytical bound. 
To make progress, we follow \cite{Demanet}, as done throughout the whole section. It is possible to define a projector $P=T(T^\dagger T)^{-1}T^\dagger$ onto the range of $T$ which has the property
\begin{equation}
(T^\dagger T)^{-1}T^\dagger P=(T^\dagger T)^{-1}T^\dagger\;.
\end{equation}
This means that we can write
\begin{equation}
\label{eq:Eeps0}
\begin{array}{lll}
 \mathbb{E}_\text{exp}\left[||(T^\dagger T)^{-1}T^\dagger\vec{\epsilon}||_2 \right]^2&=& \mathbb{E}_\text{exp}\left[||(T^\dagger T)^{-1}T^\dagger P\vec{\epsilon}||_2 \right]^2\\
 &\leq&K^2\cdot \mathbb{E}_\text{exp}\left[|| P\vec{\epsilon}||_2 \right]^2\\
 &\leq& K^2\cdot \mathbb{E}_\text{exp}\left[|| P\vec{\epsilon}||^2_2 \right],
 \end{array}
\end{equation}
where, in the second step, we used Eq.~(\ref{eq:normdec}), and where we defined
\begin{equation}
 K^2=     \left[||(T^\dagger T)^{-1}T^\dagger||_2\right]^2\;.
\end{equation}
We also used $\mathbb{E}_\text{exp}[X]^2\leq \mathbb{E}_\text{exp}[X^2]$ for a generic random variable $X$ (special case of Jensen's inequality), in the last step. Now, since $P$ projects onto the range $R_T$ of $T$, we can write
\begin{equation}
\label{eq:Eeps}
\renewcommand{\arraystretch}{1.3}
\begin{array}{lll}
\mathbb{E}_\text{exp}\left[|| P\vec{\epsilon}||_2 \right]^2&=& \mathbb{E}_\text{exp}\left[||\sum_{\vec{v}\in R_T}\ketbra{\vec{v}}{\vec{v}}\vec{\epsilon}||_2 \right]\\
&=& \mathbb{E}_\text{exp}\left[||\sum_{\vec{v},\vec{v}'\in R_T}\bra{\vec{\epsilon}}\ketbra{\vec{v}'}{\vec{v}'}\ketbra{\vec{v}}{\vec{v}}\vec{\epsilon}||_2 \right]\\
&=& \mathbb{E}_\text{exp}\left[||\sum_{\vec{v}\in R_T}|\braket{\vec{\epsilon}}{\vec{v}} |^2\right]\\
&=& \mathbb{E}_\text{exp}\left[||\sum_{\vec{v}\in R_T}\sum_{j=0}^N|\epsilon^j \vec{v}^j |^2\right]\\
&=&\sigma_\text{exp}^2\mathbb{E}_\text{exp}\left[||\sum_{\vec{v}\in R_T}\sum_{j=0}^N|\vec{v}^j |^2\right]\\
&=&\sigma_\text{exp}^2\mathbb{E}_\text{exp}\left[||\sum_{\vec{v}\in R_T}1\right]\\
&=&(M+1)\sigma_\text{exp}^2\;,
 \end{array}
\end{equation}
where $\vec{v}\in R_T$ are orthonormal vectors and where we used a notation borrowed from quantum mechanics to indicate vectors and their duals as kets and bras, respectively. Note that, in the last step, we assumed that $M<N$, i.e., that the number of measurements is bigger than the degree of the interpolating polynomial, as it is the regime relevant for us. In that case, since $T$ is a $(N+1)\times(M+1)$ matrix, its rank cannot be bigger than $(M+1)$, justifying the last step.\\
Using Eq.~(\ref{eq:Eeps}) into Eq.~(\ref{eq:Eeps0}), we obtain
\begin{equation}
\label{eq:Eeps2}
\mathbb{E}_\text{exp}\left[||(T^\dagger T)^{-1}T^\dagger\vec{\epsilon}||_2 \right]\leq \sigma_\text{exp}\sqrt{M+1} ||(T^\dagger T)^{-1}T^\dagger||_2\;.
\end{equation}
Now, using Eqs.~(\ref{eq:tmp1}), (\ref{eq:Eeps2}), and Eq.~(\ref{eq:1divsigma}), we obtain
\begin{equation}
\label{eq:ttt}
\begin{array}{lll}
\mathbb{E}_\text{exp}\left[||(T^\dagger T)^{-1}T^\dagger\vec{\epsilon}||_\infty \right]&\leq&\sqrt{N+1}E\left[||(T^\dagger T)^{-1}T^\dagger\vec{\epsilon}||_2 \right]\\
&\leq&\displaystyle\sigma_\text{exp}\frac{\sqrt{N+1}\sqrt{M+1}}{\text{min}(\sigma_T)}\;.
\end{array}
\end{equation}
Now, using Eq.~(\ref{eq:T_bound}), Eq.~(\ref{eq:ttt}), and the first of Eq.~(\ref{eq:tmp4}) into Eq.~(\ref{eq:Err2Err2}), we get
\begin{equation}
\begin{array}{lll}
\text{Err}_2&\leq&\displaystyle\sigma_\text{exp}\frac{\sqrt{N+1}\sqrt{M+1}}{\text{min}(\sigma_T)}\frac{1-\rho_z^{M+1}}{1-\rho_z}\;.
\end{array}
\end{equation}
Note that, using the bound in Eq.~(\ref{eq:alt_bound}) instead of the one in Eq.~(\ref{eq:tmp4}) results in the cleaner but less tight bound
\begin{equation}
\begin{array}{lll}
\text{Err}_2&\leq&\displaystyle\sigma_\text{exp}\frac{\sqrt{N+1}(M+1)^{3/2}}{\text{min}(\sigma_T)}\rho_z^M\;,
\end{array}
\end{equation}
which can be compared to the expression in \cite{Demanet}, corollary 3.

\section{Useful definitions and identities}
\label{sec:leastsquares}
Here, we present identities which are used in other sections.
\subsection{Definitions and identities on norms}
\label{sec:norms}
The vector space $\mathbb{C}^n$ can be endowed with norms such that
\begin{equation}
\label{eq:norms}
\begin{array}{lll}
\displaystyle||\vec{v}||_n&=&\displaystyle\left(\sum_i|v_i|^n\right)^{1/n}\\
\displaystyle||\vec{v}||_\infty&=&\displaystyle\max(|v_i|)\;,
\end{array}
\end{equation}
where $\vec{v}\in\mathbb{C}^n$. All the norms above are equivalent for a finite-dimensional vector space. We consider, specifically, the following properties 
\begin{equation}
\label{eq:ABC}
\begin{array}{lll}
||\vec{v}||_1&=&\displaystyle\sqrt{\left(\sum_i |v_i|\right)^2}\leq\displaystyle\sqrt{n\sum_i v_i^2}=\sqrt{n}\;||\vec{v}||_2\;,
\end{array}
\end{equation}
where to derive the second step, we used the Holder inequality
\begin{equation}
|x^\text{T}y|\leq||x||_p||y||_q\;,
\end{equation}
for $1/p+1/q=1$ (the case $p=q=2$ being the Cauchy-Schwartz inequality). We have
\begin{equation}
\begin{array}{lll}
\displaystyle\left(\sum_{i=1}^n  x_i\right)^2&=&\displaystyle\left(\sum_{i=1}^n 1\cdot x_i\right)^2=|x^\text{T}\vec{1}|^2\\
&\leq&\displaystyle ||x||_2^2||\vec{1}||_2^2=n^2\sum_{i=1}^n x_i^2\;,
\end{array}
\end{equation}
where $\vec{1}$ is a reference vector whose entries are all ones. We note that this identity has been used in  Eq.~(\ref{eq:ABC}). We also have
\begin{equation}
||\vec{v}||_\infty=\max|v_i|\leq\sum_i|v|_i=||\vec{v}||_1
\end{equation}
which, together with Eq.~(\ref{eq:ABC}) implies
\begin{equation}
\label{eq:norm_inf_2}
||\vec{v}||_\infty\leq\sqrt{n}||\vec{v}||_2\;.
\end{equation}
The definitions above regarding the vector spaces $\mathbb{C}^n$ can be used to define induced norms in the vector space of operators $A:\mathbb{C}^q\rightarrow\mathbb{C}^p$ as
\begin{equation}
\begin{array}{lll}
||A||_n&=&\displaystyle\sup_{\vec{x}\neq 0}\frac{||A \vec{x}||_n}{||\vec{x}||_n}=\sup_{||\vec{x}||_n=1}||A \vec{x}||_n\;.
\end{array}
\end{equation}
Note that it is possible to extend the previous definition to $n\rightarrow\infty$. Note that, when $||\vec{x}||_n\neq 0$, the previous definition implies
\begin{equation}
\label{eq:normdec}
||A||_n ||\vec{x}||_n \geq ||A\vec{x}||_n\;.
\end{equation}
It is interesting to explicitly compute $||A||_2$. To do this, we notice that $A$ can be represented by a $p\times q$ matrix $A_{pq}$ which, using singular value decomposition, can be written as $A_{p\times q}=V_{p\times p}\Sigma_{p\times q}U_{q\times q}$ in terms of unitary matrices $V$ and $U$ and an upper diagonal $\Sigma$. We have
\begin{equation}
\label{eq:sigmaA}
\begin{array}{lll}
||A||_2&=&\displaystyle\sup_{||\vec{x}||_2=1}||A \vec{x}||_2\\
&=&\displaystyle\sup_{||\vec{x}||_2=1}||V \Sigma \left(U \vec{x}\right)||_2\\
&=&\displaystyle\sup_{||\vec{x}'||_2=1}||V \Sigma  \vec{x}'||_2\\
&=&\displaystyle(\max{\sigma_A})\sup_{||\vec{x}'||_2=1}||V  \vec{x}'||_2\\
&=&\displaystyle\max{\sigma_A}\;,
\end{array}
\end{equation}
where $\sigma_A$ are the singular values of the matrix $A$ (square root of the eigenvalues of $A^\dagger A$).\\
Note that other choices of norms for operators are possible such as the Shatten norms 
\begin{equation}
||A||^S_n=\left(\tr{|A|^n}\right)^{1/n}\;,
\end{equation}
where $|A|=\sqrt{{A^\dagger} A}$.
For future reference, we also write here the following identities:
\begin{equation}
\label{eq:geometric_series}
\begin{array}{lll}
\displaystyle\sum_{n=0}^A r^n=\frac{1-r^{A+1}}{1-r},\;\;\;\sum_{n=A}^\infty r^n=\displaystyle\frac{r^{A}}{1-r}\;,
\end{array}
\end{equation}
with $|r|<1$ for the second identity to hold. 
\subsection{Least Squares}
\label{app:leastSquares}
We want to define the best estimate for a vector $\vec{c}\in\mathbb{C}^q$ such that its image under a linear mapping $T:\mathbb{C}^q\mapsto\mathbb{C}^p$ is as close as possible to a given vector $\vec{f}\in\mathbb{C}^p$ under the $||\cdot||_2$ norm. Equivalently, we want to find the vector $\vec{c}$ which minimizes the function
\begin{equation}
\label{eq:L_least}
L=||T\vec{c}-\vec{f}||^2_2\;.
\end{equation}
We have
\begin{equation}
\begin{array}{lll}
L&=&(T\vec{c}-\vec{f})^\dagger(T\vec{c}-\vec{f})\\
&=&\displaystyle\sum_{i,j,k}(T^*_{ij}c^*_j-f^*_i)(T_{ik}c_k-f_i)\;,
\end{array}
\end{equation}
which allows to write
\begin{equation}
\begin{array}{lll}
\partial_{c_\alpha}L&=&\displaystyle\sum_{i,j}(T^*_{ij}c^*_j-f^*_i)T_{i\alpha}\\
\partial_{c^*_\alpha}L&=&\displaystyle\sum_{i,k}T^*_{i\alpha}(T_{ik}c_k-f_i)\;.
\end{array}
\end{equation}
The equations $\partial_{c_\alpha}L=0$ and $\partial_{c^*_\alpha}L=0$ are equivalent and lead to
\begin{equation}
\label{eq:minimization_c}
T^\dagger T\vec{c}=T^\dagger \vec{f}\;.
\end{equation}
When $T^\dagger T$ is invertible, this leads to
\begin{equation}
\label{eq:minimization_c2}
\vec{c}=(T^\dagger T)^{-1}T^\dagger \vec{f}\;,
\end{equation}
which is the explicit expression for the vector $\vec{c}$  minimizing the least squares distance $L$ in Eq.~(\ref{eq:L_least}).
\end{document}